\documentclass[11pt,a4paper]{article}
\pdfoutput = 1
\usepackage{jheppub}
\usepackage{enumitem}
\usepackage{amsmath}
\usepackage{amsfonts}
\usepackage{graphicx}
\usepackage{caption}

\title{The Thermal Scalar and Random Walks in $AdS_3$ and $BTZ$}

\author[a]{Thomas G. Mertens,}
\author[a]{Henri Verschelde}
\author[b]{and Valentin I. Zakharov}
\affiliation[a]{Ghent University, Department of Physics and Astronomy\\
Krijgslaan, 281-S9, 9000 Gent, Belgium}
\affiliation[b]{ITEP, B. Cheremushkinskaya 25, Moscow, 117218 Russia,\\
Moscow Inst Phys \& Technol, Dolgoprudny, Moscow Region, 141700 Russia , \\
School of Biomedicine, Far Eastern Federal University, Sukhanova str 8, 
Vladivostok 690950 Russia
}
\emailAdd{thomas.mertens@ugent.be}
\emailAdd{henri.verschelde@ugent.be}
\emailAdd{vzakharov@itep.ru}
\abstract{We analyze near-Hagedorn thermodynamics of strings in the WZW $AdS_3$ model. We compute the thermal spectrum of all primaries and find the thermal scalar explicitly in the string spectrum using CFT twist techniques. Then we use the link to the Euclidean WZW BTZ black hole and write down the Euclidean BTZ spectrum. We give a Hamiltonian interpretation of the thermal partition function of angular orbifolds where we find a reappearance of discrete states that dominate the partition function. Using these results, we discuss the nature of the thermal scalar in the WZW BTZ model. As a slight generalization of the angular orbifolds, we discuss the $AdS_3$ string gas with a non-zero chemical potential corresponding to angular momentum around the spatial cigar. For this model as well, we determine the thermal spectrum and the Hagedorn temperature as a function of chemical potential. Finally the nature of $\alpha'$ corrections to the $AdS_3$ thermal scalar action is analyzed and we find the random walk behavior of highly excited strings in this particular $AdS_3$ background. }
\keywords{Black Holes in String Theory, Conformal Field Models in String Theory, Tachyon Condensation, Long strings}

\begin{document}

\maketitle

\section{Introduction}
\label{pathderiv}
String theory models on WZW $AdS_3$ and BTZ have been fruitful toy models to study string dynamics on non-trivial target spaces. There has been a great deal of work on this topic: see e.g. \cite{Balog:1988jb}\cite{Petropoulos:1989fc}\cite{Hwang:1990aq}\cite{Hwang:1991ana}\cite{Bars:1995mf}\cite{Bars:1999ik}\cite{de Boer:1998pp}\cite{Giveon:1998ns}\cite{Kutasov:1999xu} for early treatments. Progress on the topic was hampered by confusion about unitarity in these models. A no-ghost theorem was proven by \cite{Evans:1998qu} but the issues with the model were only fully resolved in the work of \cite{Maldacena:2000hw}. The holographic interpretation of these theories combined with their solvability have shown to be explicit tests for the AdS/CFT correspondence. It is our interest to study string thermodynamics in these backgrounds and in particular learn about the behavior of thermodynamical quantities in geometrically non-trivial spaces. Previous studies on the thermodynamics in these spaces include \cite{Berkooz:2007fe} and \cite{Lin:2007gi}. \\
It is known that flat space string theory has a limiting Hagedorn temperature \cite{Hagedorn}\cite{Atick:1988si}\cite{Horowitz:1997jc}. Near this temperature, the string gas recombines itself into one (or possibly multiple) long, highly excited string(s) \cite{Mitchell:1987hr}\cite{Mitchell:1987th}\cite{Deo:1989bv}\cite{Bowick:1989us}. The spatial form of these constituents can be thought of as random walks \cite{Barbon:2004dd}. An equivalent way of thinking about this is through the thermal scalar. This string state is in the perturbative spectrum on the thermal manifold and captures the critical thermodynamics at temperatures sufficiently close to the Hagedorn temperature. The one-loop contribution of only this state dominates the critical free energy of the entire string gas on the Lorentzian manifold. We have analyzed the analogous picture in curved spacetimes in detail in \cite{theory}. Due to the exact CFT description of Wess-Zumino-Witten models, it seems worthwhile to study this picture also in $AdS_3$ and BTZ spacetimes. The $AdS_3$ spacetime is a mild generalization of flat space, since the thermal time circle still is topologically stable for winding strings. The BTZ black hole on the other hand presents a temporal cigar geometry where strings can simply slip off. We have already analyzed similar geometries in \cite{Mertens:2013zya} where we studied the string gas in Rindler spacetime. The Hagedorn transition in $AdS$ spacetime has also been related to the confinement/deconfinement phase transition in the dual gauge theory (see for instance \cite{Sundborg:1999ue}\cite{Aharony:2003sx}).\\
Let us first give a brief summary of the results of \cite{theory}. There we have given a path integral picture of the thermal scalar in general curved backgrounds (following the derivation of \cite{Kruczenski:2005pj}). The strategy is to perform a $\tau \to -1/\tau$ modular transformation on the torus path integral on the modular strip. After this, the large $\tau_2$ limit is taken. This reduces the string path integral to a particle path integral given by:
\begin{equation}
\label{pint} 
Z_p = 2\int_0^\infty \frac{d \tau_2}{2\tau_2} \int \left[ \mathcal{D}X \right] \sqrt{\prod_{t} \det G_{ij}} \exp - S_p(X) 
\end{equation}
where 
\begin{equation}
\label{act}
S_p = \frac{1}{4\pi \alpha'}\left[ \beta^2 \int_0^{\tau_2} dt G_{00} + \int_0^{\tau_2} dt G_{ij} \partial_t X^i \partial_t X^j\right].
\end{equation}
This realizes the random walk picture of the thermal scalar directly in the path integral language. The full string partition function has been reduced to a partition function for a non-relativistic particle moving on the purely spatial submanifold. The time evolution of the particle in its random walk is identified with the spatial form of the long highly excited string. The free energy of a gas of strings can then be identified with the single string partition function as \cite{Polchinski:1985zf}
\begin{equation}
F = -\frac{1}{\beta} Z_p,
\end{equation}
which in the critical regime is approximated by expression (\ref{pint}). Unfortunately there are corrections to the above particle action. We saw that these can be deduced from the field theory of the thermal scalar. The correction terms can be subdivided in three different categories:
\begin{itemize}
\item{There is a correction term coming from the mass of the flat space tachyon. It has the following form
\begin{equation}
\Delta S = -\frac{\beta_{H}^2\tau_2}{4\pi\alpha'}.
\end{equation}
For bosonic strings $\beta_{H}^2 = 16\pi^2\alpha'$ and for type II superstrings $\beta_{H}^2 = 8\pi^2\alpha'$.}
\item{Secondly there is a correction coming from the $G_{00}$ metric component as explained in \cite{theory}: 
\begin{equation}
\label{corr}
\Delta S = \frac{1}{4\pi\alpha'}\int_{0}^{\tau_2}dt 4\pi^2\alpha'^2 \left(-\frac{3}{16}\frac{G^{ij}\partial_iG_{00}\partial_jG_{00}}{G_{00}^2} + \frac{\nabla^2 G_{00}}{4G_{00}}\right).
\end{equation}
}
\item{Finally we could have order-by-order $\alpha'$ correction terms of the lowest order $\alpha'$ thermal scalar action.}
\end{itemize}
We also presented a simple extension to include a background NS-NS field. 
The particle action (\ref{act}) has the following extra contribution
\begin{equation}
\label{KRexten2}
S_{extra} = \mp i\frac{\beta}{2\pi\alpha'}\int_{0}^{\tau_2}dt B_{0i}(X)\partial_{t}X^{i},
\end{equation}
which represents a minimal coupling of a point particle to a vector potential $A_{i} = B_{0i}$. \\

\begin{figure}[h]
\centering
\includegraphics[width=10cm]{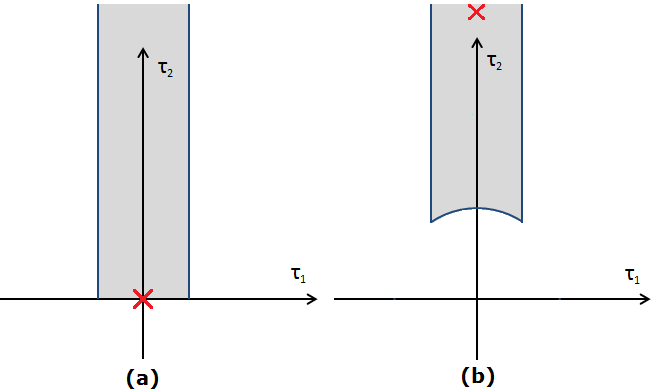}
\caption{(a) Critical thermodynamics from the strip domain. The cross depicts the critical limit. This is the approach used in \cite{Berkooz:2007fe} and \cite{Lin:2007gi}. (b) Critical thermodynamics from the fundamental domain. In this picture, the critical regime is determined by a perturbative state in the thermal spectrum: the thermal scalar. One finds this regime by taking $\tau_2 \to \infty$.}
\label{approach}
\end{figure}

\noindent Our primary goal in this paper is to study the above picture for the specific case of the $AdS_3$ and BTZ WZW models. Our study focuses on the approach to the critical thermodynamics through the thermal manifold. The different canonical approaches are depicted in figure \ref{approach}.

\noindent Our objectives can be summarized as follows.
\begin{itemize}
\item[I.]{Determine the thermal spectrum on the $AdS_3$ and $BTZ$ model.}
\item[II.]{Analyze the critical near-Hagedorn behavior of thermodynamical quantities.}
\item[III.]{Find out to what extent field theory results reproduce this.}
\item[IV.]{Determine the complete random walk picture of a highly excited string gas in these spacetimes.}
\end{itemize}
The paper is organized as follows. In section \ref{WZWmodel} we present an exact computation of the thermal string spectrum on $AdS_3$. We utilize twisting techniques on the worldsheet to determine the states. We then search for the thermal scalar in the resulting spectrum. In section \ref{BTZsection} we make the transition to the WZW BTZ thermal black hole and study the same questions. After that, in section \ref{conical} we take a look at conical orbifolds, due to their relevance for thermodynamics. Section \ref{chemical} treats a slight generalization of the results of section \ref{WZWmodel} in which we include an angular chemical potential for the string gas. Finally in section \ref{fieldtheory}, we look at the naive lowest order $\alpha'$ thermal scalar action and study to what extent it captures the critical behavior of the thermal scalar. We end with some conclusions in section \ref{conclusion}. We present background material on WZW models and more elaborate calculations in the appendices.

\section{Exact $AdS_3$ WZW model}
\label{WZWmodel}
\subsection{Random walks in $AdS_3$}
\label{randwalksads}
We consider $AdS_3$ spacetime with a non-zero Kalb-Ramond background field. The metric is given in the global coordinates of the $AdS$ spacetime as:
\begin{equation}
ds^2 = \alpha'k\left(-\cosh(\rho)^2dt^2+d\rho^2+\sinh(\rho)^2d\phi^2\right)
\end{equation}
where $\phi \sim \phi + 2\pi$ and the space includes a NS-NS two-form:
\begin{equation}
B = -\alpha'k\sinh(\rho)^2dt \wedge d\phi.
\end{equation}
The crucial aspect of this string background is that it is an exact (up to all orders in $\alpha'$) CFT because it can be written as a Wess-Zumino-Witten (WZW) model: it is the $SL(2,\mathbb{R})$ WZW model. This causes the string spectrum to be composed of irreducible representations of the affine Lie algebra underlying the WZW model. The string spectrum in this background was determined in \cite{Maldacena:2000hw}. Performing a Wick rotation on this model, yields another WZW model: the $SL(2,\mathbb{C})/SU(2)$ model. We identify the Euclidean time coordinate as $\tau \sim \tau + \beta$. Note that this time coordinate is dimensionless in these conventions. For more information regarding these WZW models and their link through analytic continuation, we refer the reader to Appendix \ref{WZWapp}.
Moreover, the full string path integral on the thermal $AdS_3$ manifold can be exactly computed and is given by \cite{Maldacena:2000kv}
\begin{align}
Z = \frac{\beta\sqrt{k-2}}{8\pi}\int_{E}\frac{d\tau_1d\tau_2}{\tau_2^{\frac{3}{2}}}&e^{4\pi\tau_2\left(1-\frac{1}{4(k-2)}\right)}\sum_{h,\overline{h}}D(h,\overline{h})e^{2\pi i \tau(h+\overline{h})} \nonumber \\
&\times \sum_{m=1}^{+\infty}e^{-\frac{(k-2)m^2\beta^2}{4\pi\tau_2}}\frac{\left|\eta(\tau)\right|^4}{\left|\vartheta_{1}\left(-\frac{im\beta}{2\pi},\tau\right)\right|^2},
\end{align}
where $E$ denotes the modular strip region and the sum over $h$ and $\bar{h}$ corresponds to the internal CFT, required to make the space a valid string background. The Hagedorn temperature in this background was determined directly from this partition function in \cite{Berkooz:2007fe}\cite{Lin:2007gi} and was found to be
\begin{equation}
\label{hag}
\beta_{H}^{2} = \frac{4\pi^2}{k}\left(4-\frac{1}{k-2}\right).
\end{equation}
The results from section \ref{pathderiv} predict that the critical behavior of the free energy of the string gas is determined by 
\begin{equation}
F = -\frac{1}{\beta}\sum_{w=\pm 1}\int_{0}^{+\infty}\frac{d\tau_2}{2\tau_2}\int\left[\mathcal{D}X\right]\exp\left(-S_p\right)
\end{equation}
where $S_p$ is given by
\begin{align}
\label{randwalk}
S_p = \frac{k}{4\pi}\int_{0}^{\tau_2}dt\left[(\partial_t \rho)^2 + (\beta^2\cosh(\rho)^2-\beta_{H,flat}^2)+\sinh(\rho)^2(\partial_t\phi)^2 + 2w \frac{\beta}{2\pi\alpha'}\sinh(\rho)^2\partial_t\phi \right.\nonumber \\
\left. + \frac{4\pi^2}{k^2}\left\{\frac{3}{4} + \frac{1}{4\cosh(\rho)^2}\right\}\right].
\end{align}
This represents a particle moving in a two-dimensional curved space in a potential determined by $\cosh(\rho)^2$ and interacting with a specific electromagnetic field. The first two corrections to the random walk discussed in the previous section have already been included. Firstly, the tachyon mass correction was included, e.g. for the bosonic string $\alpha'k\beta_{H,flat}^2 = 16\pi^2\alpha'$. Secondly, the extra final term in the action comes from the $G_{00}$ metric component and represents a mild potential that slightly damps paths that come close to the origin $\rho=0$.\footnote{In \cite{theory} (appendix D) we discussed that another term should be incorporated in the particle action when considering non-zero NS-NS flux. However, one readily checks that this term vanishes in this case using the explicit form of the metric and NS-NS field.}  We remark that we have not considered possible $\alpha'$ corrections to the thermal scalar action so there might be more contributions to the particle action (\ref{randwalk}) that have been neglected. We will turn to this problem next. 
 
\subsection{General analysis of $\alpha'$ corrections}
\begin{figure}[h!!!!]
\centering
\includegraphics[width=11cm]{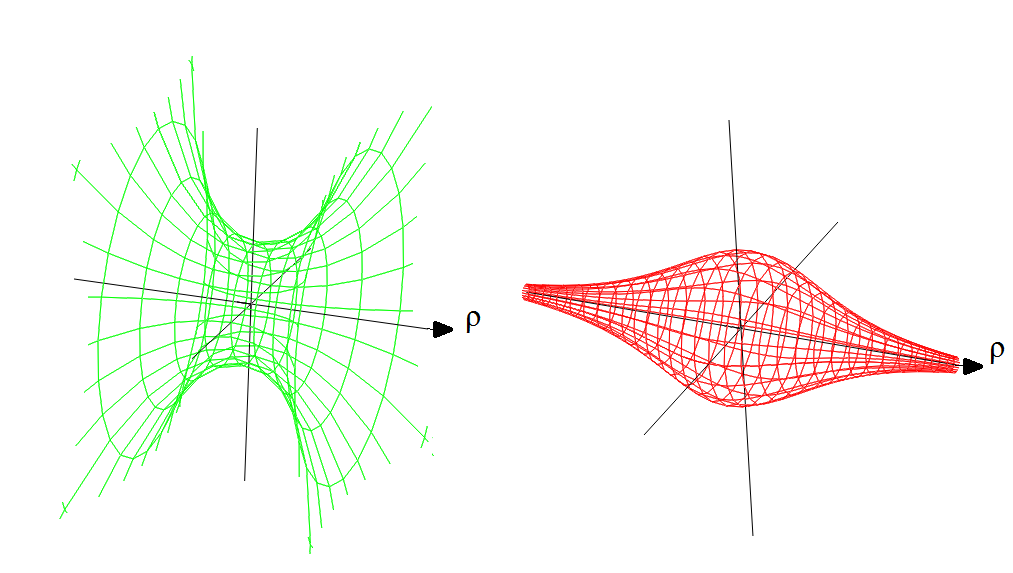}
\caption{Left figure: size of the thermal circle in $AdS$ space as a function of radial distance. The center of the figure is at $\rho=0$. Right figure: size of the thermal circle in the T-dual of $AdS$ space as a function of radial distance.}
\label{thermcircle}
\end{figure}

Now we ask whether the above random walk action is $\alpha'$-exact. Let us therefore first analyze possible corrections in general and see whether there is at least a regime in which they can be neglected. We know from previous work \cite{Mertens:2013zya} that this is not the case for the Euclidean Rindler string. In this section only, we rescale the coordinates such that they are not dimensionless.\footnote{$k\alpha'\tau^2 \to \tau^2$, $k\alpha'\rho^2 \to \rho^2$ and $k\alpha'\phi^2 \to \phi^2$. The $AdS$ length has been introduced as $l^2=k\alpha'$. The Euclidean time coordinate is obtained as $t \to i\tau$.} The Euclidean metric and Kalb-Ramond field are
\begin{equation}
ds^2 = \cosh(\rho/l)^2d\tau^2+d\rho^2+\sinh(\rho/l)^2d\phi^2
\end{equation}
and
\begin{equation}
B = -i\sinh(\rho/l)^2d\tau \wedge d\phi.
\end{equation}
We identify $\tau \sim \tau + \beta$ and note that this differs from the temperature we have introduced earlier by a factor of $l$, the $AdS$ length. The thermal scalar action consists of diffeomorphism invariants constructed with T-dual quantities. 
The T-dual Ricci tensor components are given by
\begin{equation}
\tilde{R}^{00} = 0, \quad  \tilde{R}^{\rho\rho} = \frac{2}{l^2\cosh(\rho/l)^2}, \quad \tilde{R}^{\phi\phi} = \frac{2}{l^2\sinh(\rho/l)^2},
\end{equation}
and all components with mixed indices vanish. The T-dual Ricci scalar is given by
\begin{equation}
\tilde{R} = \frac{4}{l^2\cosh(\rho/l)^2},
\end{equation}
and the T-dual dilaton has the expression:
\begin{equation}
\partial_{\rho}\tilde{\Phi} = - \frac{1}{l} \tanh(\rho/l).
\end{equation}
A peculiarity of this background is that $\tilde{B}_{\mu\nu} = 0$. Some possible terms that could appear in the thermal scalar action are given by:
\begin{align}
m^2 TT^{*} &= -\frac{4}{\alpha'}TT^{*}, \quad \text{bosonic}\quad \text{or} \quad m^2 TT^{*} = -\frac{2}{\alpha'}TT^{*}, \quad \text{type II},\\
\tilde{G}^{\mu\nu}\partial_{\mu}T\partial_{\nu}T^{*} &= \frac{\beta^2}{4\pi^2\alpha'^2}TT^{*} + \partial_\rho T \partial_\rho T^{*} - \frac{\beta}{2\pi\alpha'}(T\partial_\phi T^{*}- T^{*}\partial_\phi T) \nonumber \\
&\quad + \frac{1}{\sinh(\rho/l)^2}\partial_\phi T \partial_\phi T^{*},\\
\tilde{R} TT^{*} &= \frac{4}{\cosh(\rho/l)^2}\frac{1}{l^2} TT^{*},\\
\label{higher1}
\alpha'\tilde{R}^{\mu\nu}\partial_\mu T \partial_\nu T^{*} &=  \frac{2}{\cosh(\rho/l)^2}\frac{\alpha'}{l^2}\partial_\rho T \partial_\rho T^{*} +\frac{2}{\sinh(\rho/l)^2}\frac{\alpha'}{l^2}\partial_\phi T \partial_\phi T^{*},
\end{align}
and
\begin{align}
\partial_\mu \tilde{\Phi} \partial^{\mu} \tilde{\Phi} TT^{*} &= \frac{1}{l^2}\tanh(\rho/l)^2 TT^{*}, \\
\label{higher2}
\alpha'\partial^{\mu}\tilde{\Phi} \partial^{\nu} \tilde{\Phi} \partial_{\mu}T\partial_{\nu}T^{*} &= \frac{\alpha'}{l^2}\tanh(\rho/l)^2\partial_\rho T \partial_\rho T^{*}. 
\end{align}
In this case, all terms originating from higher order corrections such as (\ref{higher1}) and (\ref{higher2}) are suppressed as $\alpha'/l^2$. This ratio is suppressed since the T-dual geometry is only slowly varying with $\rho$. This is in sharp contrast to the black hole case, where a curvature singularity in the T-dual spaces sets in at the event horizon \cite{Mertens:2013zya}. The thermal circles for $AdS_3$ and its T-dual are sketched in figure \ref{thermcircle}. 
In \cite{Mertens:2013zya} we argued that in general two conditions need to be met in order to suppress higher $\alpha'$ terms. The first condition is the one discussed above. The second requirement is that the temperature is of order the string scale. This condition is not met for black holes where the temperature equals the Hawking temperature. In our case, this condition is not required but this is again a peculiarity of this specific model. Considering for instance the same background but with the Kalb-Ramond field turned off, one readily finds that $\beta^2/l^2 \ll 1$ is also necessary to suppress all higher $\alpha'$ corrections. \\
From these generalities we conclude, that we can find a regime (large $l$ in string units and (although not necessary here) string-scale temperatures) where we can neglect all possible $\alpha'$ corrections, if they are present in the first place. We will come back to this issue further on.

\subsection{Thermal string spectrum from a $SL(2,\mathbb{R})$ point of view}
\label{spectrumsection}
Let us now try to answer a different question. In order for the starting point of our story to be valid, we need to find a winding tachyon in the string spectrum on the thermal manifold that becomes massless at the Hagedorn temperature. The spectrum in both the Lorentzian and the Euclidean model are known, so all that is left to do is to compactify the imaginary time direction and see how the spectrum changes. In \cite{Berkooz:2007fe} a mini-superspace analysis is used to find the thermal tachyon. The authors of \cite{Argurio:2000tb}\cite{Rangamani:2007fz}\cite{Martinec:2001cf} developed methods to find the spectrum in general orbifold CFTs by introducing twist operators \cite{Dixon:1986qv}. We will show that this approach can be succesfully applied in this context. \\
We first cite the results for the string spectrum on the Lorentzian signature $SL(2,\mathbb{R})$ model \cite{Maldacena:2000hw}. \\
The spectrum in (Lorentzian) $AdS_3$ is built on two types of $SL(2,\mathbb{R})$ representations. In what follows, the quantum number $m$ is the eigenvalue of the $J^{3}_{0}$ operator in the zero-mode Lie algebra. 
\begin{itemize}
\item{$\mathcal{D}_{j}$ where $\frac{1}{2} < j < \frac{k-1}{2}$. These are the so-called principal discrete representations. These can be further classified in lowest weight principal discrete representations given by
\begin{equation}
\mathcal{D}_{j}^{+} = \left\{\left|j,m\right\rangle, m = j, j+1, j+2, ...\right\}, 
\end{equation}
and highest weight principal discrete representations
\begin{equation}
\mathcal{D}_{j}^{-} = \left\{\left|j,m\right\rangle, m = -j, -j-1, -j-2, ...\right\}.
\end{equation}
}
\item{$\mathcal{C}_{j,\alpha}$ where $j = \frac{1}{2} + is$ ($s \in \mathbb{R}$) and $0 \leq \alpha < 1$. These are the continuous representations. In this case the representations are given by
\begin{equation}
\mathcal{C}_{j,\alpha} = \left\{\left|j,\alpha,m\right\rangle , m = \alpha, \alpha \pm 1, \alpha \pm 2, ...\right\}.
\end{equation}
}
\end{itemize}
CFT primaries are labeled by these quantum numbers. Descendants can then be constructed by applying the affine algebra raising operators. The primary and its descendants in one $SL(2,\mathbb{R})$ represenation of the zero-mode algebra together with all their affine algebra descendants form a $\widehat{SL(2,\mathbb{R})}$ representation. The full affine algebra has an automorphism called \emph{spectral flow} given by:
\begin{equation}
\tilde{J}^{3}_{n} = J^{3}_{n} - \frac{k}{2}w\delta_{n,0}, \quad \tilde{J}^{\pm}_n = J^{+}_{n\pm w},
\end{equation}
which preserves the commutation relations and maps one representation into another. The amount of spectral flow is labeled by an integer $w \in \mathbb{Z}$. The Virasoro operators associated to the $\tilde{J}^{a}_n$ are given by the Sugawara construction as
\begin{equation}
\tilde{L}_n = L_n + wJ^{3}_n - \frac{k}{4}w^2 \delta_{n,0}.
\end{equation}
For compact groups this does not lead to new representations whereas for non-compact groups (such as the one we have here) this results in new representations that should be incorporated. For more details, we refer the reader to \cite{Maldacena:2000hw}. The strategy is to start with a representation of the $\tilde{J}^{a}_n$ algebra whose primary satisfies
\begin{align}
\tilde{J}^{\pm}_n \left|\tilde{j},\tilde{m}\right\rangle &= 0, \quad \tilde{J}^{3}_n\left|\tilde{j},\tilde{m}\right\rangle = 0, \quad n \geq 1, \\
&\tilde{J}^{3}_0 \left|\tilde{j},\tilde{m}\right\rangle = \tilde{m}\left|\tilde{j},\tilde{m}\right\rangle.
\end{align}
One then finds the conformal weight of this state by applying $L_0$. The conformal dimensions of the primaries of Lorentzian $AdS_3$ are given by
\begin{equation}
h^{w}_{jm\overline{m}} = -\frac{\tilde{j}(\tilde{j}-1)}{k-2}- \tilde{m}w - \frac{kw^2}{4} + h_{int},
\end{equation}
where $w$ denotes the spectral flow used to generate all primary states and $h_{int}$ is the conformal weight of an internal CFT needed to obtain the right central charge. An analogous expression holds for the antiholomorphic components. \\
The Lorentzian $AdS_3$ metric is given in global coordinates by
\begin{equation}
ds^2 = \alpha'k\left(-\cosh(\rho)^2dt^2+d\rho^2+\sinh(\rho)^2d\phi^2\right)
\end{equation}
with the periodic identification $\phi \sim \phi + 2\pi$. To obtain the thermal manifold we should also impose periodicity in imaginary time: $t \sim t + i\beta$. We will use the Lorentzian signature vertex operators and impose periodicity in imaginary time on these. The reader might feel a bit uneasy about this, but we will nonetheless obtain the expected result. Moreover, in the next section we will rederive this from a fully Euclidean point of view as well. The Lorentzian spectrum consists of normal affine $\widehat{SL(2,\mathbb{R})}$ representations and the spectral flowed ones. The latter can also be obtained by twisting the CFT by the twist operator associated with the $\phi$ identification \cite{Argurio:2000tb}. So we now utilize this method and twist the CFT in \emph{both} the $\phi$ and the $t$ direction. \\
The operators generating (spacetime) time translations and angular rotations are given by (as proven in appendix \ref{WZWapp})
\begin{eqnarray}
Q_{t} = J^{3}_{0} - \overline{J}^{3}_{0}, \\
Q_{\phi} = J^{3}_{0} + \overline{J}^{3}_{0}.
\end{eqnarray}
It is important to note that both are generated by the same set of operators.\\
We demand that states respect the periodicity of spacetime, so translation by $2\pi$ in the angular direction should reproduce the same state. In other words
\begin{equation}
\exp\left(i2\pi Q_{\phi}\right) = 1.
\end{equation}
This implies that 
\begin{equation}
\label{con1}
m + \overline{m} \in \mathbb{Z}.
\end{equation}
Analogously for the imaginary time periodicity we get
\begin{equation}
\label{con2}
\frac{i\beta}{2\pi}(m-\overline{m}) \in \mathbb{Z}.
\end{equation}
String states need to respect these conditions. However, we know that a consistent string theory should also include twisted states, so we are not finished yet. We can do this by constructing local operators that implement the above restrictions and hence twist the preceding (still inconsistent!) CFT \cite{Argurio:2000tb}. We remark that we determined the above restrictions for untwisted sectors only. The twisted sector states could have different restrictions imposed on their quantum numbers. We will come back to this later.\\ 
To proceed, we use a parafermionic representation of the current algebra by diagonalizing the $J^3$ operator: 
\begin{equation}
J^{3} = -\sqrt{\frac{k}{2}}\partial{X}, \quad J^{\pm} = \psi^{\pm} e^{\pm\sqrt{2/k}X}
\end{equation}
where the $X$ and $\psi^{\pm}$ satisfy
\begin{equation}
X(z)X(w) \sim -\ln(z-w), \quad \psi^{+}(z)\psi^{-}(w) \sim \frac{k}{(z-w)^{2+2/k}}, \quad \psi^{\pm}(z)\psi^{\mp}(w) \sim 0,
\end{equation}
and with analogous relations for the antiholomorphic copy of the current algebra. The (untwisted) primaries are then represented as
\begin{equation}
\Phi_{jm\overline{m}} = \Psi_{jm\overline{m}}e^{\sqrt{\frac{2}{k}}(mX + \overline{m}\overline{X})},
\end{equation}
where the $\Psi_{jm\overline{m}}$ are uncharged under $J^3$ and $\overline{J}^{3}$.\\
Now we construct the twist operators and demand mutual locality of the OPEs (this will correspond to projecting onto invariant states) and we demand closure of the OPE (this corresponds to inclusion of twisted sectors). In analogy with \cite{Argurio:2000tb}\cite{Rangamani:2007fz}\cite{Martinec:2001cf}, the twist operators are given by
\begin{align}
t^{\phi}_{w} &= e^{\sqrt{\frac{k}{2}}w(X-\overline{X})}, \\
t^{t}_{p} &= e^{\sqrt{\frac{k}{2}}\frac{i\beta p}{2\pi}(X+\overline{X})},
\end{align}
where $w$ denotes the twisting in the $\phi$ direction (this is the same as the spectral flow parameter $w$ used in \cite{Maldacena:2000hw}) and $p$ denotes the twisting in the imaginary time direction. Let us consider the OPE of a twist operator and an untwisted primary. For a boson field satisfying $X(z)X(w) \sim -\ln(z-w)$ the following OPE holds
\begin{equation}
e^{\alpha X(z)}e^{\beta X(w)} \sim (z-w)^{-\alpha\beta}e^{(\alpha+\beta)X(w)} + (z-w)^{-\alpha\beta+1}\alpha :\partial X (w) e^{(\alpha+\beta)X(w)}: + \hdots
\end{equation}
and higher powers of $z-w$ as dictated by the Taylor series expansion of $e^{\alpha X(z)}$ around $z=w$. This OPE holds for general complex values of both $\alpha$ and $\beta$. Obviously, depending on these values, the number of singular terms varies. Important to note is that all powers of $z-w$ are integrally shifted from $\alpha\beta$ and so it is this combination that provides restrictions on the quantum numbers as we now show. The OPEs of the twist operators with the untwisted primaries are given by:
\begin{align}
\label{eq1}
t^{\phi}_{w}(z,\bar{z})\Phi_{jm\overline{m}}(w,\bar{w})&\sim (z-w)^{-wm}(\bar{z}-\bar{w})^{w\overline{m}}\Psi_{jm\overline{m}}e^{\sqrt{\frac{2}{k}}\left[(m+w\frac{k}{2})X + (\overline{m}-w\frac{k}{2})\overline{X}\right]} + \hdots, \\
\label{eq2}
t^{t}_{p}(z,\bar{z})\Phi_{jm\overline{m}}(w,\bar{w})&\sim (z-w)^{-\frac{i\beta p}{2\pi}m}(\bar{z}-\bar{w})^{-\frac{i\beta p}{2\pi}\overline{m}}\Psi_{jm\overline{m}}e^{\sqrt{\frac{2}{k}}\left[(m+\frac{i\beta p}{2\pi}\frac{k}{2})X + (\overline{m}+\frac{i\beta p}{2\pi}\frac{k}{2})\overline{X}\right]} + \hdots.
\end{align}
The OPE of a twist operator with a primary generates new operators that must be included in the operator spectrum to close the OPE. These are the twisted primaries and these can be written as
\begin{align}
:t^{\phi}_{w}(z,\bar{z})t^{t}_{p}(z,\bar{z})\Phi_{jm\overline{m}}(z,\bar{z}):
\end{align}
for general $w$ and $p$. The most general primary vertex operator is then
\begin{equation}
\Phi^{wp}_{jm\overline{m}} = \Psi_{jm\overline{m}}e^{\sqrt{\frac{2}{k}}\left[(m+\frac{k}{2}w+\frac{k}{2}\frac{i\beta}{2\pi}p)X + (\overline{m}-\frac{k}{2}w+\frac{k}{2}\frac{i\beta}{2\pi}p)\overline{X}\right]}
\end{equation}
with conformal weight
\begin{equation}
h^{wp}_{jm\overline{m}} = -\frac{j(j-1)}{k-2} + \frac{m^2}{k} - \frac{\left(m+\frac{k}{2}w + \frac{kp}{2}\frac{i\beta}{2\pi}\right)^2}{k} + h_{int}.
\end{equation}
Next we need to determine the range of the quantum numbers $m$ and $\overline{m}$ and it is at this point that an important subtlety sets in. The conserved charges determined above are in fact in general not correct for the twisted sector states. This is related to an ambiguity of the Noether current: adding a divergence of an antisymmetric tensor gives the same conservation equation, although the conserved charge is different in topologically non-trivial sectors. 
To avoid branch cuts (mutual locality), the $m$ and $\overline{m}$ quantum numbers of the untwisted primaries are restricted from (\ref{eq1}) and (\ref{eq2}) by the following two conditions:
\begin{eqnarray}
m+\overline{m} \in \mathbb{Z}, \\
\frac{i\beta}{2\pi}\left(m-\overline{m}\right) \in \mathbb{Z},
\end{eqnarray}
which are indeed the conditions required for projecting on invariant states which we wrote down in equations (\ref{con1}) and (\ref{con2}). One can also (too naively) apply the twist operators to already twisted vertex operators. This would give us then\footnote{We denoted the $J^{3}_0$ eigenvalue as $m_J$ to distinguish it with the $m$ quantum number. These are not equal for twisted sectors. Analogous comments hold for the antiholomorphic sector.}
\begin{eqnarray}
\label{naive}
m_J+\overline{m}_J \in \mathbb{Z}, \\
\label{naive2}
\frac{i\beta}{2\pi}\left(m_J-\overline{m}_J\right) \in \mathbb{Z},
\end{eqnarray}
with 
\begin{align}
\label{relation}
m_J = m + \frac{kw}{2} + \frac{i\beta k p}{4\pi}, \\
\label{relation2}
\overline{m}_J = \overline{m} - \frac{kw}{2} + \frac{i\beta k p}{4\pi}.
\end{align}
This is however wrong: it is known that the conserved charges can be different in the twisted sectors. We will demonstrate that this is indeed the case here by using two different arguments. This failure of the twist operator construction was also previously observed in a different context in \cite{Parsons:2009si} where extremal Lorentzian BTZ black holes were considered. \\
As a first argument, let us consider the level-matching condition: $L_0 - \bar{L}_0 \in \mathbb{Z}$. It is known from studies in the past concerning heterotic 4d black holes \cite{Giddings:1993wn} and rotating WZW BTZ black holes \cite{Natsuume:1996ij} that this condition provides us with the correct projection operator \cite{Hemming:2002kd}. The general primaries have weights given by:
\begin{align}
\label{weig1}
h^{wp}_{jm\overline{m}} = -\frac{j(j-1)}{k-2} + \frac{m^2}{k} - \frac{\left(m+\frac{k}{2}w + \frac{kp}{2}\frac{i\beta}{2\pi}\right)^2}{k} + h_{int}, \\
\label{weig2}
\bar{h}^{wp}_{jm\overline{m}} = -\frac{j(j-1)}{k-2} + \frac{\overline{m}^2}{k} - \frac{\left(\overline{m}-\frac{k}{2}w + \frac{kp}{2}\frac{i\beta}{2\pi}\right)^2}{k} + \bar{h}_{int},
\end{align}
and we assume that the internal CFT is on its own level-matched: $h_{int} - \bar{h}_{int} \in \mathbb{Z}$. This gives for the level-matching condition:
\begin{align}
h - \bar{h} &= -w(m+\overline{m}) -\frac{pi\beta}{2\pi}(m-\overline{m}) - \frac{ki\beta}{2\pi}pw \in \mathbb{Z} \\
\label{levelmatch}
&= -w(m_{J}+\overline{m}_{J}) -\frac{pi\beta}{2\pi}(m_{J}-\overline{m}_{J}) + \frac{ki\beta}{2\pi}pw \in \mathbb{Z}
\end{align}
and it is clear that this is in contradiction with (\ref{naive}) and (\ref{naive2}) unless $pw=0$ which is impossible to satisfy in general since interactions of states having $pw=0$ could in principle create states that have $pw\neq0$.
Noether ambiguities can spoil the projection condition by additional terms only present in twisted sectors. Let us keep an open mind and consider the general deformation (in a non-technical sense) of the conserved charges:
\begin{align}
Q_{\phi} &= J^3_0 + \overline{J}^{3}_0 + f(w,p), \\
Q_{t} &= J^3_0 - \overline{J}^{3}_0 + g(w,p),
\end{align}
with $f$ and $g$ functions of the twists with the property that $f(0,0) = g(0,0)=0$. Now we will determine these functions using what we already know. 
Consider first the sector $p=0$ which coincides, up to the projection onto invariant states, with the non-thermal Lorentzian $AdS_3$. It was shown in \cite{Maldacena:2000kv}\cite{Maldacena:2000hw} that the energy and angular momentum really are measured by $J^3_0 \mp \overline{J}^3_0$. Applying this result here, we obtain $f(w,p) = f(p)$ and $g(w,p)=g(p)$.\footnote{Note that we neglect the possibility that $f$ (or $g$) include `mixing' terms such as $pw$. Such terms are unnatural when considering the resulting conformal weights and we will find agreement with other arguments further on.} When considering equation (\ref{levelmatch}), it is clear that choosing $g=0$ and $f=-\frac{ki\beta}{2\pi}p$ satisfies the level-matching condition which reduces to
\begin{equation}
-w\left(m_{J}+\overline{m}_{J}-\frac{ki\beta}{2\pi}p\right) \in \mathbb{Z},
\end{equation}
and which is different than the requirement (\ref{naive}). Note that this line of thought is not a rigorous derivation, but merely demonstrates that the above solution is the most natural one to choose.\\
One can also use a different argument to demonstrate this result by taking the flat space $k \to \infty$ limit defined by keeping $k\rho^2$ fixed. This argument was used by the authors of \cite{Rangamani:2007fz} to obtain the correct spacetime energy of twisted states on the Lorentzian BTZ manifold. The generator of $\phi$-translations was naively identified as 
\begin{equation}
Q_{\phi} = J^{3}_{0} + \overline{J}^{3}_{0}.
\end{equation}
The Lorentzian currents are determined in appendix \ref{WZWapp} and the relevant components are given by
\begin{align}
J^{3}(z) &= i k\cosh(\rho)^2 \partial t - ik\sinh(\rho)^2 \partial \phi, \\
\overline{J}^{3}(\bar{z}) &=  -ik\cosh(\rho)^2 \bar{\partial} t - ik\sinh(\rho)^2 \bar{\partial} \phi.
\end{align}
In the large $k$ limit (keeping $k\rho^2$ fixed) these currents become
\begin{align}
J^{3}(z) &= i k\partial t - ik\rho^2 \partial \phi, \\
\overline{J}^{3}(\bar{z}) &=  -ik\bar{\partial} t - ik\rho^2 \bar{\partial} \phi.
\end{align}
We can thus rewrite the conserved $\phi$ charge as
\begin{equation}
J^{3}_0 + \overline{J}^{3}_0 = \oint dz J^{3}(z) - \oint d\bar{z}\overline{J}^{3}(\bar{z}) = ik\oint dz \partial t + ik\oint d\bar{z} \bar{\partial} t + ik \oint dz \rho^2 \partial \phi - ik\oint d\bar{z} \rho^2\bar{\partial}\phi,
\end{equation}
where $1/(2\pi i)$ factors are left implicit in the contour integrals. The final two terms are giving us the angular rotation we seek. The first two terms however are not what we want, but these are dominant in the large $k$ limit. These can be explicitly written as
\begin{equation}
\frac{ik\beta p}{2\pi},
\end{equation}
which corresponds to a winding contribution. The generator of angular rotations can then be obtained by subtracting this part as
\begin{equation}
Q_{\phi} = J^{3}_0 + \overline{J}^{3}_0 - \frac{ik\beta p}{2\pi}.
\end{equation}
Note that for $Q_t$ on the other hand, the winding contribution is subleading in the large $k$ limit and can be neglected \cite{Rangamani:2007fz}. This agrees with the expression we determined above using the level-matching argument.\\
We conclude that the projection conditions are 
\begin{align}
\label{cond1}
m_J+\overline{m}_J - \frac{ik\beta p}{2\pi} &\in \mathbb{Z}, \\
\label{cond2}
\frac{i\beta}{2\pi}\left(m_J-\overline{m}_J\right) &\in \mathbb{Z}.
\end{align}
The two conditions (\ref{cond1}) and (\ref{cond2}) can be solved and this gives together with (\ref{relation}) and (\ref{relation2}):
\begin{align}
m &= \frac{q}{2} + i\frac{\pi n}{\beta} -\frac{kw}{2}, \\
\overline{m} &= \frac{q}{2} - i\frac{\pi n}{\beta}  +\frac{kw}{2},
\end{align}
where $q,n \in \mathbb{Z}$. Substituting these in (\ref{weig1}) and (\ref{weig2}), we finally obtain the conformal weights of the primaries:
\begin{align}
\label{ads3spectrum1prelim}
h^{wp}_{jqn} &= -\frac{j(j-1)}{k-2} -\frac{qw}{2} -\frac{i\pi nw}{\beta} + \frac{kw^2}{4}- i\frac{qp\beta}{4\pi} + \frac{ p n}{2} + \frac{kp^2\beta^2}{4(2\pi)^2} + h_{int}, \\
\label{ads3spectrum2prelim}
\bar{h}^{wp}_{jqn} &= -\frac{j(j-1)}{k-2} + \frac{qw}{2} -\frac{i\pi nw}{\beta} + \frac{kw^2}{4} - i\frac{qp\beta}{4\pi} - \frac{ p n}{2} + \frac{kp^2\beta^2}{4(2\pi)^2} + \bar{h}_{int}.
\end{align}
Note that indeed $h - \bar{h} \in \mathbb{Z}$, provided the internal CFT is on its own level-matched.

\subsection{Comments on the Euclidean $SL(2,\mathbb{C})/SU(2)$ point of view}
We now reanalyze this result from a purely Euclidean point of view. Euclidean $AdS_3$ is the hyperbolic 3-plane and can be seen as a coset $SL(2,\mathbb{C})/SU(2)$. The isometry group is given by $SL(2,\mathbb{C})$. We are interested in $AdS_3$ (and its continuation) in the \emph{global} coordinates and the continuation in that particular time coordinate. In appendix \ref{WZWapp} we discuss this continuation in some more detail.
Expanding the symmetry current in the algebra generators (which we choose to be the same as those of $\mathfrak{sl}(2,\mathbb{R})$; the only difference is that the expansion coefficients $J^{a}(z)$ are now allowed to be arbitrary complex numbers), we can identify which symmetry current is responsible for Euclidean time translations. This is again done in appendix \ref{WZWapp} and the result is
\begin{equation}
Q_{\tau} = i(J^{3}_0 - \overline{J}^{3}_0).
\end{equation}
This differs a factor of $i$ compared to the earlier result for $Q_t$. However, the thermal identification is with parameter $\beta$ now, so in all nothing changes and the derivation of the previous subsection still holds. However, we start with the principal representations of the $\mathfrak{sl}(2,\mathbb{C})$ algebra, which do not contain discrete representations nor spectral flowed representations. This sets $w=0$ from the start and $j=1/2+is$ with real $s$. This mismatch is caused by the fact that after Wick rotating all the other representations (discrete and the spectral flowed), these do not correspond to states in the Euclidean string spectrum \cite{Maldacena:2001km}, so we have computed `too much' in our first derivation.\footnote{We want to remark one subtlety: one could set $w=0$ from the start in the entire previous derivation, and consider only one twist operator. The level-matching condition is satisfied without any deformation to the energy. To determine the spacetime angular momentum of such states (find the correct form of $Q_{\phi}$), we should then resort to the large $k$ limit as explained in the previous section. The level-matching condition is of no help here. So in this case, the large $k$ limit becomes a necessary part of the procedure (and not just an alternative).} To summarize, we give the conformal weights of all primaries:
\begin{align}
\label{ads3spectrum1}
h^{p}_{jqn} &= \frac{s^2 +1/4}{k-2} - i\frac{qp\beta}{4\pi} + \frac{ p n}{2} + \frac{kp^2\beta^2}{4(2\pi)^2} + h_{int}, \\
\label{ads3spectrum2}
\bar{h}^{p}_{jqn} &= \frac{s^2 +1/4}{k-2} - i\frac{qp\beta}{4\pi} - \frac{ p n}{2} + \frac{kp^2\beta^2}{4(2\pi)^2} + \bar{h}_{int},
\end{align}
where $q,n \in \mathbb{Z}$ and $p \in \mathbb{Z}$ denotes the winding around the Euclidean time dimension. For the antiholomorphic part, one simply changes the sign of both $p$ and $q$. From here on, we assume the internal CFT to be unitary and compact such that to analyze possible tachyons, we can restrict ourselves to $h_{int} = \bar{h}_{int} = 0$.

\subsection{Atick-Witten tachyon}
\label{AWtach}
It is beneficial to now clearly state how we will identify a tachyonic state in the string spectrum. In a general bosonic string CFT, the one-loop partition function is given by
\begin{equation}
Z = \int_{F}\frac{d\tau_1 d\tau_2}{2\tau_2} \text{Tr}\left[q^{L_0-c/24}\bar{q}^{\bar{L_0}-\bar{c}/24}\right] = \int_{F}\frac{d\tau_1 d\tau_2}{2\tau_2}\left|\eta(\tau)\right|^4(q\bar{q})^{-\frac{1}{12}}\sum_{H_{matter}}{q^{h_i-1}\bar{q}^{\bar{h_i}-1}}
\end{equation}
In the second equality, we sum over only the matter contributions (of the full $c=26$ matter CFT). We have isolated a $q\bar{q}$ combination, since this precisely compensates the ghost CFT in its asymptotic behavior, meaning
\begin{equation}
Z \to \int_{F}\frac{d\tau_1 d\tau_2}{2\tau_2}\sum_{H_{matter}}{q^{h_i-1}\bar{q}^{\bar{h_i}-1}}
\end{equation}
as $\tau_2 \to \infty$. 
A tachyonic state in bosonic string theory is thus determined if the conformal dimension $h+\bar{h}$ in the matter sector is smaller than 2 (divergence for $\tau_2 \to \infty$ in $Z$) after integrating over continuous quantum numbers.\footnote{For type II superstrings, the only modification in this definition is that the conformal dimension $h+\bar{h}$ needs to be smaller than $1$ to have a tachyonic state.} Continuous quantum numbers can give a non-vanishing contribution if they integrate into a $\tau_2$-dependent exponential. Let us make some comments regarding the above conformal weights (\ref{ads3spectrum1}) and (\ref{ads3spectrum2}). Firstly note that the conformal weights have an imaginary part when both $p$ and $q$ are non-zero. This is due to the non-unitarity of the $SL(2,\mathbb{C})/SU(2)$ model \cite{Teschner:1997ft}. Complex conformal weights are not uncommon on these Euclidean signature manifolds \cite{Hemming:2002kd} and we will see them appear again below when we study BTZ black holes. Imaginary parts of conformal weights are harmless when considering divergence properties. When considering the partition function in the fundamental modular domain, each string state makes a contribution proportional to
\begin{equation}
\label{imag}
q^{h}\bar{q}^{\bar{h}} = e^{2\pi i \tau_1 (h - \bar{h})}e^{-2\pi \tau_2 (h + \bar{h})}.
\end{equation}
Since $\tau_1$ ranges from $-1/2$ to $+1/2$, the first factor is not capable of causing divergences and thus an imaginary contribution of a conformal weight can not cause divergences.
Secondly, the term $\frac{ p n}{2}$ might appear alarming, since this could be arbitrarily large and negative apparently causing tachyonic divergences. However, one should note that the antiholomorphic part has the opposite sign and the contribution to $h+\bar{h}$ hence vanishes. As a summary, only $\Re(h + \bar{h})$ matters for determining instabilities. \\
Let us make one final remark on these conformal weights. The primaries we determined above satisfy the physicality constraint $L_0 - \bar{L}_0 \in \mathbb{Z}$ by construction. The on-shell condition $L_0 + \bar{L}_0 = 2$ is in general not satisfied for all these states (and neither is $L_0 = \bar{L}_0$). This means for instance that only a subset of these can be used as vertex operators in scattering amplitudes. However, we are interested in the one-loop vacuum amplitude and the states that circle the loop are clearly off-shell. Thus we will not apply the on-shell restriction to the conformal weights. Our definition of tachyon is rooted in the one-loop amplitude and a tachyon state can hence be an off-shell state. Note that in \cite{Rangamani:2007fz} tachyons are identified only as on-shell physical states. Their definition of tachyon is hence not completely the same as ours.\\

\noindent To find the thermal tachyon, we simply set $s$ to zero and we will find the right state where we expect it. This will be an a posteriori verification that the integration over $s$ (with the correct density of states) does not yield a $\tau_2$-dependent exponential contribution, unlike for instance the linear dilaton background that we discuss elsewhere \cite{examples}. We find for the $p=\pm1$ state:
\begin{equation}
\frac{1}{4(k-2)} + \frac{k\beta^2}{4(2\pi)^2} = 1,
\end{equation}
which determines indeed the Hagedorn temperature (\ref{hag}) given in section \ref{randwalksads}. The left hand side is equal to 1 if $p = \pm 1$, so this represents a state that becomes `marginally convergent' at the Hagedorn temperature. The state is in the twisted sector and can be interpreted as a winding 1 state. So we have found a state in the thermal spectrum that becomes tachyonic at temperatures higher than the Hagedorn temperature: this is the Atick-Witten tachyon \cite{Atick:1988si}\cite{Berkooz:2007fe}. \\
An important aspect of the above analysis is that not only the $q=n=0$ state is marginal but in fact the states with arbitrary $q$ and $n$ are all marginal simultaneously at the Hagedorn temperature. This implies that the critical limit of $Z(\tau)$ includes all of these states since they are equally dominant. Note though that we are not interested in the thermal partition function for fixed $\tau$, but instead integrate over the fundamental modular domain. Tachyonic divergences are located at large $\tau_2$ and in that region, the $\tau_1$ integral is simply from $-1/2$ to $+1/2$ and acts as a projector onto states satisfying $L_0 = \bar{L}_0$ \cite{Kutasov:1990sv}. Thus when considering the critical regime of the free energy, the $n \neq 0$ states are irrelevant, but the sum over $q$ remains. The quantum number $q$ is to be interpreted as discrete momentum along the spatial cigar. This is a priori very surprising since one generally expects states that include discrete momentum quantum numbers to be more massive than those without discrete momentum. This observation will lead to some important ramifications further on when we look at the random walk behavior of the critical free energy. Of course, for this to be valid, the integration over $s$ should not alter the critical behavior: it must not give a $\tau_2$-dependent exponential for each $q$. If this is not the case, several of these states can actually be subdominant. This however does not occur. We present quantitative arguments in favor of this in appendix \ref{dos}. In subsection \ref{numads} we will further discuss, using numerical methods, that indeed all $q \in \mathbb{Z}$ states are needed to produce the critical behavior.

\subsection{Type II Superstring in $AdS_3$ space}
\label{Hagsup}
The modification to obtain the Hagedorn temperature for type II superstrings in $AdS_3$ is the following:
\begin{equation}
\frac{\frac{1}{4}+s^2}{k}+\frac{k}{4}\frac{\beta_{H}^{2}}{4\pi^2} = \frac{1}{2}.
\end{equation}
Two things have changed: the denominator of the $SL(2,\mathbb{R})$ term is now $k$ and the r.h.s. is $1/2$ which is the condition needed to ensure convergence of the thermal partition function. This immediately leads to 
\begin{equation}
\beta_{H}^{2} = \frac{4\pi^2}{k}\left(2-\frac{1}{k}\right),
\end{equation}
which satisfies the correct flat space limit when taking $k \to \infty$. This provides confidence in our method to determine the winding tachyon. \\
Our analysis of the type II superstring is not rigorous, but looks very plausible. It is known though, that in a wide variety of models based on the $SL(2,\mathbb{R})$ WZW model \cite{Giveon:1999px}\cite{Aharony:2004xn}\cite{Israel:2003ry}\cite{Israel:2004ir}\cite{Rangamani:2007fz}, the only difference with the bosonic string is the replacement $k-2 \to k$ in the first term of the conformal weights and a modification of the unitarity constraints for discrete representations. We will assume that also in this case, these are the only modifications.\\
We want to remark that the tachyon state we found (both for the bosonic string and for the type II string) is in the continuous representation as in \cite{Berkooz:2007fe} and the quantum number $s$ can be interpreted as a measure for the radial momentum. The state is delocalized over the $AdS$ space due to the repulsion from the Kalb-Ramond background just as is the case for the long strings in (Lorentzian signature) $AdS_3$ \cite{Maldacena:2000hw}.

\subsection{Summary}
Using CFT twist techniques, we have found the thermal string spectrum on the WZW $AdS_3$ background. This required introducing two twist operators and a subtle discussion on the Noether ambiguities in the projection conditions on invariant states. The thermal scalar state is clearly visible in the spectrum and it predicts the correct Hagedorn temperature. In the following we will come back to these results several times, providing more clarifications as we go along.

\section{Euclidean BTZ model}
\label{BTZsection}
Next we look at the Euclidean BTZ black hole. This is the same background as the Euclidean $AdS_3$ background, so the results of the previous sections apply. We first briefly review the link between the thermal ensembles of both backgrounds \cite{Maldacena:1998bw}. \\
Consider the following Euclidean metric
\begin{equation}
ds^2 = l^2\left(\cosh(\rho)^2d\tau^2+d\rho^2+\sinh(\rho)^2d\phi^2\right)
\end{equation}
where $l$ is the $AdS$ length. For WZW models this is related to the string length as $l^2 = k \alpha'$. We identify $\tau \sim \tau + \beta$ and $\phi \sim \phi + 2\pi$ (fixed to avoid a conical singularity at $\rho=0$) to obtain the thermal $AdS_3$ metric.
To obtain Euclidean BTZ black holes, we set $r=r_+\cosh(\rho)$, $\varphi = \frac{l}{r_+}\tau$ and $t = \frac{l}{r_+}\phi$ to obtain
\begin{equation}
ds^2 = (r^2-r_+^2)dt^2 + \frac{l^2dr^2}{r^2-r_+^2}+r^2d\varphi^2.
\end{equation}
We then identify $t \sim t + \beta_{BTZ}$ (to avoid a conical singularity at $r=r_+$) and $\varphi \sim \varphi + 2\pi$ where
\begin{equation}
\beta_{BTZ} = \frac{2\pi l }{r_+}.
\end{equation}
To connect these, note that using the BTZ manifold coordinates we have
\begin{equation}
\varphi \sim \varphi + 2 \pi \quad \Rightarrow \quad \tau \sim \tau + \frac{2\pi r_+}{l},
\end{equation}
where the $\tau$-periodicity is equal to the inverse temperature $\beta$ from the $AdS_3$ manifold.
So we have that
\begin{equation}
\beta_{BTZ} = \frac{2\pi l }{r_+} = \frac{4\pi^2}{\beta}.
\end{equation}
In all, we obtain
\begin{equation}
Z_{BTZ}\left(\beta\right) = Z_{AdS_3}\left(\frac{4\pi^2}{\beta}\right)
\end{equation}
and this is the precise link between thermodynamics in both backgrounds.\\
The generators of Euclidean time translations and angular rotations are given by
\begin{align}
Q_{t} &= \frac{2\pi}{\beta_{BTZ}}\left(J^{3}_{0} + \overline{J}^{3}_{0}\right), \\
Q_{\varphi} &= i\frac{2\pi}{\beta_{BTZ}}\left(J^{3}_{0} - \overline{J}^{3}_{0}\right),
\end{align}
where $t \sim t + \beta_{BTZ}$ and $\varphi \sim \varphi + 2\pi$. With the same caveats and modifications as in the $AdS_3$ background, this leads to the restrictions (for untwisted primaries):
\begin{align}
m+\overline{m} &\in \mathbb{Z}, \\
\frac{i 2\pi}{\beta_{BTZ}}\left(m-\overline{m}\right) &\in \mathbb{Z}.
\end{align}
This is indeed simply the substitution $\beta \to \frac{4\pi^2}{\beta}$ as discussed above. The Euclidean BTZ spectrum is thus obtained simply by evaluating the thermal $AdS_3$ spectrum at a temperature $\frac{4\pi^2}{\beta}$:
\begin{equation}
h^{w}_{jm\overline{m}} = \frac{s^2+1/4}{k-2} + \frac{m^2}{k} - \frac{\left(m + \frac{kw}{2}\frac{ir_+}{l}\right)^2}{k} + h_{int}.
\end{equation}
where the winding $w$ is around the $\varphi$ direction which is the $\tau$-direction in the $AdS_3$ language. The $m$ and $\overline{m}$ quantum numbers are similarly given by
\begin{align}
m &= \frac{q}{2} + i\frac{\beta n}{4\pi}, \\
\overline{m} &= \frac{q}{2} - i\frac{\beta n}{4\pi},
\end{align}
where $q,n \in \mathbb{Z}$. 
To sum up, the conformal weights of the primaries are given by
\begin{align}
h^{w}_{jqn} &= \frac{s^2+1/4}{k-2} - qwi\frac{r_+}{2l} + \frac{wn}{2} + \frac{kw^2r_{+}^2}{4l^2} + h_{int}, \\
\bar{h}^{w}_{jqn} &= \frac{s^2+1/4}{k-2} - qwi\frac{r_+}{2l} - \frac{wn}{2} + \frac{kw^2r_{+}^2}{4l^2} + h_{int},
\end{align}
with $q,n \in \mathbb{Z}$ and $w \in \mathbb{Z}$ denotes the winding around the angular $\varphi$ direction. These conformal weights were obtained in \cite{Hemming:2002kd}, but there the projection operation on invariant states (which determines the $m$ and $\overline{m}$ quantum numbers) was not considered. \\
Let us now discuss some peculiarities of this spectrum.

\subsection{Thermal tachyons}
\begin{figure}[h]
\centering
\includegraphics[width=15cm]{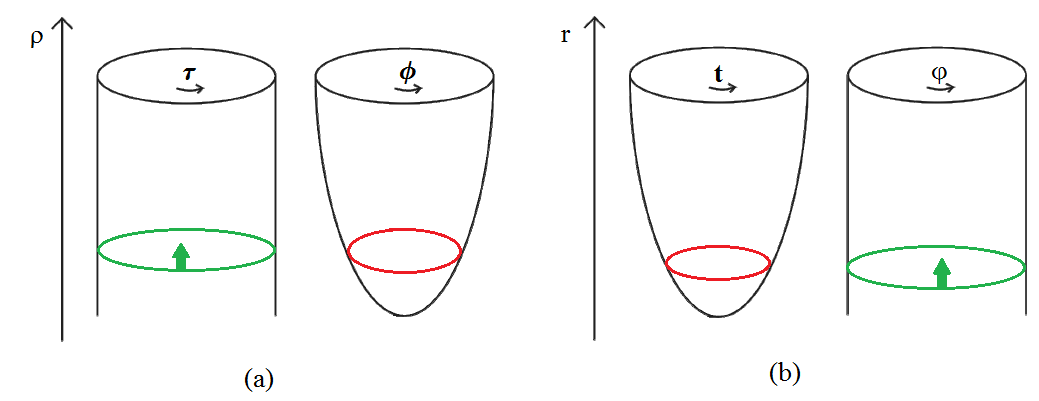}
\caption{(a) Thermal circle and spatial cigar for thermal $AdS_3$. The arrows denote the Kalb-Ramond repulsion. String states can wind the cilinder where the gravitational attraction to the center precisely compensates the Kalb-Ramond repulsion such that the winding strings are in the continuous representations. States that would wind the cigar are not allowed: these states are simply absent from the spectrum. (b) The same situation for the thermal BTZ black hole. Strings cannot wind the temporal cigar in this case.}
\label{adsbtzfigure}
\end{figure}

Firstly (and most importantly for our purposes) there are no string states that wind the $t$ direction for the same reason that the Euclidean $AdS_3$ manifold does not have cigar-winding states. Thus the thermal scalar is not even in the spectrum!\footnote{In \cite{Lin:2007gi} it was argued that this BTZ partition function might be incomplete but we believe this not to be the case as we discuss further on.} Even though such states are normalizable (they can be determined numerically from a field theory point of view as discussed in section \ref{fieldtheory}), they are simply absent from the spectrum. There are also no discrete states bound to the black hole horizon. This is a consequence of the repulsive NS-NS background field. The situation is sketched in figure \ref{adsbtzfigure}.
These results show that the WZW BTZ model is markedly different from the standard behavior of strings near uncharged black holes where we expect a zero-mode localized at string length from the horizon \cite{Mertens:2013zya}. 
We can be more explicit about this. One can consider the $AdS_3$ partition function (as for instance given in equation (\ref{partfunct})) and rewrite this in terms of the characters of $\widehat{SL(2,\mathbb{R})}$. This has been done in terms of twisted characters in \cite{Gawedzki:1991yu}\cite{Gawedzki:1988nj}. In appendix \ref{toy} and \ref{Ham1} we go one step further and rewrite these in terms of normal characters where the conformal weights of the primaries are those we determined above. The result is that no spectral flowed states (in the angular $\phi$ dimension) are present: the partition function is entirely reproduced by the primaries (both untwisted and twisted in the Euclidean time dimension) and their descendants. For the BTZ partition function (obtained by simply setting $\beta \to 4\pi^2/\beta_{BTZ}$ in the $AdS_3$ partition function (\ref{partfunct})), one obtains the same result upon switching the interpretation of time and angular: thermal winding states are not present in the BTZ partition function.

\subsection{Cylinder-winding tachyons}
Secondly, there is the possibility of a winding tachyon in the $\varphi$ direction. The winding states in the $\varphi$ direction depend on $r_+$, the horizon location. 
Let us look at a $q=n=0$ state:
\begin{equation}
h^{w}_{j00} = \frac{s^2+1/4}{k-2}  + \frac{kw^2\left(\frac{r_+}{l}\right)^2}{4} + h_{int}.
\end{equation}
To have convergence for the BTZ partition function, we need
\begin{equation}
\frac{1}{4(k-2)}  + \frac{kw^2\left(\frac{r_+}{l}\right)^2}{4} \geq 1.
\end{equation}
Decreasing the BTZ temperature, decreases $r_+$. So there is a critical BTZ temperature below which a $\varphi$-winding tachyon appears.
The location of the horizon at this critical temperature is given by
\begin{equation}
r_+^2 = \alpha'\left(4-\frac{1}{k-2}\right),
\end{equation}
and the size of the black hole is clearly string size ranging from $r_+ = \sqrt{3}\sqrt{\alpha'}$ (for the limiting $k=3$ case) to $r_+ = 2 \sqrt{\alpha'}$ (as $k \to \infty$). So this tachyon is a consequence of shrinking the horizon to string scale and has nothing to do with the thermal scalar that we are interested in. The change in $\beta_{BTZ}$ that we have considered here in this paragraph is an on-shell change (changing the temperature changes the black hole size), whereas for the thermal scalar we are interested in an off-shell change of the temperature and the ensuing introduction of conical singularities. The $\varphi$-winding state causes the $AdS_3/BTZ$ black hole condensation process as was discussed in \cite{Berkooz:2007fe}. \\
For the type II superstring, the results of this and the preceding subsection do not change qualitatively.

\subsection{Summary}
Let us summarize the BTZ WZW black hole. There is no zero-mode surrounding the black hole. This is in sharp contrast with the generic uncharged black hole where such a zero-mode is present. It is only for small black holes that a stringy state becomes marginal, but this state is a Lorentzian state that wraps the cylindrical $\varphi$ dimension. It signals the $AdS_3/BTZ$ transition and it is not the high temperature thermodynamical state that we seek.

\section{$AdS_3$ orbifolds: conical spaces}
\label{conical}
In the previous section, we saw that the Euclidean BTZ WZW model does not contain cigar-winding string states in the thermal spectrum. For thermodynamical purposes, conical orbifolds of the cigar-shaped subspace are also important since these correspond to the string gas at a temperature different than the Hawking temperature. An intriguing possibility would be that the winding state is not present for the black hole itself, but when considering conical spaces the state might reappear. This could then possibly still give an important effect. In this section we analyze the thermal spectrum on such conical spaces, and in particular consider the question whether the thermal scalar is present or not. Our primary focus is again on the lowest weight state.
\subsection{Thermal spectrum}
Let us take a closer look at the orbifolds obtained by creating conical deficits with opening angle $\frac{2\pi}{N}$ at the tip of the cigar. First we remark that the prodedure of section \ref{spectrumsection} can be applied to this case and leads simply to $w \to \frac{a}{N}$ with $a\in\mathbb{Z}$. Hence one considers \emph{fractional} winding numbers \cite{Martinec:2001cf}. We have however no control on which of these states actually appear in the spectrum. To analyze this, a Hamiltonian analysis of the exact partition function is needed, to which we turn now. The thermal partition function of these orbifolds (for $AdS_3$ or BTZ) has the schematic form \cite{Son:2001qm}
\begin{equation}
Z = \frac{1}{N}\sum_{a,b}Z_{ab},
\end{equation}
where $a$ runs over the twisted states and the sum over $b$ realizes the projection on invariant states. The sum over $b$ ranges from $0$ to $N-1$. The individual partition functions that are summed over are given by
\begin{align}
\label{numer}
Z_{ab}(\tau) = \frac{\beta\sqrt{k-2}}{8\pi\sqrt{\tau_2}}\sum_{l,p}\frac{e^{-k\beta^2\left|l-p\tau\right|^2/4\pi\tau_2 + 2\pi \Im(U_{lp})^2/\tau_2}e^{\frac{\pi\tau_2}{2}}}
{\left|\sin(\pi U_{lp})\right|^2\left|\prod_{r=1}^{+\infty}(1-q^{r})(1-q^re^{2\pi i U_{lp}})(1-q^{r}e^{-2\pi i U_{lp}})\right|^2},
\end{align}
with
\begin{equation}
U_{lp} = \frac{b}{N} + \frac{a}{N}\tau_1 - i\frac{\beta}{2\pi}(p\tau_1-l) + i\frac{a}{N}\tau_2 + \frac{p\beta}{2\pi}\tau_2.
\end{equation}
In appendices \ref{Ham2}, \ref{Poisson} and \ref{Ham3} we rewrite this orbifolded partition function in a Hamiltonian way and we identify which states actually occur in the spectrum. The upshot is that in this case there are string states that wind the cigar, but only for $w=a/N$ with $\left|w\right| < \frac{1}{2}$ and integer $a$. The reason we restrict to this interval for $w$ will be explained in the next subsection.\footnote{A priori \emph{any} range of length $N$ is allowed for $a$. This follows from the fact that $Z_{a,b} = Z_{a+N,b}$ and $Z_{a,b} = Z_{a,b+N}$ which can be seen explicitly in equation (\ref{numer}). This is related to the periodicity of the Ray-Singer torsion.} In particular, one can choose the range as follows:
\begin{align}
a&= -\frac{N-1}{2} \to \frac{N-1}{2}, \quad N \text{ odd}, \\
a&= - \frac{N-2}{2} \to \frac{N}{2}, \quad N \text{ even}.
\end{align}
A crucial observation made in appendix \ref{Ham3} is that the sectors with $w \neq 0$ can include discrete states. This depends on whether $k\left|w\right|$ is larger or smaller than 1. Roughly, these discrete modes appear as follows. To rewrite the partition function in a Hamiltonian manner, one needs to employ the general Poisson summation formula. For $w \neq 0$ however, one actually needs a proper analytic continuation of Poisson's summation formula which is presented in appendix \ref{Poisson}. The naive substitution of complex arguments in the normal summation formula gives the continuous representations. This is not enough however: one needs to include extra terms in Poisson's summation formula corresponding to simple poles of the complex function. These precisely correspond to the discrete states that are otherwise completely missed.\\
For BTZ, it is clear that now the thermal spectrum does contain twisted strings that wind the temporal cigar. The situation is similar to Euclidean Rindler space and its orbifold cousins.\\
In terms of $AdS_3$ parameters, the thermal spectrum includes continuous states with conformal weights
\begin{align}
h^{wp}_{jqn} &= \frac{s^2+1/4}{k-2} +\frac{qw}{2} +\frac{i\pi nw}{\beta} + \frac{kw^2}{4}- i\frac{qp\beta}{4\pi} + \frac{ p n}{2} + \frac{kp^2\beta^2}{4(2\pi)^2}, \\
\bar{h}^{wp}_{jqn} &= \frac{s^2+1/4}{k-2} - \frac{qw}{2} +\frac{i\pi nw}{\beta} + \frac{kw^2}{4} - i\frac{qp\beta}{4\pi} - \frac{ p n}{2} + \frac{kp^2\beta^2}{4(2\pi)^2},
\end{align}
and discrete states with weights
\begin{align}
\label{disc1}
h &= -\frac{\tilde{j}(\tilde{j}-1)}{k-2} + \frac{qw}{2} - \frac{\pi i w n }{\beta} + \frac{kw^2}{4} - \frac{i \beta  pq}{4\pi} - \frac{pn}{2} + \frac{kp^2\beta^2}{4(2\pi)^2}, \\
\label{disc2}
\bar{h} &= -\frac{\tilde{j}(\tilde{j}-1)}{k-2} - \frac{qw}{2} - \frac{\pi i w n }{\beta} + \frac{kw^2}{4}  - \frac{i \beta  pq}{4\pi} + \frac{pn}{2} + \frac{kp^2\beta^2}{4(2\pi)^2},
\end{align}
where $\tilde{j} = M - l= \frac{k\left|w\right|}{2} - \frac{\left|q\right|}{2} \pm \frac{i\pi n }{\beta} - l$ and $l=0,1,2,\hdots$. We also have $q\in N\mathbb{Z}$, $n\in\mathbb{Z}$, $p\in\mathbb{Z}$ and $w = \frac{a}{N}$ with $a$ in the range determined above. The discrete states are present for $\Re(\tilde{j}) > \frac{1}{2}$. In particular for $k\left|w\right| < 1$, no discrete states are present at all. This is important in what follows. We remark further that for the thermal manifold, the discrete states really are discrete, unlike the discrete representations utilized in the Lorentzian $AdS_3$ spacetime \cite{Maldacena:2000hw} which are actually continuous because $\tilde{j}$ and $\tilde{m}$ are continuous quantum numbers there due to the fact that one considers the universal cover of the $SL(2,\mathbb{R})$ manifold.\\
Note that the discrepancy between $M$ and $\tilde{m}$ or $\tilde{\bar{m}}$ is expected and the same sort of situation occurs for the $SL(2,\mathbb{R})/U(1)$ black hole in which case it is well understood \cite{Dijkgraaf:1991ba}\cite{Aharony:2004xn}. Actually, the resemblance of the discrete weights with those of the $SL(2,\mathbb{R})/U(1)$ black hole is remarkable: in that case one finds for $M$:
\begin{equation}
M = \frac{k\left|w\right|}{2} - \frac{\left|q\right|}{2},
\end{equation}
which is of precisely the same form as above for the quantum numbers associated to the cigar-submanifold.

\subsection{Numerical analysis}
\label{numerical}
As a first method to analyze the critical large $\tau_2$ regime, we present the results of a numerical analysis of the expression (\ref{numer}). To start, we drop the infinite product present in the denominator. As discussed in appendix \ref{Ham1}, we expect this product to arise from oscillator states. Further on we will confirm numerically that this product does not influence the critical behavior. So we analyze numerically the expression
\begin{equation}
\label{EE}
E = \lim_{\tau_2 \to \infty} \sum_{l\in \mathbb{Z}}\frac{e^{(2-k)\frac{\pi l^2}{\tau_2}\frac{\beta^2}{4\pi^2} + 4\pi l w\frac{\beta}{2\pi} + 2 \pi w^2\tau_2}}{\left|\sin(\pi\left(iw\tau_2 + il\frac{\beta}{2\pi}\right))\right|^2}.
\end{equation}
This expression gives the asymptotic limit of part of the partition function (since $w$ is fixed) for the case $\tau_1=0$, $b=0$ and $p=0$. We will comment on the more general cases we are interested in below.\\
Firstly (crucially!), notice that we can not bring the large $\tau_2$ limit into the summation over $l$. \\
Numerically we can analyze the expression $E$ by truncating the series and then taking $\tau_2$ sufficiently large but still sufficiently small compared to the truncation index.\footnote{By `sufficiently' we mean that we varied the values of these parameters up to the point where the numerical result is neglegibly influenced by any further variation.} Using numerics as a guidance, we found the following asymptotic behavior.
\begin{itemize}
\item{
If $k\left|w\right| < 1$, one finds the following asymptotic behavior\footnote{Obtained by taking the logarithm of the numerical computation and then determing the slope of the resulting line.}
\begin{equation}
\label{expone}
E \propto e^{-k\pi w^2 \tau_2}.
\end{equation}
This is the expected behavior corresponding to a continuous state. Note that indeed one finds here that the density of states does not correct the critical behavior.
The prefactor of (\ref{expone}) has a periodicity in $\tau_2$ which equals\footnote{This was determined numerically and is not manifest in expression (\ref{EE}).}
\begin{equation}
\tau_2 \to \tau_2 + \frac{N\beta}{2\pi w}, \quad N \in \mathbb{Z}
\end{equation}
and indeed, this is the symmetry one expects from the CFT point of view of the critical behavior:
\begin{equation}
\sum_{n \in \mathbb{Z}} \rho(n) e^{\frac{4\pi^2 i n w \tau_2}{\beta}},
\end{equation}
with $\rho$ the density of string states.}

\item{
If $k\left|w\right| > 1$, we find
\begin{equation}
E \propto e^{-k\pi w^2 \tau_2}e^{\frac{\pi(k\left|w\right|-1)^2}{k-2}\tau_2}.
\end{equation}
More precisely, the prefactors can be determined as
\begin{equation}
E \to \frac{8\pi}{\beta}\sqrt{\frac{\tau_2}{k-2}} e^{-k\pi w^2 \tau_2}e^{\frac{\pi(k\left|w\right|-1)^2}{k-2}\tau_2}.
\end{equation}
We note that this behavior of the prefactor is also correct even when $\tau_1 \neq 0$ or $p\neq 0$ or $b\neq0$. One can readily see that this asymptotic behavior corresponds exactly to a discrete state including the prefactors.}
\end{itemize}
Let us now briefly discuss how the numerics change when we consider the more general case. Firstly, let us set $b \neq 0$. Numerically we checked that $b\neq0$ gives precisely the same asymptotics.\\
The case with $p\neq0$ is also easily analyzed. We consider now
\begin{equation}
\lim_{\tau_2 \to \infty} \sum_{l\in \mathbb{Z}}\frac{e^{(2-k)\frac{\pi l^2}{\tau_2}\frac{\beta^2}{4\pi^2} + 4\pi l w\frac{\beta}{2\pi} + 2 \pi w^2\tau_2-k\pi p^2\tau_2\frac{\beta^2}{4\pi^2}}}{\left|\sin(\pi\left(p\tau_2\frac{\beta}{2\pi} + iw\tau_2 + il\frac{\beta}{2\pi}\right))\right|^2}.
\end{equation}
The result gives the expected extra correction $\sim e^{-k\pi p^2\tau_2}$ but no mixing between $w$ and $p$ is generated by this, as indeed our analytical results also predict.\\
The above expressions do not fully coincide with what we are interested in: from the point of view of the spectra written down in the previous subsection, we want to set $q=0$, which is enforced by the $\tau_1$ integral. Simply setting $\tau_1$ equal to zero still gives us the sum over $q$ 
with the $q$-dependent density of states. We can extend the above numerical analysis to include $\tau_1$ dependence by studying instead
\begin{equation}
\lim_{\tau_2 \to \infty} \sum_{l\in \mathbb{Z}}\frac{e^{(2-k)\frac{\pi l^2}{\tau_2}\frac{\beta^2}{4\pi^2} + 4\pi l w\frac{\beta}{2\pi} + 2 \pi w^2\tau_2}}{\left|\sin(\pi\left(w\tau_1 + iw\tau_2 + il\frac{\beta}{2\pi}\right))\right|^2}.
\end{equation}
This expression was written down with $b=0$ and $p=0$. Numerical analysis yields the same asymptotic behavior as before, irrespective of the value of $\tau_1$ (it does however influence the prefactors). Integrating $\tau_1$ from $-1/2$ to $+1/2$ hence does not alter the asymptotic form.\footnote{For instance, consider the mean value theorem from elementary integral calculus: the $\tau_1$-integral equals the length of the interval (=1) times the function value at some intermediate point. But all intermediate points display the same asymptotic form. Thus the asymptotic form cannot be influenced by the integral over $\tau_1$.} Note that the sum over $b$ becomes irrelevant since the $\tau_1$ integral enforces $q=0$ and the restriction of $q \in N\mathbb{Z}$ achieved by summing over $b$ does not influence the final result.

\subsubsection*{Infinite oscillator product}
Up to this point, we did not analyze the infinite product present in equation (\ref{numer}). Its treatment is tricky. We dismissed it somewhat carelessly in appendix \ref{Ham1} when considering the Hamiltonian picture. A delicate point is that the Taylor expansion we should use for each of the factors of the infinite product depends on the precise value of $l$. Therefore we should split the sum over $l$ in different pieces. But we needed the entire sum over $l$ to perform the Poisson resummation formula. This is related to the fact that one cannot take the large $\tau_2$ limit through the summation over $l$. This is an issue that requires more thought and we will not discuss this further here.\\
Numerically however, we can analyze this infinite product (albeit in a truncated way of course).
One finds the following. Firstly periodicity $w \to w+1$ is recovered. This symmetry is exactly present in (\ref{numer}) due to the periodicity properties of the Ray-Singer torsion, but is compromised upon dropping the infinite product. It is nice to find numerical evidence that when including a truncated version of this infinite product, the symmetry becomes more and more restored.\footnote{The symmetry restoration is apparent only for values of $\left|w\right|$ not too large. Roughly speaking, if one wants the periodicity domain to increase by an integer, one should include one more factor in the numerical treatment of the infinite product.} Secondly, for values of $w$ that satisfy $\left|w\right|<\frac{1}{2}$, the infinite product does not contribute to the large $\tau_2$ limit and one can hence trust the above expressions to yield the dominant behavior. If $w$ is outside this interval, the infinite product makes a contribution that survives the large $\tau_2$ limit, precisely to restore the periodicity $w \to w+1$. Hence we restrict $w$ to $\left|w\right|<\frac{1}{2}$ and drop the infinite product. All other intervals of $w$ should be found by periodicity. As noted before, this numerical treatment is actually the only indication we have on how the infinite product affects the critical behavior.

\subsection{A brief look back at the $AdS_3$ string gas}
\label{numads}
Before we proceed, let us briefly return to the normal $AdS_3$ space. The numerical methods discussed in the previous subsection can also be used for this (easier) case. We hence analyze
\begin{equation}
E = \lim_{\tau_2 \to \infty} \sum_{l\in \mathbb{Z}}\frac{e^{(2-k)\frac{\pi l^2}{\tau_2}\frac{\beta^2}{4\pi^2} - k\pi p^2\frac{\beta^2}{4\pi^2}\tau_2}}{\left|\sin(\pi\left(\frac{p\beta}{2\pi}\tau_2 + i\frac{\beta}{2\pi}l\right))\right|^2}.
\end{equation}
One finds
\begin{equation}
E \sim e^{-k\pi p^2\frac{\beta^2}{4\pi^2}\tau_2}
\end{equation}
with a prefactor periodic in $\tau_2$ as
\begin{equation}
\tau_2 \to \tau_2 + \frac{2\pi N}{\beta p}, \quad N \in \mathbb{Z},
\end{equation}
which is in accord with what we argued for in subsection \ref{AWtach}. There we discussed the fact that all $q\in\mathbb{Z}$ stringy states contribute to the critical regime. And indeed, their sum respects this symmetry since it gives (schematically):
\begin{equation}
\sum_{q \in \mathbb{Z}} \rho(q) e^{i q p \beta \tau_2}.
\end{equation}
It is hard to imagine how this periodicity would be generated if the $q\in\mathbb{Z}$ states would be subdominant.\footnote{For instance setting $q=0$ would not give a critical behavior that respects this symmetry.} A periodic prefactor of the critical behavior was also explicitly determined in \cite{Lin:2007gi}. \\
Note that in this case no change of dominant behavior occurs. This case is hence for all values of the parameters similar to the $k\left|w\right| < 1$ case of the orbifolds discussed above. 

\subsection{Dominant state}
\begin{figure}[h]
\centering
\begin{minipage}{.4\textwidth}
  \centering
  \includegraphics[width=\linewidth]{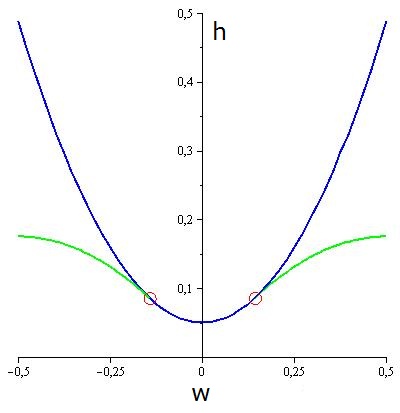}
  \caption*{(a)}
\end{minipage}%
\begin{minipage}{.4\textwidth}
  \centering
  \includegraphics[width=\linewidth]{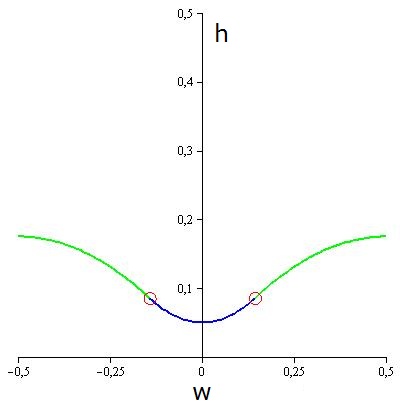}
  \caption*{(b)}
\end{minipage}
\caption{(a) Conformal weight $h$ of the most tachyonic mode of both types of representations (continuous and discrete) as a function of $w$. The blue curve originates from the continuous representations. The green curve represents the lowest mode of the discrete representations. The latter only appears when $\left|w\right| > \frac{1}{k}$, represented by the red circles. (b) Conformal weight of the overall lowest weight state relevant for stability issues. As soon as the discrete representations are present, they dominate the continuous modes.}
\label{weight}
\end{figure}

Let us now study the large $\tau_2$ limit analytically. Which state is the dominant one? This is easy to analyze. First let us take a look at the continuous states. The situation is the same as that analyzed in section \ref{AWtach} and one can repeat the entire discussion given there, including the arguments presented in appendix \ref{dos} as to why the density of states does not influence the critical weight.\\
Next, we focus on the discrete states. Consider (part of) the conformal weight $h = - \frac{\tilde{j}(\tilde{j}-1)}{k-2}$. The most negative conformal weight dominates. Since $\tilde{j}>1/2$ is required, the dominant state has the largest value of $\tilde{j}$ allowed by the constraints. This is then obviously a state which has $q=l=0$. The $n$ quantum number has an $n^2$ contribution in the conformal weight as given before. It is hence dominated by $n=0$.\\
Note that this is a major qualitative difference between the continuous states and the discrete states. For continuous states in $AdS_3$ we noted before in subsection (\ref{AWtach}) that the dominant state has arbitrary $q\in\mathbb{Z}$ and one should sum these to get the critical behavior. The same story applies here for the continuous states: arbitrary $n$ is allowed and we should sum these for the critical behavior (and in order to satisfy the periodicity in $\tau_2$ of the prefactors as discussed in the previous subsections). For discrete states on these orbifolds, $n$ and $q$ are both set to zero. Hence no summation is required to obtain the critical behavior. The reason for this peculiar behavior of the continuous states is the absence of $n^2$ and $q^2$ terms in the expressions for the conformal weights (\ref{ads3spectrum1prelim}) and (\ref{ads3spectrum2prelim}). We will discuss this further in section \ref{thscaction}. \\
For any value of $N$, one can then determine whether the system is stable or not by considering the mode with lowest conformal weight. The conformal weight $h$ as a function of $w = \frac{a}{N}$ is given in figure \ref{weight}. Even though this figure is drawn for $w$ a continuous variable, one should keep in mind that in principle we only determined this for $w= \frac{a}{N}$. From the numerical approach, we learned that we should restrict to $\left|w\right| < 1/2$. The lowest mode for a fixed value of $N$, is given by taking $w = \frac{1}{N}$. In particular, the dominant state is discrete for $N < k$ and is continuous otherwise.
We remark that the curve of the lowest conformal weight (as displayed in figure \ref{weight}), is smooth across the `joints' $\left|w\right| = \frac{1}{k}$. This analysis is in precise agreement with the numerical study presented in the previous subsection.\\
Note further that since $k > 2$, these orbifold models \emph{always} include twisted discrete states in their spectrum. \\
If we analyze solely the large $\tau_2$ behavior with $w$ arbitrary and not summed over, i.e. one of the expressions we studied numerically in the previous subsection, then the lowest conformal weight displayed in figure \ref{weight} should be periodically continued with period 1 both to the left and to the right.

\subsection{Hagedorn temperature}
\begin{figure}[h]
\centering
\includegraphics[width=7cm]{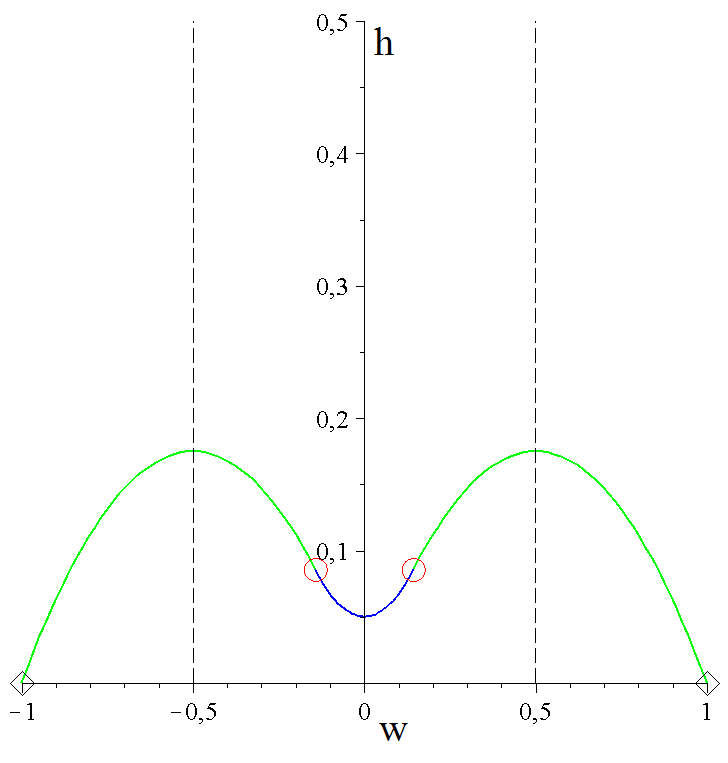}
\caption{Conformal weight of lowest weight state when continuing the expressions up to $\left|w\right| =1$. The vertical dashed lines denote the boundary $\left|w\right| = \frac{1}{2}$. The two black diamonds depict the weight of the state at $\left|w\right| =1$, where it becomes zero.}
\label{weight3}
\end{figure}

\begin{figure}[h]
\centering
\begin{minipage}{.33\textwidth}
  \centering
  \includegraphics[width=\linewidth]{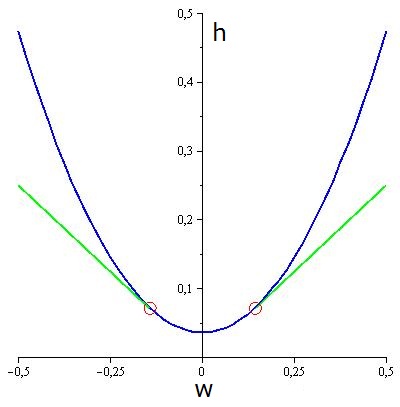}
  \caption*{(a)}
\end{minipage}%
\begin{minipage}{.33\textwidth}
  \centering
  \includegraphics[width=\linewidth]{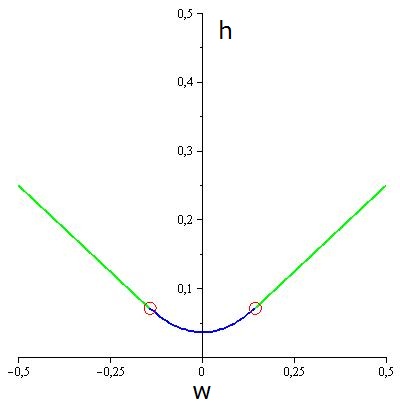}
  \caption*{(b)}
\end{minipage}
\begin{minipage}{.33\textwidth}
  \centering
  \includegraphics[width=\linewidth]{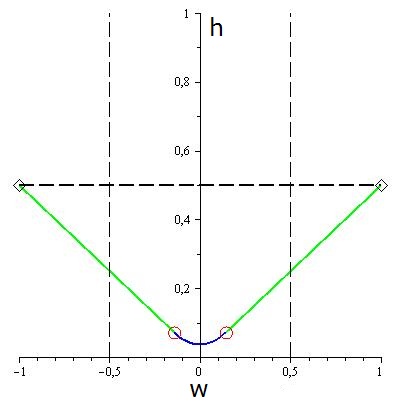}
  \caption*{(b)}
\end{minipage}
\caption{(a) and (b) represent the analogous figures for the type II superstring as figure \ref{weight} does for the bosonic string. (c) The curves are drawn all the way to $\left|w\right|=1$, where the green curve is found to precisely equal $\frac{1}{2}$. This point is depicted with black diamonds. The vertical dashed lines denote the boundary $\left|w\right| = \frac{1}{2}$.}
\label{IIweight}
\end{figure}

Can we utilize these observations to determine the Hagedorn temperature of the BTZ black hole? Analogous to the treatment of conical orbifolds of the flat plane \cite{Dabholkar:1994ai}\cite{Lowe:1994ah}\cite{Mertens:2013zya}, the $\frac{1}{N}$ parameter should be interpreted as $\frac{\beta}{\beta_{BTZ}}$, the ratio of the actual temperature and the Hawking temperature. The lowest state for each $N$ is given by choosing $\left|w\right|=\frac{1}{N}$, thus setting $a=\pm1$. This corresponds to taking $w = \frac{\beta}{\beta_{BTZ}}$. To determine whether the BTZ black hole itself is stable, we are hence interested in taking $N\to 1$. Thus we would like to continue the above expressions to $N = 1$. The discussion that follows is hence necessarily more speculative than the previous results. Unlike for the flat $\mathbb{C}/\mathbb{Z}_N$ cones, in this case this continuation in $N$ is more ambiguous: we have a piecewise definition of the lowest conformal weight and it is a priori unclear which is the correct way to proceed. A naive application of the above periodicity results would suggest then that the lowest weight state for $w=1$ is the same as that for $w=0$ and the BTZ black hole would be unstable. However, we believe this is not correct for the following reasons. Firstly, from a Lorentzian point of view, the canonical partition function is given by\footnote{Let us remark that the equality of the partition function on the thermal manifold at one loop and the Lorentzian thermal free field trace is not a settled issue for black hole spacetimes \cite{Susskind:1994sm}. Different arguments are presented below.}
\begin{equation}
\label{Lorpart}
Z = \sum_{n\in\mathcal{H}}e^{-\beta E_n}.
\end{equation}
It is clear from this formula that periodic behavior of the system as $w \to w + 1$ or $\beta \to \beta + \beta_{BTZ}$ is impossible. The reason for this periodicity symmetry can be traced back to the following. As we discuss in appendix \ref{Ham1}, the first step in evaluating the partition function using path integral methods consists of performing a coordinate transformation that makes manifest that $\phi$ is an angular coordinate. However, this coordinate transformation is 1:1 only when the range of $\phi$ is less than $2\pi$. Thus the periodicity $w \to w+1$ is an artifact of the new coordinates and is \emph{not} a symmetry of the original space. \\
A better approach, which we believe to be the correct one, is to ignore this periodicity. The presence of discrete states causes the partition function to diverge for $w \approx 1$. In figure \ref{weight3} we draw the lowest conformal weight when we continue the discrete state all the way to $\left|w\right| = 1$. Curiously, the conformal weight becomes zero and the conclusion seems to be that the bosonic BTZ WZW model is divergent at the Hawking temperature. In fact, it appears to be divergent for any value of $\beta$. The computation of the critical weight for the discrete states is actually completely the same as that for the discrete modes in Euclidean Rindler space \cite{Mertens:2013zya}. In both cases, it is very ambiguous how to correctly interpret the continuation $N\to1$ in terms of convergence or divergence of thermodynamic quantities.\footnote{Note further that even if we did take the $\beta \to \beta + \beta_{BTZ}$ periodicity seriously, our conclusion would remain unaltered: the singly wound state yields a divergence for $\left|w\right| \to 1$.}
Despite the speculative nature of the above discussion, for \emph{conical} BTZ spaces the dominant behavior is well-defined. The thermal scalar on these BTZ conical spaces for $N < k$ is characterized by $\left|w\right|=\frac{1}{N}$,\footnote{Winding $\pm 1$ on the cone is equivalent to winding $\pm \frac{1}{N}$ on the `covering' space of the cone, i.e. the unorbifolded space.} $q=0$, $l=0$, $p=0$ and $n=0$. Unlike the thermal scalar on the $AdS_3$ space which included a summation over $q$ to distill the critical random walk behavior, in this case no summation is required.\\
Let us look at the analogous formulas for the type II superstring. We did not determine these ab initio, but it seems obvious how to modify the resulting conformal weights: one simply replaces $k-2 \to k$ in the denominator of the first term of the conformal weights.\footnote{We have already argued for this before in section \ref{Hagsup}.} After integrating out the continuous quantum number $s$, the most tachyonic continuous state hence has
\begin{equation}
h = \bar{h} = \frac{1}{4k} + \frac{kw^2}{4},
\end{equation}
whereas the most tachyonic discrete state (for $k\left|w\right| > 1$) has
\begin{equation}
h = \bar{h} = - \frac{\frac{kw}{2}\left(\frac{k\left|w\right|}{2}-1\right)}{k} + \frac{kw^2}{4} = \frac{\left|w\right|}{2}.
\end{equation}
Crucially, unlike the bosonic string, the term quadratic in $w$ cancels out and we are left with a linear dependence on $w$. Moreover, and this is the intriguing part of this analysis, if we demand that $h = \frac{1}{2}$, we find precisely $\left|w\right|=1$, i.e. $\beta = \beta_{BTZ}$. The situation is drawn in figure \ref{IIweight}.
A further field-theoretic argument in favor of this continuation in $N$ will be given in section \ref{cigarwind}. We are inclined to believe these results since they are very reminiscent on the results for $SL(2,\mathbb{R})/U(1)$ black holes and its Euclidean Rindler limit. In fact, qualitatively the situation is almost the same. Bosonic strings have (save for the non-thermal closed string tachyon) a convergent free energy, but as soon as the temperature is varied, thermodynamic quantities (such as the thermal entropy) diverge. Type II superstrings precisely have $T_H = T_{Hawking}$, meaning a marginal convergence is achieved for thermodynamic quantities. The major difference is that the thermal scalar state is absent for the BTZ black hole itself.

\subsection{Summary}
In this section we analyzed conical orbifolds obtained by identifying points on the cigar after a certain rotation. The partition function contains states that wind the cigar now. A surprising result is that the spectrum includes also a set of discrete states in the twisted sectors. These states appear only for $k\left|w\right| >1$. Mathematically, the appearance of the discrete states can be traced back to a correct analytic continuation of Poisson's summation formula. These discrete states are crucial, since if they are present, they dominate the continuous states. We also presented a numerical analysis of the partition function in the large $\tau_2$ limit. The results agree with the predictions in terms of the spectrum. The two major lessons we learned from the numerical analysis are the following. Firstly, the summation over $l$ and the large $\tau_2$ limit cannot be interchanged. Secondly, the infinite product gives a subdominant contribution as long as $\left|w\right| < \frac{1}{2}$. Else, the infinite product simply restores the periodicity $w \to w + 1$ that is not present in the partition function if one simply drops the infinite product. We then discussed how the Hagedorn temperature emerges for the BTZ black holes. We continued the conformal weight of the discrete representations all the way to $\left|w\right|=1$, and found a divergence for the bosonic string BTZ black hole. For type II superstrings on the BTZ black holes however, we find a marginal convergence: $\beta_H = \beta_{BTZ}$.

\section{The inclusion of a chemical potential for the $AdS_3$ string gas}
\label{chemical}
\subsection{Thermal spectrum}
A simple generalization of the $AdS$ WZW model at finite temperature is by substituting $\beta \to \beta(1+i\mu)$ in $U_{lp}$ as defined in equation (\ref{Ulp}) in the exact path integral treatment. This corresponds physically to the introduction of a chemical potential for the angular momentum of the string gas around the cigar-shaped angular submanifold. On the Euclidean manifold this is realized as the simultaneous identification $\tau \sim \tau + \beta$ and $\phi \sim \phi + \mu\beta$. As a small reminder, this can be seen by writing down the grand-canonical partition function:
\begin{equation}
\mathcal{Z} = \text{Tr}e^{-\beta(H+i\mu Q)}
\end{equation}
with $Q$ the conserved charge. This is the generator of angular rotations in our case. Again as usual anti-periodic boundary conditions for the fermions are required. These identifications lead to skew tori as fundamental domains in the ($\phi$, $\tau$) plane as shown in figure \ref{chemi} below.
\begin{figure}[h]
\centering
\includegraphics[width=10cm]{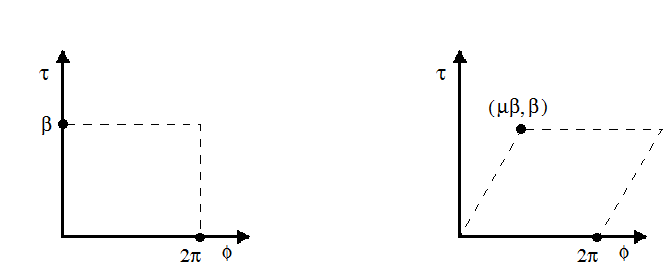}
\caption{Left figure: The identifications of the coordinates $\tau \sim \tau + \beta$ and $\phi \sim \phi + 2\pi$ define a rectangular torus as fundamental domain. Right figure: When including $\mu$, the two identifications are more involved. The first is the simultaneous identification of $\tau \sim \tau + \beta$ and $\phi \sim \phi + \mu\beta$. The second identification is again $\phi \sim \phi + 2\pi$. These define a more general skew torus as fundamental domain.}
\label{chemi}
\end{figure}
The path integral and Hamiltonian interpretation techniques that we analyzed in appendix \ref{Ham1} provide the fastest way to get the string spectrum in this case. In fact, our results on angular orbifolds in appendices \ref{Ham2}, \ref{Poisson} and \ref{Ham3} can be almost exactly copied to study this case. We present the computational details in appendix \ref{chemapp}. One finds a continuous spectrum of states with conformal weights
\begin{align}
\label{chemw1}
h^{p}_{sqn} &= \frac{s^2 +1/4}{k-2} + i\frac{\mu n p}{2} - i\frac{qp\beta}{4\pi} + \frac{ p n}{2} + \frac{kp^2\beta^2}{4(2\pi)^2}(1+\mu^2) - i \frac{\mu^2\beta qp}{4\pi}+ h_{int}, \\
\label{chemw2}
\bar{h}^{p}_{sqn} &= \frac{s^2 +1/4}{k-2} + i\frac{\mu n p}{2} - i\frac{qp\beta}{4\pi} - \frac{ p n}{2} + \frac{kp^2\beta^2}{4(2\pi)^2}(1+\mu^2) - i \frac{\mu^2\beta qp}{4\pi} + \bar{h}_{int},
\end{align}
and a set of discrete states with weights
\begin{align}
h^{p}_{jqn} &= -\frac{\tilde{j}(\tilde{j}-1)}{k-2} - i\frac{\mu n p}{2} - i\frac{qp\beta}{4\pi} - \frac{ p n}{2} + \frac{kp^2\beta^2}{4(2\pi)^2}(1+\mu^2) - i \frac{\mu^2\beta qp}{4\pi}+ h_{int}, \\
\bar{h}^{p}_{jqn} &= -\frac{\tilde{j}(\tilde{j}-1)}{k-2} - i\frac{\mu n p}{2} - i\frac{qp\beta}{4\pi} + \frac{ p n}{2} + \frac{kp^2\beta^2}{4(2\pi)^2}(1+\mu^2) - i \frac{\mu^2\beta qp}{4\pi} + \bar{h}_{int},
\end{align}
where now $ \tilde{j} = \frac{k\left|\mu p\right|\beta}{4\pi} - \frac{\left|q\right|}{2} - \frac{i\mu q}{2} \pm \frac{in\beta}{\pi} - l$. The quantum numbers take values as follows: $q \in \mathbb{Z}$, $n\in\mathbb{Z}$, $p\in\mathbb{Z}$ and $l = 0,1,2,\hdots$. The discrete states include all states that satisfy $\Re(\tilde{j}) > 1/2$. The unitarity constraint $\Re(\tilde{j}) < \frac{k-1}{2}$ is trivially satisfied for all such states, provided $k>2$.

\subsection{Dominant state and critical Hagedorn thermodynamics}
\begin{figure}[h]
\centering
\includegraphics[width=12cm]{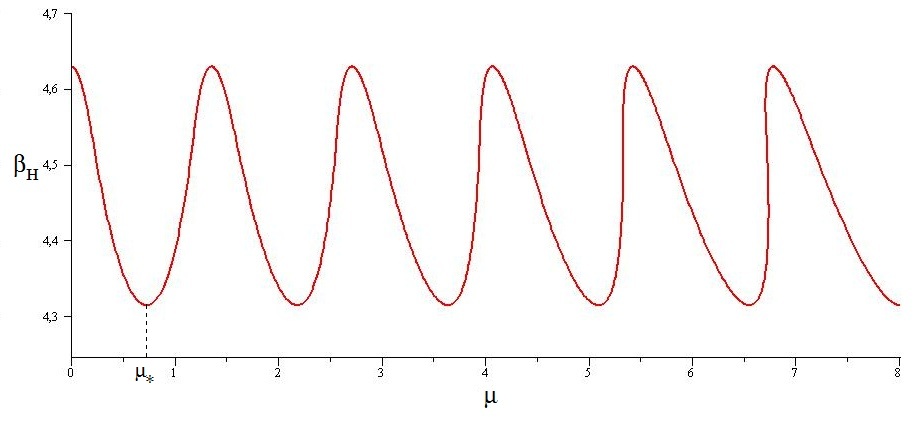}
\caption{$\beta_H$ as a function of chemical potential $\mu$ for the case $k=7$. Above the displayed curve, the system is stable. Below the curve, the system is unstable. Note that for increasing chemical potential, the curve becomes deformed.}
\label{Hagechem}
\end{figure}

\begin{figure}[h]
\centering
\includegraphics[width=6cm]{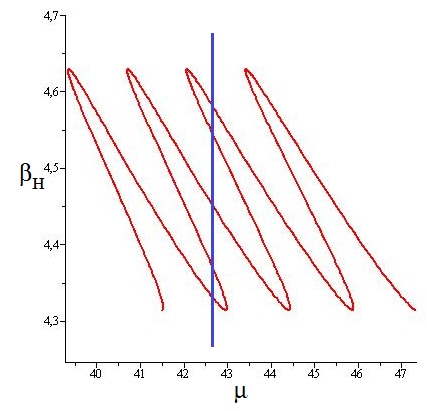}
\caption{$\beta_H$ as a function of chemical potential $\mu$ for sufficiently large $\mu$ for $k=7$. The blue line denotes a thermodynamic path one could follow in heating (or cooling) the system at fixed chemical potential. It is now apparent that multiple crossing with the critical curve are possible, indicating an alternating convergent and divergent system.}
\label{Hagechem2}
\end{figure}

\begin{figure}[h]
\centering
\includegraphics[width=7cm]{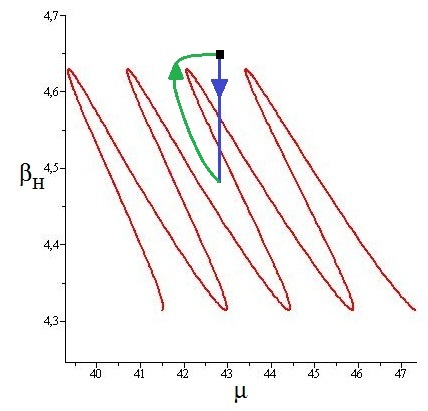}
\caption{Thermodynamical path through the ($\mu$, $\beta$) plane. We start at the black square. First the system is heated following the blue path. Then the green path is followed to finally end up at the same point from which we started.}
\label{TD}
\end{figure}

\begin{figure}[h]
\centering
\includegraphics[width=7cm]{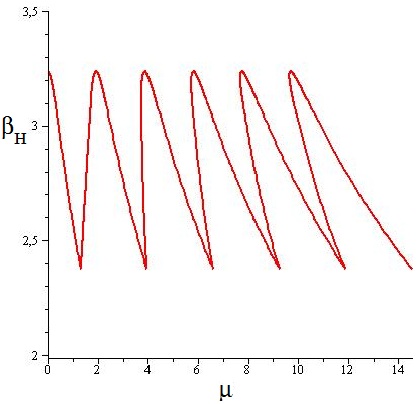}
\caption{$\beta_H$ as a function of chemical potential $\mu$ for the case $k=7$ for the type II superstring. Above the displayed curve, the system is stable. Below the curve, the system is unstable. Again the curve deforms for increasing $\mu$.}
\label{IIchem}
\end{figure}

As for the $AdS_3$ orbifolds, the dominant state can be either continuous or discrete depending on the value of $k\left|w\right|$.\\
For $k\frac{\left|\mu\right|\beta}{2\pi}<1$, the dominant state is continuous and characterized by $p=\pm1$, $n=0$ but arbitrary $q$.\\
The dominant discrete state for $k\frac{\left|\mu\right|\beta}{2\pi}>1$ is again given by considering $q=l=n=0$ and $p = \pm 1$. It is the same state as the one in the conical $AdS_3$ space characterized by $p=\pm1$, $w= \frac{\mu\beta}{2\pi} p$. \\
The numerical analysis done in the previous section can be readily extended to include this case as well.\footnote{More precisely, one should analyze
\begin{equation}
\lim_{\tau_2 \to \infty} \sum_{l\in \mathbb{Z}}\frac{e^{(2-k)\frac{\pi l^2}{\tau_2}\frac{\beta^2}{4\pi^2} + 4\pi l w\frac{\beta}{2\pi} + 2 \pi w^2\tau_2-k\pi p^2\frac{\beta^2}{4\pi^2}\tau_2}}{\left|\sin(\pi\left(\frac{p\beta}{2\pi}\tau_2 + iw\tau_2 + i\frac{\beta}{2\pi}l - \frac{\beta}{2\pi}l\mu\right))\right|^2}, 
\end{equation}
which yields results in agreement with the analytical predictions. Actually, an additional numerical consistency check can be performed. In the regime where the continuous state dominates, the prefactor of the leading behavior is expected to exhibit periodicity in $\tau_2$. According to the precise form of the weights (\ref{chemw1}) and (\ref{chemw2}), the periodicity is 
\begin{equation}
\text{lcm}\left(\frac{1}{\mu},\frac{2\pi}{\beta(1+\mu^2)}\right), 
\end{equation}
which as a sidenote exists only when the ratio of these two numbers is rational. Such a periodicity is indeed what is observed numerically.} \\
After including the spectator dimensions, the condition to find the Hagedorn temperature is given by
\begin{equation}
\frac{1}{4(k-2)} + k \frac{\beta^2}{16\pi^2}(1+\mu^2) = 1
\end{equation}
when $\frac{\left|\mu\right|\beta}{2\pi} < \frac{1}{k}$ or by
\begin{equation}
\label{discrHag}
\frac{1}{4(k-2)}-\frac{(\frac{k\left|\mu\right|\beta}{2\pi}-1)^2}{4(k-2)} + k \frac{\beta^2}{16\pi^2}(1+\mu^2) = 1
\end{equation}
when $\frac{1}{k} < \frac{\left|\mu\right|\beta}{2\pi} < \frac{1}{2}$. For even larger values of the chemical potential, periodicity of the system under $w \to w+1$ should be used. This periodicity is obvious from a Lorentzian point of view as well, since the grand-canonical partition function is given by \cite{Maldacena:2000hw}\cite{Maloney:2007ud}
\begin{equation}
\label{grandcan}
\mathcal{Z} = \sum_{n\in\mathcal{H}}e^{-\beta E_n}e^{il_n \mu\beta}
\end{equation}
for integer angular momentum $l_n$, which is manifestly periodic under
\begin{equation}
\label{periodic}
\mu \to \mu + \frac{2\pi N}{\beta}, \quad N \in\mathbb{Z}.
\end{equation}
If $\frac{\left|\mu\right|\beta}{2\pi} < \frac{1}{k}$, the Hagedorn temperature is readily found and is given by
\begin{equation}
\beta_H^2 = \frac{4\pi^2}{k(1+\mu^2)}\left(4-\frac{1}{k-2}\right),
\end{equation}
which is the expression written down in \cite{Lin:2007gi}. However, for other values of the chemical potential, the Hagedorn temperature disagrees with the above formula and should instead be determined by equation (\ref{discrHag}). For completeness, the Hagedorn temperature in this regime is given by the not-so-transparant expression:
\begin{equation}
\beta_H = -2\pi\frac{k\left|\mu\right|-\sqrt{(-7k^2\mu^2+16k\mu^2-16k^2+4k^3+16k)}}{k(-2\mu^2+k-2)}.
\end{equation}
For even larger values of the chemical potential, one needs to use the periodicity (\ref{periodic}). However, the periodic parameter is $\mu\beta$ and not $\mu$ itself. This then leads to a distortion of the $\beta_H(\mu)$ curve for larger values of $\mu$. Note that the Hagedorn temperature only depends on $\left|\mu\right|$, thus the critical curve is symmetric under $\mu \to - \mu$. The resulting critical curve is depicted in figure \ref{Hagechem}.
Note that indeed the critical curve is smooth at the points where a different formula for the Hagedorn temperature should be used, as we noticed before.
For very large values of $\mu$, the curve can even `fall over' such that for a single value of $\mu$, multiple critical temperatures exist. The interpretation is then as follows. Starting from a low-temperature gas at fixed chemical potential, we turn up the temperature. At the beginning, the system is thermodynamically stable. Then at some temperature (which is at least as high as the $\mu=0$ Hagedorn temperature), the system becomes unstable. However, when bearing through this region, one again encounters a regime where the system is stable. This stability is soon after again compromised and one re-enters the divergent Hagedorn phase. For very large temperatures, being larger than the temperature associated to $\mu_{*} = \sqrt{\frac{k(k-2)}{14k-32}}$, the system always becomes unstable. Thus there is an interval of temperatures, where depending on the chemical potentials, the system can alternate between convergent and divergent behavior. This strange zone is bounded as follows:
\begin{equation}
\frac{\sqrt{k}}{2\pi}\sqrt{\frac{k-2}{4k-9}} < T < \frac{\sqrt{k}(k-2)}{\sqrt{2}\pi\sqrt{7k^2-30k+32}},
\end{equation}
where the upper temperature is computed as $T(\mu = \mu_{*})$. This feature is illustrated in figure \ref{Hagechem2}.
Let us finally remark that the Hagedorn temperature is always at least as high as the $\mu=0$ Hagedorn temperature. This is intuitively obvious \cite{Lin:2007gi}, since we expect string states with prescribed average angular momentum to be less numerous than the general string states. Thus a higher temperature is necessary to get a sufficient number of string states to yield Hagedorn behavior.\\
The alternating divergence behavior is quite strange. In fact, in the past a similar situation arose when computing the one-loop free energy of heterotic strings in flat space \cite{O'Brien:1987pn}.\footnote{The flat space heterotic string has two critical temperatures and divergences occur only in between these two temperatures. This suggests at first sight that the heterotic string is again stable at high temperature.} It was shown \cite{Atick:1988si} that this behavior is unphysical and incompatible with the monotonicity of a canonical partition function in $\beta$. The genus zero condensate of the thermal heterotic string in flat space cannot disappear at higher temperatures. In this case however, we are using the grand-canonical partition function (\ref{grandcan}) which need not be monotonic in $\beta$. Hence the above behavior is in principle allowed. Even more so, we now present an argument that the persistence of the genus zero condensate as in \cite{Atick:1988si} cannot occur here. The crucial ingredient is the fact that we have an extra parameter $\mu$ to play with. Consider a thermodynamical path as given in figure \ref{TD}.
First we follow the blue curve by heating the system. Then we follow a suitable trajectory in the ($\mu$, $\beta$) plane (the green curve) that does not cross the critical curve. We end up at the same point. This point could have been chosen arbitrarily low in temperature and hence we expect the system to converge there in the low-temperature phase, both initially and finally after following the thermodynamical process. However, a genus zero condensate forms as soon as one crosses the critical curve the first time. If this condensate persists after crossing the critical curve the second time, then this path shows that it is possible to cool down the system without any more crossings, which implies that the low temperature system we obtain after this entire process would still contain a genus zero condensate. This is impossible.\\
Generalizing this argument, if the convergent region in the plane of parameters (here $\mu$ and $\beta$) is connected then the entire critical curve is important and the genus zero condensate can disappear again at higher temperatures. If it is not connected, one cannot a priori know what will happen at higher temperature as is the case for the heterotic string in flat space. In that case, the canonical partition function itself shows that one cannot return to a convergent region with the same degrees of freedom as the low-temperature phase.\\
For completeness, we present the analogous picture for the type II superstring (assuming the obvious replacements) in figure \ref{IIchem}.
Peculiar to note is that the smooth gluing of the piecewise defined function for the bosonic string is not present here: the critical curve exhibits points where it is non-differentiable.

\subsection{BTZ with $\mu \neq 0$}
In the previous section, we noted that the BTZ thermal spectrum does not include the thermal scalar state. We then considered conical orbifolds and found that the thermal scalar reappears. A different question one could ask is whether introducing a chemical potential for the BTZ black hole can cause the thermal scalar to appear. Let us briefly look into this.
One readily finds the generalization of the relation between $AdS_3$ and BTZ for non-vanishing chemical potential (in terms of $AdS_3$ parameters):
\begin{align}
\label{firstid}
\tau &\sim \tau + \frac{2\pi r_+ }{l}, \\
\phi &\sim \phi + 2\pi, \, \tau \sim \tau + 2\pi \mu_{BTZ}.
\end{align}
This is a situation we have not yet analyzed, both identications include the temporal dimension. The $\phi$ identification is trivial and we can drop it. Since the identification is purely in the $AdS$ time direction, we do not encounter the problems with the $w \to w + 1$ periodicity shift. The result is an $AdS_3$ where the thermal direction includes two independent identifications: a double toroidal model. More elaborate computations for this set-up will not be treated here. We just remark that cigar-winding states make a reappearance here due to the non-trivial identification $\tau \sim \tau + 2\pi \mu_{BTZ}$.
We hope to come back to this model in the future.

\subsection{Summary}
In this section, we presented results on the treatment of the string gas in $AdS_3$ spacetimes (and its orbifolds) when including a chemical potential for the angular momentum. The techniques used are very similar to those we employed in the previous section when analyzing conical orbifolds. We found the thermal spectrum and the thermal scalar for $AdS_3$ in this case. The periodicity of $\mu$ then allowed us to construct the entire critical curve in the ($\mu,\beta)$ plane. This extends the work of \cite{Lin:2007gi}. A curious feature we discovered was that for large enough chemical potentials, the system can change between stable and unstable several times over a certain range of temperatures. This is, as far as we know, an effect that has not been noted before.

\section{The $AdS_3$ field theory point of view}
\label{fieldtheory}
After the exact analysis of the primaries in the previous sections, we now take a closer look at the thermal scalar field equation and its possible $\alpha'$-corrections. Armed with the exact thermal spectrum on $AdS_3$ in (\ref{ads3spectrum1}) and (\ref{ads3spectrum2}), we look at the lowest order (in $\alpha'$) action and we analyze to what extent the eigenvalues reproduce the exact spectrum. 

\subsection{Winding states from the field theory action}
Firstly, we look into winding states. In general, the lowest order (in $\alpha'$) action for winding modes (with winding $p$) including a NS-NS background is given by
\begin{align}
S \sim \int d^{D-1}x \sqrt{G}e^{-2\Phi} \nonumber \\
\times\left(\tilde{G}^{ij}\partial_{i}T_{p}\partial_{j}T_{p}^{*}+\frac{p^2R^2\tilde{G}^{00}}{\alpha'^2}T_{p}T_{p}^{*}+ \tilde{G}^{0i}\frac{ipR}{\alpha'}\left(T_{p}\partial_{i}T_{p}^{*}- T_{p}^{*}\partial_{i}T_{p}\right)+ m^2T_{p}T_{p}^{*}\right),
\end{align}
where $\tilde{G}^{\mu\nu}$ denotes the T-dual metric. In our case, the T-dual metric and its inverse are given in matrix notation with ordering ($\tau$, $\rho$, $\phi$) by
\begin{equation}
\tilde{G}_{\mu\nu}=\left[\begin{array}{ccc} 
\frac{1}{l^2\cosh(\rho)^2} & 0 & -i\tanh(\rho)^2 \\
0 & l^2 & 0 \\
-i\tanh(\rho)^2 & 0 & l^2\sinh(\rho)^2 - l^2\sinh(\rho)^2\tanh(\rho)^2 \\
\end{array}\right]
, \quad \tilde{G}^{\mu\nu}=\left[\begin{array}{ccc} 
l^2 & 0 & i\\
0 & 1/l^2 & 0 \\
i & 0 & \frac{1}{l^2\sinh(\rho)^2} \\
  \end{array}\right],
\end{equation}
where $l^2=k\alpha'$. Note that the $\tilde{G}^{00}$ metric component is constant: the background NS-NS field has cancelled the confining $G_{00}$ potential $\propto \cosh^2(\rho)$.
So we obtain
\begin{align}
S \propto \int d^{D}x \sinh(\rho)\cosh(\rho)&\left[\frac{1}{l^2}\partial_\rho T \partial_\rho T^* + \frac{1}{l^2\sinh(\rho)^2}\partial_\phi T \partial_\phi T^* \right. \nonumber \\
&\left.- \frac{pR}{\alpha'}\left(T\partial_{\phi}T^{*}- T^{*}\partial_{\phi}T\right) + \frac{R^2p^2l^2}{\alpha'^2}TT^*+ m^2 TT^*\right].
\end{align}
Since $\phi \sim \phi + 2\pi$, we can expand in Fourier modes $\propto e^{iq\phi}$ for $q\in\mathbb{Z}$. The $q^{th}$ mode corresponds to the following eigenvalue equation:
\begin{equation}
\frac{1}{l^2}\left(-\partial_{\rho}\partial_{\rho} T - 2\coth(2\rho)\partial_{\rho} T + \frac{q^2}{\sinh(\rho)^2}T\right) + \frac{2iqpR}{\alpha'}T + \frac{R^2p^2l^2}{\alpha'^2}T - \frac{4}{\alpha'}T = \lambda T.
\end{equation}
When looking for $\delta$-normalizable eigenmodes, the set of eigenfunctions need to decay faster than $e^{-\rho}$ to compensate for the growing measure factor $\sim e^{2\rho}$ as $\rho\to\infty$. 
One can show that the eigenvalues of the first two terms are given by
\begin{equation}
\lambda = \frac{4s^2 + 1}{l^2},
\end{equation}
where $s$ is a real number. The restriction to $\lambda > 1/l^2$ arises precisely due to this restriction on normalizability of the wavefunctions. Using numerical methods, we checked that the same spectrum is obtained when including the third term containing the $q$ quantum number. The eigenvalues $\lambda$ of the above operator are hence
\begin{equation}
\lambda = \frac{4s^2 + 1}{l^2} + \frac{iqp\beta}{\pi\alpha'} + \frac{\beta^2p^2l^2}{(2\pi)^2\alpha'^2} - \frac{4}{\alpha'}.
\end{equation}
Multiplying by $\alpha'/4$, we can clearly see aspects of the exact conformal weight spectrum (\ref{ads3spectrum1}) and (\ref{ads3spectrum2}) appear:
\begin{equation}
h = \frac{s^2 + 1/4}{k} + \frac{iqp\beta}{4\pi} + \frac{k\beta^2p^2}{4(2\pi)^2} - 1.
\end{equation}
The criterion for non-negative eigenvalues is equivalent to the criterion for conformal weights larger than one. The only discrepancy is the $k \to k-2$ appearing in the denominator of the Laplacian, to which this field theory analysis is a priori insensitive.

\subsection{Exact WZW analysis for non-winding states}
Now we look into the non-winding states, but instead of using the lowest order (in $\alpha'$) spacetime action, we utilize the geometrization of the (inverse) string propagator $L_0 + \bar{L}_0$. Following the treatment of \cite{Dijkgraaf:1991ba} for the gauged WZW case, we will use this geometrization to deduce the form of the spacetime action of the non-winding modes. Using formulas given in appendix \ref{WZWapp}, one can write the zero-modes of the currents as differential operators when acting on vertex operators. We find the following form for the holomorphic currents:
\begin{align}
\hat{D}^3 &= -\frac{1}{2i}\partial_\phi + \frac{1}{2}\partial_\tau, \\
\hat{D}^{+} &= i\frac{e^{\tau-i\phi}}{2}\left[-\partial_\rho + i\coth(\rho) \partial_\phi + \tanh(\rho)\partial_\tau \right], \\
\hat{D}^{-} &= -i\frac{e^{-\tau+i\phi}}{2}\left[\partial_\rho + i\coth(\rho) \partial_\phi + \tanh(\rho)\partial_\tau \right].
\end{align}
Analogous formulas hold for the antiholomorphic sector as given in appendix \ref{WZWapp}. 
One can readily check that indeed these operators satisfy the $\mathfrak{sl}(2,\mathbb{R})$ algebra:
\begin{align}
\left[\hat{D}^3, \hat{D}^{\pm}\right] = \pm \hat{D}^{\pm}, \\
\left[\hat{D}^+, \hat{D}^-\right] = -2 \hat{D}^{3}.
\end{align}
After some more algebra, the $L_{0}$ and $\bar{L}_0$ operators can be found using the Sugawara construction. These are equal for this case and given by
\begin{equation}
L_0 = \bar{L}_0 = \frac{1}{4(k-2)}\left(-\partial_{\rho}\partial_{\rho} T - 2\coth(2\rho)\partial_{\rho} T - \frac{1}{\sinh(\rho)^2}\partial_{\phi}^2T - \frac{1}{\cosh(\rho)^2}\partial_{\tau}^2T\right). 
\end{equation}
This expression coincides with the scalar Laplacian on the group manifold. We arrive at the eigenvalue equation:
\begin{equation}
\frac{1}{4(k-2)}\left(-\partial_{\rho}\partial_{\rho} T - 2\coth(2\rho)\partial_{\rho} T - \frac{1}{\sinh(\rho)^2}\partial_{\phi}^2T - \frac{1}{\cosh(\rho)^2}\partial_{\tau}^2T\right) - T = \lambda T. 
\end{equation}
Like before, modes with $\phi$ dependence do not give a modification of the spectrum. Similarly, modes with $\tau$ dependence also do not modify the spectrum (obviously both of these do modify the eigenfunctions). The resulting eigenvalues are then 
\begin{equation}
\lambda = \frac{s^2+1/4}{k-2}.
\end{equation}
This is exactly as expected: discrete momentum modes in the $\phi$ and/or the $\tau$ directions do not alter the conformal weights.

\subsection{Thermal scalar action}
\label{thscaction}
Combining the results from the previous subsections and the exactly known spectrum, we propose the following exact form of the eigenvalue equation for primary operators on $AdS_3$:
\begin{equation}
\label{tsequation}
\frac{1}{4(k-2)}\left(-\partial_{\rho}^2 T - 2\coth(2\rho)\partial_{\rho} T + \frac{q^2}{\sinh(\rho)^2}T + \frac{n^2 4\pi^2}{\beta^2\cosh(\rho)^2}T\right) - \frac{iqp\beta}{4\pi}T + \frac{k\beta^2p^2}{4(2\pi)^2}T = T.
\end{equation}
In particular, the thermal scalar action (with $q=0$, $n=0$ and $p=\pm1$) becomes
\begin{equation}
\frac{1}{4(k-2)}\left(-\partial_{\rho}^2 T - 2\coth(2\rho)\partial_{\rho} T\right)  + \frac{k\beta^2}{4(2\pi)^2}T - T = 0.
\end{equation}
Compared to the lowest order (in $\alpha'$) thermal scalar action, the only difference is the shift $k \to k-2$ for the Laplacian term. For the type II superstring, this shift does not occur, and just like for the WZW cigar discussed in \cite{Dijkgraaf:1991ba}, the lowest order effective action is exact (for describing the on-shell conditions). \\
We see that the field theory action reproduces the string spectrum (up to $k\to k-2$ for the bosonic string). This also clearly shows the physical interpretation of the different quantum numbers we introduced in section \ref{spectrumsection}: $q$ represents the discrete momentum around the angular $\phi$-cigar, $w$ is the winding around this cigar, $n$ denotes the discrete momentum around the thermal circle whereas $p$ is the winding around the thermal circle.

\subsection{Flat space limit}
In \cite{Giveon:2013ica}\cite{Giveon:2014hfa}, the authors utilize the flat limit of the $SL(2,\mathbb{R})/U(1)$ to get information on string theory in polar coordinates (Euclidean Rindler space). Also for the thermal $AdS_3$ manifold, one has the opportunity to look at the flat limit as $k\to\infty$. What happens in this case? \\
The temporal part of the background becomes a flat toroidal dimension. However, looking back at the spectrum (\ref{ads3spectrum1}) and (\ref{ads3spectrum2}), we see that there is no $n^2$ contribution. One can see how this comes about by looking at the spacetime equation of motion (\ref{tsequation}). Discrete momentum states clearly do not influence the spectrum for any finite value of $k$, which one can see by looking at the large $\rho$ asymptotics. However, if $k\rho^2$ is held fixed, $\cosh \to 1$, and the discrete momentum term no longer vanishes in the asymptotic large $\rho$ region. Thus this term represents an additive contribution to the eigenvalue which becomes the addition $\frac{n^2\pi^2\alpha'}{\beta^2}$ to the conformal weight. This is precisely the missing term. The same story happens for the $\phi$ part. The geometry asymptotes to flat space in polar coordinates, but the discrete momentum part $q^2$ is missing. Although geometrically the space turns to flat space in polar coordinates, the potential term is still cancelled by the Kalb-Ramond background, a feature which is not present in purely flat space. Despite the resemblances with the $SL(2,\mathbb{R})/U(1)$ cigar (as discussed in section \ref{conical}), this shows that taking the large $k$ limit of this space is not a good starting point to analyze string theory in polar coordinates, unlike the $SL(2,\mathbb{R})/U(1)$ cigar CFT itself as studied by \cite{Giveon:2012kp}\cite{Giveon:2013ica}\cite{Giveon:2014hfa}. \\
It is precisely the absence of the $n^2$ and $q^2$ terms in the conformal weights that causes the simultaneous marginality of these states at the Hagedorn temperature. Unlike the majority of the pathological properties of this model, the physical reason for this is \emph{not} the Kalb-Ramond field. The ever-increasing circumference of the angular cigar and the temporal cylinder, i.e. the asymptotic geometry, is the culprit here. Naively, when a compact dimension has a large circumference, discrete momentum modes are contributing little energy. This is the analogous effect of winding modes becoming light when the compact dimension becomes very small. \\
Note that also in Euclidean Rindler space the thermal circle becomes infinitely large at infinity. However, inspection of the conformal weights \cite{Mertens:2013zya} shows that in that case the degeneracy of marginal states does not occur. The above feature is hence not generic.

\subsection{Random walk behavior in $AdS_3$}
Equiped with the field theory action corresponding to (\ref{tsequation}) for the primaries, the random walk picture can be completed now. If the only critical state present were the state with $q=n=0$ (the thermal scalar), then the random walk displayed in (\ref{randwalk}) would not be modified (except for the bosonic $k\to k-2$ shift). However, we saw in section \ref{spectrumsection} that the states with arbitrary $q$ (and $n$, but as discussed above this is irrelevant for thermodynamical quantities) all become marginal at the Hagedorn temperature. Hence all of these contribute to the critical behavior. The random walk should hence contain a sum over these quantum numbers: 
\begin{equation}
Z_p = \sum_{w=\pm 1}\sum_{q\in\mathbb{Z}}\int_{0}^{+\infty}\frac{d\tau_2}{2\tau_2}\int\left[\mathcal{D}X\right]\exp\left(-S_p\right).
\end{equation}
For definiteness, we focus on the type II superstring. Utilizing the explicit action for the primaries, the particle action becomes
\begin{align}
S_p = \frac{k}{4\pi}\int_{0}^{\tau_2}dt\left[(\partial_t \rho)^2 + (\beta^2\cosh(\rho)^2-\beta_{H,flat}^2)+\sinh(\rho)^2(\partial_t\phi)^2 + 2w \frac{\beta}{2\pi\alpha'}\sinh(\rho)^2\partial_t\phi \right.\nonumber \\
\left. + \frac{4\pi^2}{k^2}\left\{\frac{3}{4} + \frac{1}{4\cosh(\rho)^2}\right\} - \frac{i4\pi\beta qw}{k} + \frac{4\pi^2 q^2}{k^2\sinh(\rho)^2} \right].
\end{align}

\subsection{Cigar-winding states}
\label{cigarwind}
Previously we stated that the cigar-winding string states are absent for $AdS_3$ and BTZ. However, we did (formally) obtain these in section \ref{spectrumsection}. Moreover, we saw in appendix \ref{Ham2} that conical spaces reintroduce these states. It thus seems worthwile to look at the field theory equation for such states. One can obtain these simply by T-dualizing along the $\phi$ direction. The T-dual (in the $\phi$-direction) metric components are given by
\begin{equation}
\tilde{G}_{\phi\phi} = \frac{1}{l^2\sinh(\rho)^2}, \quad \tilde{G}_{\phi \tau} = \frac{B_{\phi\tau}}{G_{\phi\phi}} = i, \quad \tilde{G}_{\tau\tau} = G_{\tau\tau} + \frac{B_{\phi\tau}B_{\phi\tau}}{G_{\phi\phi}} = l^2.
\end{equation}
The T-dual metric and its inverse are hence given by (the ordering of the coordinates in the matrices is now ($\phi$, $\rho$, $\tau$)):
\begin{equation}
\tilde{G}_{\mu\nu}=\left[\begin{array}{ccc} 
\frac{1}{l^2\sinh(\rho)^2} & 0 & i \\
0 & l^2 & 0 \\
i & 0 & l^2 \\
\end{array}\right]
, \quad \tilde{G}^{\mu\nu}=\left[\begin{array}{ccc} 
l^2\tanh(\rho)^2 & 0 & -i\tanh(\rho)^2\\
0 & 1/l^2 & 0 \\
-i\tanh(\rho)^2 & 0 & \frac{1}{l^2\cosh(\rho)^2} \\
  \end{array}\right]
\end{equation}
This leads to the following eigenvalue equation for pure winding states:
\begin{equation}
\label{phiwinding}
\frac{1}{l^2}\left(-\partial_{\rho}\partial_{\rho} T - 2\coth(2\rho)\partial_{\rho} T \right) + \frac{w^2l^2}{\alpha'^2}\tanh(\rho)^2 T - \frac{4}{\alpha'}T = \lambda T.
\end{equation}
In general, the spectrum of this eigenvalue equation contains both a discrete part and a continuous part. The continuous eigenvalues are:
\begin{equation}
\lambda = \frac{4s^2+1}{k} + \frac{w^2l^2}{\alpha'^2} - \frac{4}{\alpha'}
\end{equation}
and the second term is clearly the pure cigar-winding contribution $\frac{kw^2}{4}$ to the conformal weights. \\
The field theory of point view also exhibits the discrete states with the correct eigenvalues, again modulo the substitution $k \to k-2$ in the Laplacian operator. Let us briefly analyze this in more detail. We study the eigenvalue equation
\begin{equation}
\left(-\partial_{\rho}\partial_{\rho} T - 2\coth(2\rho)\partial_{\rho} T \right) + (kw)^2\tanh(\rho)^2 T  = \lambda T
\end{equation}
which is obtained from equation (\ref{phiwinding}) by setting $\lambda_{new} = l^2\lambda_{old} + \frac{4}{\alpha'}l^2$. \\
Numerically, one finds discrete bound states when 
\begin{align}
\lambda &= 2nk\left|w\right| - (n^2-1), \quad n = 1,3,5, \hdots  \\
&= -4\left(\frac{k\left|w\right|}{2}-l\right)\left(\frac{k\left|w\right|}{2}-1-l\right) + (kw)^2, \quad l = 0,1,2, \hdots
\end{align}
and indeed, precisely for these values of $\lambda$ one encounters the discrete states of expressions (\ref{disc1}) and (\ref{disc2}).\footnote{Again, with the remark that one should take $k\to k-2$ in the term associated to the Laplacian.} The continuum starts at the eigenvalue
\begin{equation}
\lambda^{*} = 1 + (kw)^2
\end{equation}
and discrete states have eigenvalues lower than this bound since
\begin{equation}
2nk\left|w\right| - (n^2-1) < 1 + (kw)^2 \quad \Leftrightarrow \quad (k\left|w\right|-n)^2 > 0
\end{equation}
which is automatically satisfied obviously. As an illustration, we draw the wavefunctions for the case $kw=6$ in the following figures \ref{bs} and \ref{cs}. Three bound state wavefunctions are found, and indeed the inequality
\begin{equation}
\frac{k\left|w\right|}{2}-l > \frac{1}{2}, \quad l=0,1,2,\hdots
\end{equation}
has three solutions. The unitarity bound $\tilde{j} < \frac{k-1}{2}$ is also satisfied for $\left|w\right|<\frac{1}{2}$ as we have shown earlier.\\
\begin{figure}[h]
\centering
\begin{minipage}{.3\textwidth}
  \centering
  \includegraphics[width=\linewidth]{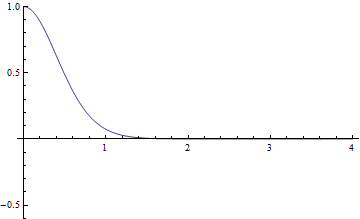}
  \caption*{(a)}
\end{minipage}
\begin{minipage}{.3\textwidth}
  \centering
  \includegraphics[width=\linewidth]{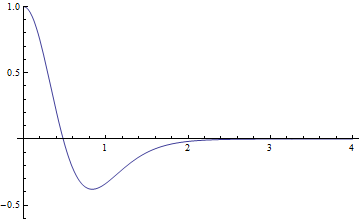}
  \caption*{(b)}
\end{minipage}
\begin{minipage}{.3\textwidth}
  \centering
  \includegraphics[width=\linewidth]{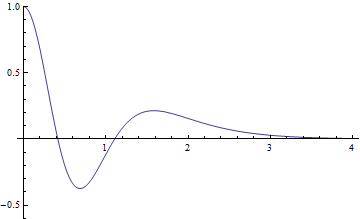}
  \caption*{(c)}
\end{minipage}
\caption{Bound state solutions for $kw=6$ as a function of radial distance $\rho$ with $T(0)=1$ chosen as normalization. (a) Lowest bound state with $\lambda=12$. (b) Second bound state with $\lambda = 28$. (c) Final bound state with $\lambda = 36$.}
\label{bs}
\end{figure}
\begin{figure}[h]
\centering
  \includegraphics[width=0.4\linewidth]{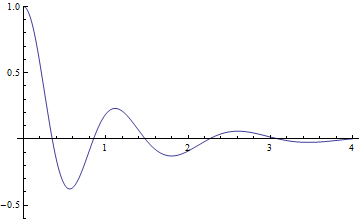}
  \caption{Example of a continuous wavefunction for $kw=6$ with $\lambda=50$ as a function of radial distance $\rho$ with $T(0)=1$ chosen as normalization.. Such states have eigenvalues larger than those of the discrete spectrum.}
  \label{cs}
\end{figure}
\noindent The equation of motion (\ref{phiwinding}) provides another argument in favor of the fact that one should not use the $w \to w+1$ periodicity of the partition function to obtain the Hagedorn divergence for the BTZ black hole. Equation (\ref{phiwinding}) determines the dominant state for $w=\frac{1}{N}$ for the $\mathbb{Z}_N$ orbifold of the BTZ black hole. We expect that one need only continue this wave equation to $N=1$ to obtain the dominant behavior for the BTZ black hole itself. This wave equation includes both continuous and discrete states (if present) and the lowest eigenvalue determines the critical behavior of the string gas. It would be very strange indeed if this equation would need to be drastically altered as soon as we are interested in $\left|w\right| > \frac{1}{2}$.

\subsection{Summary}
In this section we have taken a field theory point of view. This allowed us to clearly observe the physical meaning of the quantum numbers $q$ and $n$ that we introduced in section \ref{spectrumsection}. Since we have seen that the thermal scalar action is not modified (save for the $k\to k-2$ for bosonic strings) from the lowest $\alpha'$ effective thermal scalar action, the random walk picture exhibited in (\ref{randwalk}) is also at first sight not modified. However, there are infinitely many string states becoming marginal at the Hagedorn temperature. Hence the random walk should contain a sum over discrete momenta around the angular cigar. This random walk is not localized to the $AdS$ origin, since the Kalb-Ramond background field precisely compensates the gravitational potential. These results solve the random walk problem in the $AdS_3$ WZW target space. Cigar-winding modes exhibit explicitly a discrete part in their spectrum, precisely corresponding to the expected conformal weights.

\section{Conclusions and outlook}
\label{conclusion}
We have discussed thermal properties of the $AdS_3$ and BTZ WZW models from the thermal spectrum. Firstly we briefly analyzed the form of $\alpha'$ corrections to the thermal scalar action. Then we used CFT twisting techniques to fully determine the string spectrum on the thermal manifolds. We explicitly found the thermal scalar in the string spectrum with the correct mass. The thermal scalar is not localized to the $AdS$ origin but instead fluctuates all over space (it is in a continuous representation of the symmetry group). The reason is the Kalb-Ramond flux whose repulsion precisely compensates the gravitational attraction. For the BTZ black hole, the thermal scalar is not present in the string spectrum. The state however reappears when considering sufficiently conical spaces. We discovered that the twisted sectors on the cone also exhibit discrete modes. From a mathematical perspective, these are found by properly performing the analytic continuation of Poisson's summation formula. We discovered that these discrete states, if they are present, dominate the continuous states. Hence they are important for the critical thermodynamics. After that, using techniques we employed to analyze the temporal BTZ orbifolds, we have analyzed the generalization of the $AdS_3$ string gas by including a chemical potential corresponding to the angular momentum on the cigar. We found the critical curve in the ($\mu,\beta$) plane for the $AdS_3$ string gas and discovered a peculiar effect where the string gas alternated between stable and unstable in a certain temperature interval. In the final section, we looked at the field theory point of view. The lowest order $\alpha'$ thermal scalar action reproduces the correct conformal weights for the type II superstring and we expect it to be exact. Bosonic string actions are nearly exact, the only difference is the shift $k\to k-2$ in the kinetic term. With all these techniques, we have written down the random walk picture of the string on $AdS_3$ spacetime. A strange feature is that arbitrary $q$ quantum numbers are all simultaneously dominant. \\
Throughout this paper, we have presented four different methods to analyze the thermal spectrum on $AdS_3$ space and its orbifolds. Let us compare these.
\begin{itemize}
\item{
The first method we presented in section \ref{spectrumsection} utilized worldsheet CFT twisted operator methods to twist the non-thermal Euclidean CFT into the thermal one. The approach is elegant and computationally easy to carry out, though it has some subtle points in that the derivation is not airtight: we needed to take a plausible guess when considering the conserved charges in the twisted sectors. In spite of this, this approach leads to a solid prediction of the thermal spectrum whose quantum numbers can be directly related to physical quantities. However, this approach is not sensitive to some constraints on the quantum numbers. We view it as a blueprint of the exact result: the precise thermal spectrum needs to be taylored to the form predicted using this vertex operator method.
}
\item{
A second approach we followed utilized the (lowest order in $\alpha'$) field theory equations of motion in the curved background (using the $SL(2,\mathbb{R})$ Laplacian). This approach is a priori not sensitive to possible $\alpha'$ corrections, and it misses the $k \to k-2$ shift in the denominator of the kinetic Laplacian term. For type II superstrings, this approach does not have this problem. Modulo this difficulty, one correctly predicts all possible string states. This method clearly demonstrated the possibility of discrete states when $w\neq0$. The downside is that, just like the vertex operator method, one is not sensitive to constraints, although normalizability $\tilde{j}>\frac{1}{2}$ can be directly checked and this constraint is most transparent when using this method. 
}
\item{
Thirdly, we looked at the exact partition function using numerical methods (i.e. truncating the series and taking the large $\tau_2$ limit cautiously). This approach, when utilized with care, is the most failsafe option we have. Of course, one is limited to certain questions only, for instance the large $\tau_2$ limit is accessible but moderate values of $\tau_2$ are hard to analyze. We detected the presence of discrete states with this and this was our only way to analyze what the infinite product contributes to the final result. It is of course difficult to analyze what one finds if we do not know where to look in advance: the physical meaning of the results are obscured.
}
\item{
Finally, we turned to the Hamiltonian formulation of the one-loop partition function. This approach is in principle the best one, since one can distill all states and the unitarity constraints should be visible. However, this approach is the most cumbersome to use. Moreover, in order to rewrite this in the desired way, one needs to know in advance what the conformal weights of the states might look like. It is at this point that the vertex operator method is ideally suited to provide insight.
}
\end{itemize}
In this paper, we used a combination of all four of these methods to analyze the thermal spectrum and the critical Hagedorn behavior. \\
Some open avenues that can be further explored are for instance a more elaborate treatment of the type II superstring on these spaces. We have merely displayed the expected changes, though a more thorough analysis would be welcome. \\
Another possibility is to try to get further insight in the Hagedorn temperature for the BTZ black hole. The continuation in the orbifold integer $N$ is somewhat dubious in this case, more so than for the Euclidean Rindler case or the $SL(2,\mathbb{R})/U(1)$ cigar CFT.\\
A further front on which progress can be made is the treatment of the infinite product of oscillators in (\ref{numer}). Out of the four methods discussed above, the only approach we have successfully applied to treat this product is the numerical approach. The difficulty is that the factors in the infinite product must in principle be series-expanded in different series depending on the value of $l$. But the sum over $l$ needs to be treated all at once to be able to apply the Poisson summation techniques to it. We leave a better treatment of this as an open issue.\\
Another path that can be explored, is the treatment of the $AdS_3$ string gas with chemical potential using saddle point methods along the lines of \cite{Lin:2007gi} and see whether the critical Hagedorn curve can be fully explained.\\
We have also merely provided a starting point for the treatment of BTZ black holes when including chemical potentials. Also here progress can be made.\\
It is also known for quite some time now that the WZW $AdS_3$ model is related to other CFTs by exactly marginal perturbations on the worldsheet \cite{oai:arXiv.org:hep-th/9407198}\cite{oai:arXiv.org:hep-th/9303016}\cite{Israel:2003ry}. It would be interesting to learn how the thermal spectrum and the critical Hagedorn thermodynamics of the string gas behaves along this marginal line of CFTs.\\
To conclude, we have illustrated the general results of \cite{theory} in a concrete non-trivial example. The apparently marginal behavior of all $q \in \mathbb{Z}$ states and the absence of the thermal scalar for BTZ black holes studied here, urges us to be careful when considering non-trivial geometries, especially spaces with topologically trivial thermal circles. 

\section*{Acknowledgements}
The authors would like to thank David Dudal for several valuable discussions. TM thanks the UGent Special Research Fund for financial support. The work of VIZ was partially supported by the RFBR grant 14-02-01185.

\appendix

\section{\boldmath $SL(2,\mathbb{R})$ and $SL(2,\mathbb{C})/SU(2)$ WZW models}
\label{WZWapp}

In this appendix we provide some background material concerning WZW models and in particular the two models relevant for the $AdS_3$ background. We also establish our conventions and provide several formulas for later reference.

\subsection{$SL(2,\mathbb{R})$ model}
We follow the conventions of \cite{Martinec:2001cf}\cite{Rangamani:2007fz}. The Wess-Zumino-Witten (WZW) model is given by
\begin{equation}
S = \frac{k}{8\pi}\int d^2\sigma \text{Tr}\left(g^{-1}\partial_{\mu}g g^{-1} \partial^{\mu}g\right) + k \Gamma_{WZ},
\end{equation}
where 
\begin{equation}
\Gamma_{WZ} = \frac{i}{12\pi}\int_{M^3}\text{Tr}\left(\omega \wedge \omega \wedge \omega\right).
\end{equation}
The Maurer-Cartan 1-form is denoted by $\omega = g^{-1}dg$. The first term of the WZW action is actually the traced square of the Maurer-Cartan 1-form and is as such the natural metric to put on a group manifold (originating from the Cartan-Killing metric on the algebra).\footnote{This holds for semi-simple groups. For non-semi-simple groups, different bilinear symmetric forms exist other than the Cartan-Killing form. The Cartan-Killing form itself is degenerate in this case and is not a valid starting point to construct a metric on the group manifold. See \cite{Nappi:1993ie}.} From this we conclude that the string model has a metric background equal to the Cartan-Killing metric. The second Wess-Zumino term is necessary to ensure conformal invariance at the quantum level and can be interpreted as a background Kalb-Ramond background for the string model. 
In our case $g$ is a $SL(2,\mathbb{R})$ matrix. We choose the following basis of generators for the $\mathfrak{sl}(2,\mathbb{R})$ Lie algebra
\begin{equation}
\tau^{1} = \frac{i}{2}\sigma^{3}, \quad \tau^{2} = \frac{i}{2}\sigma^{1}, \quad \tau^{3} = \frac{1}{2}\sigma^{2}.
\end{equation}
where the $\sigma^{i}$ are the Pauli matrices:
\begin{equation}
\sigma^{1} = \left[ 
\begin{array}{cc}
0 & 1 \\
1 & 0  \end{array} 
\right],
\quad
\sigma^{2} = \left[ 
\begin{array}{cc}
0 & -i \\
i & 0  \end{array} 
\right],
\quad
\sigma^{3} = \left[ 
\begin{array}{cc}
1 & 0 \\
0 & -1  \end{array} 
\right].
\end{equation}
These generators satisfy the Lie algebra
\begin{equation}
\left[\tau^{a},\tau^{b}\right]=i{\epsilon^{ab}}_{c}\tau^{c}
\end{equation}
where $\epsilon^{123} = 1$ and indices are raised and lowered with the metric $\eta^{ab} = $diag$(+1,+1,-1)$.
The Cartan-Killing metric is not proportional to the unit matrix in this case:
\begin{equation}
\text{Tr}(\tau^{a}\tau^{b})= -\frac{1}{2}\eta^{ab}.
\end{equation}
It is not negative definite due to the non-compactness of the $SL(2,\mathbb{R})$ manifold. The $SL(2,\mathbb{R})$ group element can be written in general as\footnote{For convenience, we now rescale the coordinate fields by $\sqrt{\alpha'}$ to make them dimensionless.}
\begin{eqnarray}
&g = e^{i\frac{t+\phi}{2}\sigma_2}e^{\rho\sigma_3}e^{i\frac{t-\phi}{2}\sigma_2} \\
&= \left[
\begin{array}{cc}
\cos(t)\cosh(\rho)+\cos(\phi)\sinh(\rho) & \sin(t)\cosh(\rho)-\sin(\phi)\sinh(\rho) \\
-\sin(t)\cosh(\rho)-\sin(\phi)\sinh(\rho) & \cos(t)\cosh(\rho)-\cos(\phi)\sinh(\rho) \end{array} 
\right].
\end{eqnarray}
This is a parameterization of the group manifold in coordinates ($t$, $\rho$, $\phi$). Notice that the WZW model is written in a manifestly coordinate invariant way (intrinsic on the group manifold). Coordinate transformations simply correspond to choosing a different parameterization of the element $g$.

\subsubsection{String background field from WZW action}
Let us first compute the background fields by identifying with the non-linear sigma model.
With the above parametrization of $g$, we can evaluate the WZW action here explicitly. We need to read off the background metric and NS field by comparing with the standard non-linear sigma model
\begin{equation}
S = \frac{1}{4\pi\alpha'}\int d^{2}\sigma \left(\delta^{ab}G_{\mu\nu} + i \epsilon^{ab} B_{\mu\nu}\right) \partial_a X^{\mu}\partial_b X^{\nu}
\end{equation}
where $\epsilon^{12} = 1$ and a flat worldsheet metric was chosen. \\
The first term in the WZW action corresponds indeed to the background metric since 
\begin{equation}
\text{Tr}\left(g^{-1} \partial^{a} g g^{-1} \partial_a g\right) = \text{Tr}\left(g^{-1} \frac{\partial g}{\partial X^{\mu}} g^{-1} \frac{\partial g}{\partial X^{\nu}}\right) \frac{\partial X^{\mu}}{\partial \sigma_a}\frac{\partial X^{\nu}}{\partial \sigma_a}.
\end{equation}
Using the explicit parameterization of the $SL(2,\mathbb{R})$ group element given above, we can read off the metric as
\begin{equation}
ds^2 = \alpha'k \left(-\cosh^2(\rho) dt^2 + d\rho^2 + \sinh^2(\rho) d\phi^2\right).
\end{equation}
We next focus on the WZ term. First we compute
\begin{align}
\text{Tr}\left(\omega^3\right) &= \text{Tr} \left(g^{-1}\frac{\partial g}{\partial X^{\mu}}g^{-1}\frac{\partial g}{\partial X^{\nu}}g^{-1}\frac{\partial g}{\partial X^{\sigma}}\right)\frac{\partial X^{\mu}}{\partial \sigma_i}\frac{\partial X^{\nu}}{\partial \sigma_j}\frac{\partial X^{\sigma}}{\partial \sigma_k} d\sigma^{i} \wedge d\sigma^{j} \wedge d\sigma^{k} \\
&= d\left(6\sinh^2(\rho)\frac{\partial \phi}{\partial \sigma_j}\frac{\partial t}{\partial \sigma_k} d\sigma^{j} \wedge d\sigma^{k}\right)
\end{align}
from which we can read off the Wess-Zumino term. The background Kalb-Ramond field is given by
\begin{equation}
B_{t \phi} = - \alpha' k \sinh^2(\rho) \quad \text{or} \quad B = - \alpha' k \sinh^2(\rho) dt \wedge d\phi.
\end{equation}
The corresponding $H$-field is then
\begin{equation}
H = dB = - \alpha' k \sinh(2\rho) d\rho \wedge dt \wedge d\phi.
\end{equation}

\subsubsection{Currents, Ward identities and OPEs}
From now on we go to complex worldsheet coordinates ($\sigma_1$,$\sigma_2$) $\to$ ($z$,$\bar{z}$) as usual. The general WZW model is invariant under 
\begin{equation}
g(z,\bar{z}) \to g'(z,\bar{z}) = \Omega(z)g(z,\bar{z})\overline{\Omega}(\bar{z})^{-1}
\end{equation}
where in this case $\Omega$ and $\overline{\Omega}$ are two (independent) $SL(2,\mathbb{R})$ matrices. This corresponds to an isometry of the metric (and Kalb-Ramond background) since it states that $g'$ parametrized by the transformed coordinates ($t'$, $\rho'$, $\phi'$) gives the same metric as $g$ parametrized by ($t$, $\rho$, $\phi$). The isometry group is thus $SL(2,\mathbb{R}) \times SL(2,\mathbb{R})$.
Infinitesimal transformations give $g \to \omega g - g \overline{\omega}$ where $\omega(z)$ is traceless and real. The symmetry currents corresponding to these symmetries are proportional to 
\begin{equation}
J(z) \propto \partial g g^{-1}, \quad \overline{J}(\bar{z}) \propto g^{-1} \bar{\partial} g.
\end{equation}
Following \cite{DiFrancesco:1997nk}, we choose them as
\begin{equation}
J(z) = -\frac{k}{2} \partial g g^{-1}, \quad \overline{J}(\bar{z}) = \frac{k}{2} g^{-1} \bar{\partial} g.
\end{equation}
The sign of the antiholomorphic currents is chosen differently than in \cite{Martinec:2001cf} and these currents give hence an extra minus sign in the flat space $k \to \infty$ limit compared to those in \cite{Martinec:2001cf}. This symmetry entails a corresponding Ward identity for a general field $A$ given by 
\begin{equation}
\label{ward}
\delta A = -\oint_{w} \frac{dz}{2\pi i}\omega_{a}J^{a}(z)A(w,\bar{w}) + \oint_{w} \frac{d\bar{z}}{2\pi i}\overline{\omega_{a}}\overline{J^{a}}(z)A(w,\bar{w})
\end{equation}
where we have expanded the functions in the Lie algebra basis
\begin{equation}
J(z) = J_{a}\tau^{a}, \quad \overline{J}(\bar{z}) = \overline{J_{a}}\tau^{a}, \quad \omega(z) = \omega_{a}\tau^{a}, \quad \overline{\omega}(\bar{z}) = \overline{\omega_{a}}\tau^{a}
\end{equation}
and we should be careful with indices since upper and lower indices are not equal. Note that $\omega_{a}$ is an imaginary number. Multiplying the $J(z)$ expansion by $\tau^{b}$ and tracing gives
\begin{align}
\label{currents}
J^{a} = k \text{Tr}\left(\tau^{a}\partial g g^{-1}\right), \\
\label{currents2}
\overline{J}^{a} = -k \text{Tr}\left(\tau^{a} g^{-1} \bar{\partial} g\right).
\end{align}
If we now take a WZW primary field for which $\delta A = \omega(z)A - A \overline{\omega}(\overline{z})$, we can match this with the Ward identity and read off the following OPEs
\begin{equation}
J^{a}(z)A(w,\bar{w}) \sim -\frac{\tau^{a}A(w,\bar{w})}{z-w}, \quad\quad \overline{J}^{a}(\bar{z})A(w,\bar{w}) \sim \frac{A(w,\bar{w})\tau^{a}}{\bar{z}-\bar{w}},
\end{equation}
which immediately lead to the commutation relations of the zero mode of the current with the field $A$ 
\begin{equation}
\left[J^{3}_{0}, A\right] = -\tau^{3} A, \quad\quad \left[\overline{J}^{3}_{0}, A\right] = A\tau^{3}.
\end{equation}
The currents we have constructed satisfy the following Kac-Moody algebra relations:
\begin{align}
J^{a}(z)J^{b}(w) \sim \frac{k\eta^{ab}}{2(z-w)^2} + \frac{i{f^{ab}}_cJ^{c}(w)}{z-w}, \\
\overline{J}^{a}(\bar{z})\overline{J}^{b}(\bar{w}) \sim \frac{k\eta^{ab}}{2(\bar{z}-\bar{w})^2} + \frac{i{f^{ab}}_c\overline{J}^{c}(\bar{w})}{\bar{z}-\bar{w}}.
\end{align}
The reason we work with this sign-convention for the antiholomorphic currents is that in this case we can identify them directly with the $\overline{\Omega}$ transformations (and not their inverses). \\
Next, we identify the generator of spacetime time translations. This generator is defined by
\begin{equation}
\delta_{t} A = -i\delta t \left[Q_{t}, A\right].
\end{equation}
Since in the general parameterization it holds that
\begin{eqnarray}
\delta_{t} A &= \frac{i \delta t}{2} \sigma^{2} A +\frac{i \delta t}{2} A \sigma^{2} \nonumber\\
&= i \delta t \tau^{3} A + i \delta t A \tau^{3} \nonumber\\
&= -i \delta t \left[J^{3}_{0} - \overline{J}^{3}_{0}, A \right]
\end{eqnarray}
we see that $Q_{t} = J^{3}_{0} - \overline{J}^{3}_{0}$. Analogously one shows that $Q_{\phi} = J^{3}_{0} + \overline{J}^{3}_{0}$.
For later reference, we state the currents in terms of the global coordinates:
\begin{align}
J^{3} &= ik\left(\cosh(\rho)^2\partial t - \sinh(\rho)^2 \partial \phi\right)\\
J^{1} &= ik\left(\sin(\phi+t)\cosh(\rho)\sinh(\rho)\partial t -\sin(\phi+t)\cosh(\rho)\sinh(\rho)\partial \phi +\cos(t+\phi)\partial \rho\right) \nonumber \\
J^{2} &= ik\left(\cos(\phi+t)\cosh(\rho)\sinh(\rho)\partial t -\cos(\phi+t)\cosh(\rho)\sinh(\rho)\partial \phi -\sin(t+\phi)\partial \rho\right) \nonumber \\
\overline{J}^{3} &= -ik\left(\cosh(\rho)^2\bar{\partial} t + \sinh(\rho)^2 \bar{\partial} \phi\right)\\
\overline{J}^{1} &= -ik\left(\sin(t-\phi)\cosh(\rho)\sinh(\rho)\bar{\partial} t +\sin(t-\phi)\cosh(\rho)\sinh(\rho)\bar{\partial} \phi +\cos(t-\phi)\bar{\partial} \rho\right) \nonumber \\
\overline{J}^{2} &= ik\left(\cos(t-\phi)\cosh(\rho)\sinh(\rho)\bar{\partial} t +\cos(t-\phi)\cosh(\rho)\sinh(\rho)\bar{\partial} \phi -\sin(t-\phi)\bar{\partial} \rho\right). \nonumber
\end{align}

\subsection{$SL(2,\mathbb{C})/SU(2)$ model}
We describe the analytic continuation of this model and its relation to the $SL(2,\mathbb{C})/SU(2)$ model. Analytically continuing $t \to i\tau$ immediately gives
\begin{eqnarray}
&g = e^{i\frac{i\tau+\phi}{2}\sigma_2}e^{\rho\sigma_3}e^{i\frac{i\tau-\phi}{2}\sigma_2} \\
&=\left[
\begin{array}{cc}
 \cosh(\tau)\cosh(\rho)+\cos(\phi)\sinh(\rho) & i\sinh(\tau)\cosh(\rho)-\sin(\phi)\sinh(\rho) \\
 -i\sinh(\tau)\cosh(\rho)-\sin(\phi)\sinh(\rho) & \cosh(\tau)\cosh(\rho)-\cos(\phi)\sinh(\rho)\end{array} 
\right].
\end{eqnarray}
This group element obviously still has unit determinant, but is no longer real. It is however Hermitian. This identifies the continuation in global coordinates as the $SL(2,\mathbb{C})/SU(2)$ model. It has previously been noted that analytic continuation in the Poincar\'e patch time coordinate of the $AdS_3$ manifold corresponds to going from the $SL(2,\mathbb{R})$ to the $SL(2,\mathbb{C})/SU(2)$ WZW model \cite{Hemming:2002kd}. \\
This WZW model is invariant under
\begin{equation}
g(z,\bar{z}) \to g'(z,\bar{z}) = \Omega(z)g(z,\bar{z})\overline{\Omega}(\bar{z})^{-1}
\end{equation}
where $\Omega$ and $\overline{\Omega}$ are $SL(2,\mathbb{C})$ matrices such that $g'$ is a Hermitian matrix. This immediately implies $\Omega = \left(\overline{\Omega}^{\dagger}\right)^{-1}$. On an infinitesimal level, this means that $\omega(z) = - \overline{\omega}(\bar{z})^{\dagger}$. So the symmetry group is $SL(2,\mathbb{C})$. \\
In general $\delta_\tau g = -i\delta \tau \left[Q_{\tau},g\right]$. An infinitesimal Euclidean time translation corresponds in our parameterization to 
\begin{equation}
g \to g -\frac{\delta\tau}{2} \sigma_2 g - \frac{\delta\tau}{2} g \sigma_2,
\end{equation}
which we can rewrite as
\begin{equation}
g \to g + \delta\tau  \left[J^{3}_0-\overline{J}^{3}_0,g\right].
\end{equation}
This identifies the Euclidean time translation operator as
\begin{equation}
Q_{\tau} = i(J^{3}_0 - \overline{J}^{3}_0).
\end{equation}
Let us now take a closer look at the link between the $SL(2,\mathbb{R})$ and the $SL(2,\mathbb{C})/SU(2)$ models in terms of the currents. To see the link, we first take a step back and consider the $SL(2,\mathbb{C})$ WZW model. As a Lie algebra basis we choose the same three generators as for the $SL(2,\mathbb{R})$ model and allow for complex expansion parameters. The infinitesimal transformation $\omega(z)$ is a complex traceless matrix and the $\omega_{a}$ are complex numbers. The currents are still given by (\ref{currents}) and (\ref{currents}). So for arbitrary infinitesimal transformations the Ward identity still reads as in equation (\ref{ward}), but now with complex $\omega_{a}$ and with the current $J^{a}$ calculated with the Wick-rotated $g$ matrix. This last step is simply the analytic continuation in the currents directly. \\
The Euclidean model is however, not the general $SL(2,\mathbb{C})$ model but a right coset of this. So the left and right moving infinitesimal transformation are linked according to $\omega(z) = - \overline{\omega}(\bar{z})^{\dagger}$. This implies for the $\omega_{a}$
\begin{equation}
\omega_{1} = \overline{\omega}_{1}, \quad \omega_{2} = \overline{\omega}_{2}, \quad \omega_{3} = -\overline{\omega}_{3}.
\end{equation}
So in all, we double the number of effective currents by going to fully complex expansion parameters, but we then retain only half of these due to the left-right identification.

\subsection{WZW currents as differential operators}
In this subsection we discuss how to associate differential operators to the action of the Lie algebra currents.
Vertex operators of the WZW model are functions of the field $g$ which in turn is parametrized by the group manifold coordinates. The zero-mode symmetry operators have an action on functions as
\begin{equation}
J^{a}_0 (f(g)) = i\left.\frac{\partial}{\partial t} f \left(e^{it \tau^a} g \right)\right|_{t = 0}
\end{equation}
and
\begin{equation}
\overline{J}^{a}_0 (f(g)) = i\left.\frac{\partial}{\partial \bar{t}} f \left( g e^{-i\bar{t} \tau^{a}}\right)\right|_{\bar{t} = 0}.
\end{equation}
This is in fact simply the action of vector fields as differential operators on functions. We will denote the corresponding operators as $\hat{D}^a$ and $\hat{\overline{D}}^a$. Their action is defined through the infinitesimal group transformations. This operator has to satisfy
\begin{equation}
\hat{D}^{a} f(g) = \frac{df}{d g}(g) \left(\hat{D}^{a} g\right) = \frac{df}{d g}(g) \left( - \tau^{a} g\right),
\end{equation}
or 
\begin{align}
\hat{D}^{a} g &=  - \tau^{a} g, \\
\hat{\overline{D}}^{a} g &= g \tau^{a}.
\end{align}
One should compare this with the OPEs we derived earlier for the currents:
\begin{equation}
J^{a}(z)g(w,\bar{w}) \sim -\frac{\tau^{a}g(w,\bar{w})}{z-w}, \quad\quad \overline{J}^{a}(\bar{z})g(w,\bar{w}) \sim \frac{g(w,\bar{w})\tau^{a}}{\bar{z}-\bar{w}},
\end{equation}
and we conclude that indeed the normalization of the currents is precisely such that they are the algebra generators in the sense of the operator equations above. These formulas identify the differential operators as the dual vectors of the right (respectively left) invariant Maurer-Cartan 1-forms, with an extra minus sign for the right-invariant vector. \\
This suggests a first method to compute these differential operators: find the Maurer-Cartan forms and then dualize these into vector fields. We will however follow a more pedestrian path and simply compute the operators from the above conditions using some Pauli matrix algebra. From the above formula, one can see that these differential operators satisfy the zero-mode Lie-algebra since
\begin{align}
\left[\hat{D}^a, \hat{D}^b\right] g &= - \left[\tau^a, \tau^b\right] g = -i{f^{ab}}_c \tau^{c} g = i{f^{ab}}_c (\hat{D}^c g), \\
\left[\hat{\overline{D}}^a, \hat{\overline{D}}^b\right] g &= g \left[\tau^a, \tau^b\right] = i{f^{ab}}_c g \tau^{c} = i{f^{ab}}_c (\hat{\overline{D}}^c g).
\end{align}
The same algebra is satisfied when applying these operators on arbitrary functions on the group manifold.
The on-shell equation for a string state is simply $L_0 + \bar{L}_0 = 2$ and when one rewrites this using the Sugawara construction, we have the full stringy wave equation for the state. \\
Let us discuss this in full detail for the holomorphic part of the symmetry algebra. The general $SL(2,\mathbb{C})/SU(2)$ element was parametrized as 
\begin{equation}
g = e^{(-\tau+i\phi)\tau^3}e^{-2i\rho\tau^1}e^{(-\tau-i\phi)\tau^3}.
\end{equation}
This leads to
\begin{align}
\partial_\tau g &= -\tau^3 g - g\tau^3, \\
\partial_\phi g &= i\tau^3 g - i g\tau^3, \\
\partial_\rho g &= -2i e^{(-\tau+i\phi)\tau^3} \tau_1 e^{-2i\rho\tau^1} e^{(-\tau-i\phi)\tau^3}
\end{align}
Our goal now is to rewrite this in a form where all algebra generators are in front of the group element. This requires some rearranging of the Pauli-matrices using the following three lemmas:
\begin{equation}
e^{A\tau^3}\tau^1 = \left(\cosh(A) \tau^1 + i\sinh(A) \tau^2 \right) e^{A\tau^3},
\end{equation}
\begin{equation}
e^{B\tau^1}\tau^3 = \left(\cos(B) \tau^3 - i\sin(B) \tau^2 \right) e^{B\tau^1},
\end{equation}
\begin{equation}
e^{C\tau^3}\tau^2 = \left(\cosh(C) \tau^2 - i\sinh(C) \tau^1 \right) e^{C\tau^3}.
\end{equation}
Using these, we obtain
\begin{align}
\partial_\tau g &= -\tau^3 g - \cos(2i\rho)\tau^3 g - i \sin(2i\rho)\cosh(\tau - i \phi) \tau^2 g + \sin(2i\rho)\sinh(\tau - i \phi) \tau^1 g , \\
\partial_\phi g &= i\tau^3 g - i\cos(2i\rho)\tau^3 g + \sin(2i\rho)\cosh(\tau - i \phi) \tau^2 g + i\sin(2i\rho)\sinh(\tau - i \phi) \tau^1 g, \\
\partial_\rho g &= -2i\cosh(\tau - i \phi) \tau^1 g - 2 \sinh(\tau - i \phi) \tau^2 g. 
\end{align}
After solving the equalities $\hat{D}^{a} g = -\tau^{a} g$ for a general first-order differential operator, we obtain the unique solution for the differential operators:
\begin{align}
\hat{D}^3 &= -\frac{1}{2i}\partial_\phi + \frac{1}{2}\partial_\tau, \\
\hat{D}^{1} &= \frac{1}{2}\left[i \sinh(\tau-i\phi)\tanh(\rho)\partial_\tau - \sinh(\tau-i\phi)\coth(\rho)\partial_\phi - i\cosh(\tau-i\phi)\partial_\rho\right] ,\\
\hat{D}^{2} &= \frac{1}{2}\left[\cosh(\tau-i\phi)\tanh(\rho)\partial_\tau + i\cosh(\tau-i\phi)\coth(\rho)\partial_\phi - \sinh(\tau-i\phi)\partial_\rho\right],
\end{align}
and for the $+$ and $-$ operators, defined as $ \hat{D}^{\pm} = \hat{D}^{1} \pm i \hat{D}^{2}$, we obtain
\begin{align}
\hat{D}^{+} &= i\frac{e^{\tau-i\phi}}{2}\left[-\partial_\rho + i\coth(\rho) \partial_\phi + \tanh(\rho)\partial_\tau \right], \\
\hat{D}^{-} &= -i\frac{e^{-\tau+i\phi}}{2}\left[\partial_\rho + i\coth(\rho) \partial_\phi + \tanh(\rho)\partial_\tau \right].
\end{align}
One can explicity check using the above parametrization of $g$ that indeed
\begin{equation}
\hat{D}^{3} g = -\tau^3 g, \quad \hat{D}^{+} g = - (\tau^{1} + i\tau^2) g, \quad \hat{D}^{-} g = - (\tau^1 - i \tau^2 ) g.
\end{equation}
One can also check that
\begin{align}
\hat{\overline{D}}^3 &= -\frac{1}{2i}\partial_\phi - \frac{1}{2}\partial_\tau, \\
\hat{\overline{D}}^{+} &= i\frac{e^{-\tau-i\phi}}{2}\left[\partial_\rho - i\coth(\rho) \partial_\phi + \tanh(\rho)\partial_\tau \right], \\
\hat{\overline{D}}^{-} &= -i\frac{e^{\tau+i\phi}}{2}\left[-\partial_\rho - i\coth(\rho) \partial_\phi + \tanh(\rho)\partial_\tau \right],
\end{align}
satisfy
\begin{equation}
\hat{\overline{D}}^3 g = g \tau_3 ,\quad \hat{\overline{D}}^{+} g =  g (\tau_1 +i\tau_2), \quad \hat{\overline{D}}^{-} g =  g (\tau_1 - i\tau_2).
\end{equation}
We already saw above from the explicit construction that there is a unique solution to these conditions.

\section{Argument why the density of states does not influence the critical behavior}
\label{dos}
In this appendix we argue that the density of states can not alter our conclusions about the critical temperature.\\
So far in the main text we have not written down an explicit expression for the density of states. The correct expression is given in equation (\ref{dosorig}) \cite{Maldacena:2000kv}. \\ 
A first argument as to why the $s$-integration does not make a state more tachyonic is the following. In general, the integration over $s$ is of the following form:
\begin{equation}
\int_{0}^{+\infty}ds \rho(s) e^{-\tau_2 s^2} \, \to \, 0 \quad \text{for} \quad \tau_2 \to \infty
\end{equation}
so this integral cannot yield a contribution that behaves as $e^{A\tau_2}$ for positive $A$ as long as $\rho(s)$ is well-behaved near $s=0$ which it is.\footnote{This follows largely from equation (\ref{dosorig}) with $n=0$: for any fixed $q$ the limit $s \to 0$ is zero.} \\
A more general argument goes as follows. Firstly, as remarked before, the $\tau_1$-integral forces $np = 0$, which implies $n=0$ for our purposes. Important to note is that the contribution from $q$ is an imaginary exponential. Consider the following general expression\footnote{The $q$ quantum number and its prefactors are rescaled into a new number $q$.}:
\begin{equation}
\int_{0}^{+\infty}ds \sum_{q\in\mathbb{Z}}\rho(s,q)e^{iq\tau_2}e^{-s^2\tau_2}
\end{equation}
where $\rho(s,q)$ denotes the density of states with $n=0$.
Performing the sum first, we can rewrite this as
\begin{equation}
\int_{0}^{+\infty}ds F(s,\tau_2)e^{-s^2\tau_2}
\end{equation}
where $F$ has Fourier coefficients $\rho$ and is periodic in $\tau_2$. We are only interested in whether the integral is capable of producing a $\tau_2$-dependent exponential after integration (like $e^{\pm C \tau_2}$ for some constant $C$). Laplace's method gives
\begin{equation}
\int_{0}^{+\infty}ds F(s,\tau_2)e^{-s^2\tau_2} \approx \sqrt{\frac{\pi}{\tau_2}}\frac{F(0,\tau_2)}{2} + \hdots
\end{equation}
since the periodic function does not correct the leading behavior. One simple way to see this is to take a discrete limit with steps precisely equal to the periodicity of $F$ in $\tau_2$. In this case $F$ becomes effectively independent of $\tau_2$ and one can use the textbook Laplace method to get the above result. The $\tau_2$-dependence of the final result may indicate that the limit itself is ill-defined but regardless it cannot influence the critical behavior and we only care about this.
Note that the assumption that the integration over $s$ does not influence the tachyonic nature of a state, was made implicitly by the authors of \cite{Maldacena:2000hw} and \cite{Rangamani:2007fz}.

\section{Hamiltonian description of thermal $AdS_3$ and its orbifolds}
In this lengthy appendix, we describe in detail how the Hamiltonian description of the spectrum is obtained. For clarity, we first use a trivial toy model to illustrate the strategy that we will employ.

\subsection{Flat space toy model to illustrate the strategy}
\label{toy}
Consider the 1d flat space Laplacian. The operator commutes with $\hat{J} = i\partial_x$. We choose eigenfunctions
\begin{equation}
\label{eigf}
\psi_k(x) \propto e^{ikx},
\end{equation}
satisfying
\begin{align}
\label{rel1}
\Delta \psi &= -k^2 \psi, \\
\label{rel2}
\hat{J} \psi = i \partial_x \psi &= -k \psi.
\end{align}
We then evaluate
\begin{equation}
\text{Tr}\left[e^{t\Delta}e^{2\pi i U \hat{J}}\right]
\end{equation}
for some fixed number $U$. Firstly, in configuration space this equals 
\begin{equation}
\int dx \left\langle x\right|e^{t\Delta}\left|x + 2\pi U\right\rangle = V\frac{1}{2\sqrt{\pi t}}e^{-\frac{\pi^2U^2}{t}},
\end{equation}
where we used $ e^{i a \hat{J}}\left|x\right\rangle = \left|x + a\right\rangle$ and the flat space heat kernel. Secondly, we can evaluate it using the eigenfunctions (\ref{eigf}) and the relations (\ref{rel1}) and (\ref{rel2}) as:
\begin{equation}
\int dk \delta(0) e^{-tk^2}e^{-2\pi i U k} = \frac{V}{2\pi}\sqrt{\pi/t}e^{-\frac{\pi^2U^2}{t}}.
\end{equation}
We see that both expressions are manifestly the same. Note that the density of states is present in the form of $\delta(0)$. It is the second description that we are after, since then the trace over the quantum numbers (in this case only $k$) is apparent. In the following subsections, we will apply this same idea to the much more complicated $AdS_3$ space and its orbifolds.

\subsection{$AdS_3$ Thermal partition function}
\label{Ham1}
In this subsection we give a Hamiltonian description of the thermal $AdS_3$ partition function. With a suitable substitution of parameters, this is also the thermal BTZ partition function as discussed in section \ref{BTZsection}. The partition function is given by
\begin{align}
\label{partfunct}
Z(\tau) = \frac{\beta\sqrt{k-2}}{8\pi\sqrt{\tau_2}}\sum_{l,p}\frac{e^{-k\beta^2\left|l-p\tau\right|^2/4\pi\tau_2 + 2\pi \Im(U_{lp})^2/\tau_2}e^{\frac{\pi\tau_2}{2}}}
{\left|\sin(\pi U_{lp})\right|^2\left|\prod_{r=1}^{+\infty}(1-q^{r})(1-q^re^{2\pi i U_{lp}})(1-q^{r}e^{-2\pi i U_{lp}})\right|^2}
\end{align}
where $q=e^{2\pi i \tau}$ and\footnote{This differs in two ways from the expression written down in \cite{Maldacena:2000kv}: firstly the $(q\bar{q})^{-3/24}$ was inserted in the above expression. Secondly, we utilize the complex conjugate of the $U_{lp}$ as defined in \cite{Maldacena:2000kv}. This is in accord with the Ray-Singer torsion \cite{Ray:1973sb} and this was noticed in \cite{Son:2001qm} by one of the authors themselves.} 
\begin{equation}
\label{Ulp}
U_{lp} = -\frac{i\beta}{2\pi}(p\tau-l).
\end{equation}
The quantum number $p$ will correspond to thermal winding, whereas the quantum number $l$ will be Poisson-resummed into the discrete momentum. Our goal is to rewrite this in the form
\begin{equation}
\label{gen}
Z(\tau) = \sum_i N_{i\tilde{i}}\chi_{i}(\tau) \chi_{\tilde{i}}^{*}(\tau) = \text{Tr} q^{L_0 - c/24}\bar{q}^{\bar{L}_0 - c/24},
\end{equation}
with the characters
\begin{equation}
\chi_i(\tau) = \text{Tr}_i q^{L_0 - c/24}.
\end{equation}
The second equality in (\ref{gen}) traces over all primaries and their secondaries in the string spectrum. In these expressions the conformal weights $h$ and $\bar{h}$ are possibly a subset of those we determined in section \ref{spectrumsection} using CFT arguments. \\
First of all, we have that
\begin{equation}
\sum_{N=1}^{+\infty}\sum_{\mathcal{P} \in P(N)}q^{N}e^{2\pi i U_{lp} O(\mathcal{P})} = \prod_{r=1}^{+\infty}\frac{1}{1 - q^r e^{2\pi i U_{lp}}}
\end{equation}
where $P(N)$ denotes the different partitions of the integer $N$ and $O(\mathcal{P})$ denotes the size of the partition $\mathcal{P}$. It is clear that the infinite product in (\ref{partfunct}) corresponds to the different oscillator states and we will not be interested in this. Note though that this is a bit naive since the Taylor expansion we should use depends on the precise value of $U_{lp}$. Nonetheless, as a first step towards understanding the partition function (\ref{partfunct}) we choose to drop the infinite product. Comments on this are provided in the main text. \\
The method to proceed was developed by \cite{Gawedzki:1991yu}\cite{Gawedzki:1988nj} and we adapt it for our purposes.\footnote{Note that in \cite{Gawedzki:1991yu}, the author only considers twisting in one torus direction. Here we consider both.} We first evaluate the following trace (for fixed $l$ and $p$):
\begin{equation}
\label{trace}
\text{Tr}\left[\exp\left(-\tau_2\frac{\beta^2 k p^2}{4\pi}\right)\exp\left(4\pi\tau_2\frac{\Delta}{k-2}\right)\exp(2\pi i (U_{lp} J^{3}_0 + \bar{U}_{lp} \overline{J}^{3}_0))\right]
\end{equation}
where we trace over a basis of all normalizable functions $\psi_a(g)$ on $H_3^+$ and $J^{3}_0$ and $\overline{J}^{3}_0$ are differential operators acting on these functions. As a basis, we choose eigenfunctions of $\Delta$, $J^{3}_0$ and $\overline{J}^{3}_0$. The explicit form of these eigenfunctions will not be needed; the interested reader can take a closer look at appendix A of \cite{Teschner:1997fv} to find elaborate expressions. More explicitly, consider the three operators $\Delta$, $J_0^3 + \overline{J}_0^3 = i\partial_{\phi}$ and $J_0^3 - \overline{J}_0^3 = - \partial_\tau$ in coordinates $\tau,\phi,\rho$. These mutually commute so let us choose functions $\psi_{s,m,q}$ such that
\begin{align}
\label{eigf1}
\Delta \psi = (-s^2 - 1/4)\psi, \\
\label{eigf2}
(J_0^3 + \overline{J}_0^3)\psi = q \psi, \\
\label{eigf3}
-i(J_0^3 - \overline{J}_0^3)\psi = m \psi.
\end{align} 
It can be shown that when $m$ is real and $q$ is an integer, $\Delta$ is a Hermitian operator w.r.t. the standard inner product on $L_2(H_3^+)$. Therefore the $\psi_{s,m,q}$ form a basis when restricting to these quantum numbers. The starting point is then the evaluation of
\begin{equation}
\sum_{s,m,q}\left\langle \psi_{s,m,q}\right|\hat{\mathcal{O}}\left|\psi_{s,m,q}\right\rangle
\end{equation}
with $s$ a real positive number, $m$ a real number and $q$ an integer. So explicitly
\begin{equation}
\label{summations}
\sum_{s,m,q} = \int_{\mathbb{R}^{+}}ds \int_{\mathbb{R}}dm \sum_{q\in\mathbb{Z}}.
\end{equation}
In our case, the operator $\hat{\mathcal{O}}$ is given as:
\begin{equation}
\hat{\mathcal{O}} = e^{-\tau_2\frac{\beta^2 k p^2}{4\pi}}e^{4\pi\tau_2 \frac{\Delta+\frac{1}{4}}{k-2}}e^{2\pi i (U_{lp}J_0^3 + \bar{U}_{lp}\overline{J}_0^3)}.
\end{equation}
We rewrite this as 
\begin{equation}
\int dg \sum_{s,m,q}\psi_{s,m,q}(g)^{*}\psi_{s,m,q}(g) \lambda_{s,m,q} = \sum_{s,m,q}\delta(0) \lambda_{s,m,q},
\end{equation}
where $g$ denotes a group element of the $H_3^+ = SL(2,\mathbb{C})/SU(2)$ group manifold and $\lambda_{s,m,q}$ is the eigenvalue of $\hat{\mathcal{O}}$. Here $\delta(0) = \rho(s,m,q)$ is the density of states and depends on $s,m$ and $q$. The operator $\hat{\mathcal{O}}$ is labeled by $p,l$ quantum numbers and the entire expression is then summed over $l$ and $p$. 
The expression for the density of states on such spaces was written down in \cite{Maldacena:2000kv}. We use a slight modification of this expression:
\begin{equation}
\label{dosh3}
\mbox{{\small{$\displaystyle\rho(s,m,q) = 2\frac{L}{2\pi}\left[\frac{1}{2\pi}2\log(\epsilon) + \frac{1}{2\pi i}\frac{d}{2ds}\log\left(\frac{\Gamma(\frac{1}{2} - is - q/2 - im/2)\Gamma(\frac{1}{2} - is -q/2 + im/2)}{\Gamma(\frac{1}{2} + is -q/2- im/2)\Gamma(\frac{1}{2} + is -q/2+ im/2)}\right)\right]$}}}.
\end{equation}
The expression in square brackets is the usual one, used in \cite{Maldacena:2000kv}. The parameter $\epsilon$ is an IR regulator ($\epsilon \to 0$) corresponding to the radial direction. We here multiply this expression by a further factor of $\frac{L}{2\pi}$ where $L$ is the IR regulator ($L \to \infty$) of the Euclidean time direction. The reason that we include it here, is that we are considering the Euclidean space, for which the density of states also includes the temporal direction on the same foot as the other space directions. A proper definition of $L$ follows shortly. We also include an extra factor of $2$.\footnote{The reason is that we consider the entire prefactor of the $s$-integral and this includes an extra factor of 2. See e.g. \cite{Hanany:2002ev}\cite{Maldacena:2000kv} where this factor is written explicitly.} \\
We need to sum the trace (\ref{trace}) over $l$ and $p$ and divide by the range of this summation. In this case, $l$ and $p$ run over $\mathbb{Z}$ so the range is infinite (let us call it $P$), although this infinity will cancel with $L$ further on to yield a finite contribution. \\
As in the previous subsection \ref{toy}, we will evaluate this operator trace in two different ways. Let us first focus on the configuration space evaluation. We choose the free-field coordinates ($\Phi$, $v$, $\bar{v}$) introduced in \cite{Gawedzki:1991yu} to parametrize the group element:
\begin{equation}
g = \left[ 
\begin{array}{cc}
e^{\Phi}(1+v\bar{v})^{1/2} & v \\
\bar{v} & e^{-\Phi}(1+v\bar{v})^{1/2} \end{array} 
\right].
\end{equation}
for real $\Phi$ and complex $v$ (and $\bar{v}$). In these coordinates it was shown in \cite{Gawedzki:1991yu} that the zero-modes of the currents have the following form:
\begin{align}
J^{3}_0 - \overline{J}^{3}_0 &= -v\partial_v + \bar{v}\partial_{\bar{v}} = i \partial_{\phi}, \\
J^{3}_0 + \overline{J}^{3}_0 &= - \partial_{\Phi} = -\partial_{\tau},
\end{align}
where the first equality is in terms of the coordinates above, and the second equality is in terms of the global coordinates.\footnote{A few technicalities are in order here. The free-field coordinates ($\Phi$, $v$, $\bar{v}$) are related to the global $AdS$ coordinates as
\begin{align}
\label{cotrans}
\begin{cases}
v &= \sinh(\rho)e^{i\phi} \\
\bar{v} &= \sinh(\rho)e^{-i\phi} \\
\Phi &= t - 2\text{log}\cosh(\rho)
\end{cases} 
\end{align}
The $\Phi$ coordinate used in \cite{Gawedzki:1991yu} is related to that used in \cite{Maldacena:2000kv} (denoted $\tilde{\Phi}$ here) by $\Phi = \tilde{\Phi} + \frac{1}{2}\text{log}(1+\left|v\right|^2)$, which explains the factor of $2$ appearing in the final line of (\ref{cotrans}).} To conform to our conventions, we change the sign of the antiholomorphic sector, such that:
\begin{align}
J^{3}_0 + \overline{J}^{3}_0 &= -v\partial_v + \bar{v}\partial_{\bar{v}} = i \partial_{\phi}, \\
-i(J^{3}_0 - \overline{J}^{3}_0) &= i \partial_{\Phi} = i\partial_{\tau}.
\end{align}
Then these generators are indeed the angular and Euclidean time generators defined before. \\
From the form of the generator in terms of differential operators, we have that
\begin{align}
\exp(2\pi i U_{lp} J^{3}_0) g &= \exp(-\pi i U_{lp} \sigma^{3}) g, \\
\exp(2\pi i \bar{U}_{lp} \overline{J}^{3}_0) g &=  g \exp(\pi i \bar{U}_{lp} \sigma^{3}).
\end{align}
We can rewrite the trace as
\begin{align}
\sum_{a}&\left\langle \psi_a\right|\exp\left(-\tau_2\frac{\beta^2 k p^2}{4\pi}\right)\exp\left(4\pi\tau_2\frac{\Delta}{k-2}\right)\exp\left(2\pi i (U_{lp} J^{3}_0 + \bar{U}_{lp} \overline{J}^{3}_0)\right)\left|\psi_a\right\rangle \\
&= \int d g \left\langle g\right|\exp\left(-\tau_2\frac{\beta^2 k p^2}{4\pi}\right)\exp\left(4\pi\tau_2\frac{\Delta}{k-2}\right)\left|\exp(-\pi i U_{lp} \sigma^{3})g\exp(\pi i \bar{U}_{lp} \sigma^{3})\right\rangle.
\end{align}
In the last line, we integrate the heat kernel over the group manifold, but with twisted boundary conditions. Next we explicitly perform the integration on the group manifold, i.e. over the $v$, $\bar{v}$ and $\Phi$ coordinates. The measure is given by $dg = d\Phi dv d\bar{v}$. The group metric (to which the Laplacian above is associated) is given by
\begin{equation}
ds^2 = d\tilde{\Phi}^2 + (dv + v d\tilde{\Phi})(d\bar{v} + \bar{v}d\tilde{\Phi})
\end{equation} 
and is independent of $\tilde{\Phi}$ (or $\Phi$ itself). Thus integrating over $\Phi$ can be done by using this fact: since the heat kernel is a sum over paths between two points, it is independent of the `center of mass' $\Phi$ coordinate of both points.\footnote{Note also that both $J^{3}_0$ and $\overline{J}^{3}_0$ are independent of the $\Phi$ coordinate, which is a necessary condition for this statement.} We define $\int d\Phi = L$ and combining it with the $\frac{1}{P}$ present in the partition sum, it produces $\beta$; just like it does in the flat toroidal case. \\
The heat kernel on $H_3^+$ is given by (see e.g. \cite{David:2009xg})
\begin{equation}
e^{t\Delta}(g_1,g_2) = (\pi t)^{-3/2}\frac{d}{\sinh d}e^{-t/4-d^2/t},
\end{equation}
where $d$ is the geodesic (= hyperbolic) distance between the 2 points. In particular, between $\exp(-\pi i U_{np} \sigma^{3})g\exp(\pi i \bar{U}_{np} \sigma^{3})$ and $g$ with $\Phi = 0$, this is given by:
\begin{equation}
\cosh d = \left(1+\left|v\right|^2\right)\cosh(2\pi U_2) - \left|v\right|^2\cos(2\pi U_1).
\end{equation}
The integration can then be done by some simple substitutions and leads to
\begin{align}
&\int d g \left\langle g\right|\exp\left(4\pi\tau_2\frac{\Delta}{k-2}\right)\left|\exp(-\pi i U_{lp} \sigma^{3})g\exp(\pi i \bar{U}_{lp} \sigma^{3})\right\rangle \\
&= L \frac{\sqrt{k-2}}{8\pi\sqrt{\tau_2}}e^{-\pi \tau_2/(k-2)}e^{-\pi (k-2)\Im(U_{lp})^2/\tau_2}\left|\sin(\pi U_{lp})\right|^{-2}.
\end{align}
When we put everything together, we obtain
\begin{align}
\label{fulltrace}
\frac{1}{P}\text{Tr}&\left[\exp\left(-\tau_2\frac{\beta^2 k p^2}{4\pi}\right)\exp\left(4\pi\tau_2\frac{\Delta+1/4}{k-2}\right)\exp(2\pi i (U_{lp} J^{3}_0 + \bar{U}_{lp} \overline{J}^{3}_0))\right] \\
 &= \frac{\beta\sqrt{k-2}}{8\pi\sqrt{\tau_2}}\exp\left(-\tau_2\frac{\beta^2 k p^2}{4\pi}\right)e^{-\pi (k-2)\beta^2 (p\tau_1-l)^2/4\pi^2\tau_2}\left|\sin(\pi U_{lp})\right|^{-2} \\
 \label{fulltrace2}
 &=\frac{\beta\sqrt{k-2}}{8\pi\sqrt{\tau_2}}e^{- k\beta^2\left|l-p\tau\right|^2/4\pi\tau_2 + 2\pi \Im(U_{lp})^2/\tau_2}\left|\sin(\pi U_{lp})\right|^{-2}.
\end{align}
To fully agree with (\ref{partfunct}) we should multiply the trace by $e^{\frac{\pi\tau_2}{2}}$ and sum this expression over $l$ and $p$. This concludes the first computation: the partition function has been rewritten in terms of a trace of some operator $\hat{\mathcal{O}}$. \\ 
Next we would like to rewrite it fully in terms of the quantum numbers of the states we wrote down in section \ref{spectrumsection}. So we start afresh with the operator trace (\ref{fulltrace}) and we evaluate the trace by using the basis of eigenfunctions that we discussed around equations (\ref{eigf1}), (\ref{eigf2}) and (\ref{eigf3}). To proceed, we focus on the integral over $m$ in equation (\ref{summations}) and the summation over $l$. The part of the trace that depends on $l$ gives us
\begin{align}
\label{projexpo}
\frac{1}{P}\int_{\mathbb{R}}dm\sum_{l\in\mathbb{Z}}\exp\left(2\pi i \left(\frac{\beta l}{2\pi}m\right)\right) &= \frac{1}{P}\frac{2\pi}{\beta}\int_{\mathbb{R}}dm\sum_{n\in\mathbb{Z}}\delta\left(m-\frac{2\pi n }{\beta}\right) \\
 &= \frac{2\pi}{L}\int_{\mathbb{R}}dm\sum_{n\in\mathbb{Z}}\delta\left(m-\frac{2\pi n }{\beta}\right).
\end{align}
Note the appearance of a prefactor $\frac{2\pi}{L}$. Hence the integral over $m$ only has contributions for $m = \frac{2\pi n }{\beta}$ for integer $n$.
With this value of $m$ and the fact that
\begin{equation}
J^{3}_0 + \overline{J}^{3}_0  = q
\end{equation} 
on the eigenfunctions $\psi$, we obtain using (\ref{Ulp}) for the remaining parts of $U_{lp}$:
\begin{align}
\beta p \tau\left(\frac{q}{2} + \frac{i\pi n }{\beta} \right) &\subset 2\pi i U_{lp} J^{3}_0, \\
-\beta p \bar{\tau}\left(\frac{q}{2} - \frac{i\pi n }{\beta} \right) &\subset 2\pi i \bar{U}_{lp} \overline{J}^{3}_0.
\end{align}
This determines all different factors of the operator trace (\ref{fulltrace}) in terms of the quantum numbers we are interested in.\\
The factor of $e^{\pi\tau_2/2}$ that we manually added after equation (\ref{fulltrace2}) can be written as
\begin{equation}
e^{\pi\tau_2/2} = (q\bar{q})^{-3/24}.
\end{equation}
Also multiplying (and dividing) the trace by $e^{\frac{-\pi\tau_2}{k-2}}$, we have precisely rewritten the partition function as
\begin{equation}
\text{Tr}q^{L_0 - c/24}\bar{q}^{\bar{L}_0-c/24},
\end{equation}
where
\begin{align}
h &= \frac{s^2 +1/4}{k-2} - i\frac{qp\beta}{4\pi} + \frac{ p n}{2} + \frac{kp^2\beta^2}{4(2\pi)^2}, \\
\bar{h} &= \frac{s^2 +1/4}{k-2} - i\frac{qp\beta}{4\pi} - \frac{ p n}{2} + \frac{kp^2\beta^2}{4(2\pi)^2},
\end{align}
and with $c = 3 + 6/(k-2)$, the central charge of the $SL(2,\mathbb{R})$ model. We trace only over the continuous states with the density of states:
\begin{equation}
\label{dosorig}
\rho(s,n,q) = 2\left[\frac{1}{2\pi}2\log(\epsilon) + \frac{1}{2\pi i}\frac{d}{2ds}\log\left(\frac{\Gamma(\frac{1}{2} - is - q/2 - i\frac{\pi n }{\beta})\Gamma(\frac{1}{2} - is -q/2 + i\frac{\pi n }{\beta})}{\Gamma(\frac{1}{2} + is -q/2- i\frac{\pi n }{\beta})\Gamma(\frac{1}{2} + is -q/2+ i\frac{\pi n }{\beta})}\right)\right].
\end{equation}
The factor $\frac{L}{2\pi}$ has dropped out in this expression: this makes sense, since this direction has become compact due to the thermal identification and compact dimensions do not give volume-scaling prefactors when considering the Hamiltonian formulation of the partition function (see for instance any textbook on string theory).\\
No discrete states are present and there are also no states that wind the angular cigar. 

\subsection{Angular orbifolds}
\label{Ham2}
We consider orbifolds obtained by identifying $\phi \sim \phi + \frac{2\pi}{N}$. These angular orbifolds were extensively studied in \cite{Son:2001qm} and \cite{Martinec:2001cf}. The thermal partition function was computed in \cite{Son:2001qm}. It was shown there that the thermal partition function on such orbifolds has the form
\begin{equation}
\label{pforbi}
Z = \frac{1}{N}\sum_{a,b}Z_{ab},
\end{equation}
where each $Z_{ab}$ is obtained from the untwisted partition function (\ref{partfunct}) by the simple substitution
\begin{equation}
U_{lp} \to U_{lp} + \frac{a}{N}\tau + \frac{b}{N}.
\end{equation}
These parameters are hence given by
\begin{equation}
U_{lp} = \frac{b}{N} + \frac{a}{N}\tau_1 - i\frac{\beta}{2\pi}(p\tau_1-l) + i\frac{a}{N}\tau_2 + \frac{p\beta}{2\pi}\tau_2,
\end{equation}
whose imaginary part equals
\begin{equation}
\Im(U_{lp}) = -\frac{\beta}{2\pi}(p\tau_1-l) + \frac{a}{N}\tau_2.
\end{equation}
Performing the above heat kernel computation again, we obtain
\begin{align}
\text{Tr}&\left[\exp\left(-\tau_2\frac{\beta^2 k p^2}{4\pi}\right)\exp\left(4\pi\tau_2\frac{\Delta+1/4}{k-2}\right)\exp(2\pi i (U_{lp} J^{3}_0 + \bar{U}_{lp} \overline{J}^{3}_0))\right] \\
 &=\frac{\beta\sqrt{k-2}}{8\pi\sqrt{\tau_2}}e^{- k\beta^2\left|l-p\tau\right|^2/4\pi\tau_2 + 2\pi \Im(U_{lp})^2/\tau_2 -\frac{\pi k}{\tau_2}\left(-(p\tau_1-l)\frac{a}{N}\frac{\beta}{\pi}\tau_2 + \frac{a^2}{N^2}\tau_2^2\right)}\left|\sin(\pi U_{lp})\right|^{-2}.
\end{align}
By a slight rearrangement of this expression, we get 
\begin{align}
\label{expre}
\text{Tr}&\left[\exp\left(-(p\tau_1-l) k\frac{a}{N}\beta+ \pi k\frac{a^2}{N^2}\tau_2\right)\exp\left(-\tau_2\frac{\beta^2 k p^2}{4\pi}\right)\right. \nonumber \\
&\quad\quad\quad\quad \left.\times\exp\left(4\pi\tau_2\frac{\Delta+1/4}{k-2}\right)\exp(2\pi i (U_{lp} J^{3}_0 + \bar{U}_{lp} \overline{J}^{3}_0))\right] \\
 &=\frac{\beta\sqrt{k-2}}{8\pi\sqrt{\tau_2}}e^{- k\beta^2\left|l-p\tau\right|^2/4\pi\tau_2 + 2\pi \Im(U_{lp})^2/\tau_2 }\left|\sin(\pi U_{lp})\right|^{-2}.
\end{align}
Let us again interpret this from a CFT point of view. Firstly we determine and solve the analogous conditions as (\ref{projexpo}) for this case. These will be called the projection conditions in what follows. These are given by
\begin{align}
\sum_{b}\exp\left(2\pi i \frac{b}{N}(J^{3}_0 + \overline{J}^{3}_0)\right), \\
\sum_{l}\exp\left(2\pi i \left(\frac{i\beta}{2\pi} l J^{3}_0 - \frac{i\beta}{2\pi} l\overline{J}^{3}_0\right) + lkw\beta\right),
\end{align}
where $w= \frac{a}{N}$. We already have that $m_{J} + \overline{m}_J \in \mathbb{Z}$ from the covering space. Then the sum over $b$ gives us
\begin{equation}
\frac{1}{N} \sum_{b=0}^{N-1}e^{2\pi i \frac{b}{N} (m_{J} + \overline{m}_J)} = 1 \quad \text{iff} \quad m_{J} + \overline{m}_J \in N \mathbb{Z}
\end{equation}
and it vanishes in the other cases. The sum over $l$ is more problematic: the $e^{lkw\beta}$ contribution is real. We will nevertheless utilize formally the same strategy as in the previous subsection. The treatment we present here is not rigorous. We will come back to this in the next few sections, but for now let us continue this line of thought.
The above conditions project the values of $m_J$ and $\overline{m}_J$ on a discrete set given by
\begin{align}
m_{J} &= \frac{q}{2} + \frac{i\pi n }{\beta} + \frac{kw}{2}, \\
\overline{m}_{J} &= \frac{q}{2} - \frac{i\pi n }{\beta} - \frac{kw}{2},
\end{align}
for $n, w \in \mathbb{Z}$ and $q \in N\mathbb{Z}$. The prefactor of the Poisson summation in $l$ again precisely cancels the $\frac{L}{2\pi}$ present in the density of states. We remark that for the twisted sectors, the above $J^3_0$ operators are not the same as the $J^3_0$ operators we used to determine the spectrum in section \ref{spectrumsection}. Thus the above operators are not the actual $J$ operators and should be better denoted by $J'$ but we refrain from doing this. With the above form for $U_{lp}$ and these eigenvalues of the $J^3_0$ and $\overline{J}^3_0$ operators, one finds
\begin{align}
\label{ads3spectrum3}
h^{wp}_{jqn} &= \frac{s^2+1/4}{k-2} +\frac{qw}{2} +\frac{i\pi nw}{\beta} + \frac{kw^2}{4}- i\frac{qp\beta}{4\pi} + \frac{ p n}{2} + \frac{kp^2\beta^2}{4(2\pi)^2}, \\
\label{ads3spectrum4}
\bar{h}^{wp}_{jqn} &= \frac{s^2+1/4}{k-2} - \frac{qw}{2} +\frac{i\pi nw}{\beta} + \frac{kw^2}{4} - i\frac{qp\beta}{4\pi} - \frac{ p n}{2} + \frac{kp^2\beta^2}{4(2\pi)^2},
\end{align}
which, upon setting $q \to -q$, $n \to -n$ and $p \to -p$, coincides with equations (\ref{ads3spectrum1prelim}) and (\ref{ads3spectrum2prelim}) 
where $w = \frac{a}{N}$. Which values of $a$ should we sum over? The partition function itself (\ref{pforbi}) is periodic under $a \to a +N$. This symmetry is absent in our analysis since we dropped the infinite product. This issue is settled in subsection \ref{numerical} where we take a numerical approach to analyze the infinite product. The result is that one should restrict to $\left|w\right| < \frac{1}{2}$, which is indeed an interval of length 1. Only in this interval does the infinite product not yield a contribution that corrects the conformal weights of the primaries. This implies the following range for $a$:
\begin{align}
a&= -\frac{N-1}{2} \to \frac{N-1}{2}, \quad N \text{ odd}, \\
a&= - \frac{N-2}{2} \to \frac{N}{2}, \quad N \text{ even}.
\end{align}
Strings that are wound more times than this are not in the spectrum. Discrete momentum on the cigar on the other hand is present for all $q \in N\mathbb{Z}$. As a consistency check, note that the resulting spectrum satisfies $h - \bar{h} \in \mathbb{Z}$. To arrive at the Euclidean BTZ orbifold string spectrum, one should simply replace $\beta \to \frac{4\pi^2}{\beta_{BTZ}}$ as discussed in section \ref{BTZsection}.\\
Is this the end of the story? Not quite, our analysis of the projection exponential was not complete. In principle, the summation over $l$ gives a divergent result on its own. Afterwards we integrate $m$ over the real axis which is only sensitive to dirac-poles at real values. \\
These two operations, while separately nonsense, are given meaning by Poisson's summation formula, in which we naively substitute complex arguments instead of real ones. The answer then turns out to be related to the \emph{proper} analytic continuation of Poisson's summation formula to which we now turn.

\subsection{Interlude: analytic continuation of Poisson's summation formula}
\label{Poisson}
\begin{figure}[h]
\centering
\includegraphics[width=7cm]{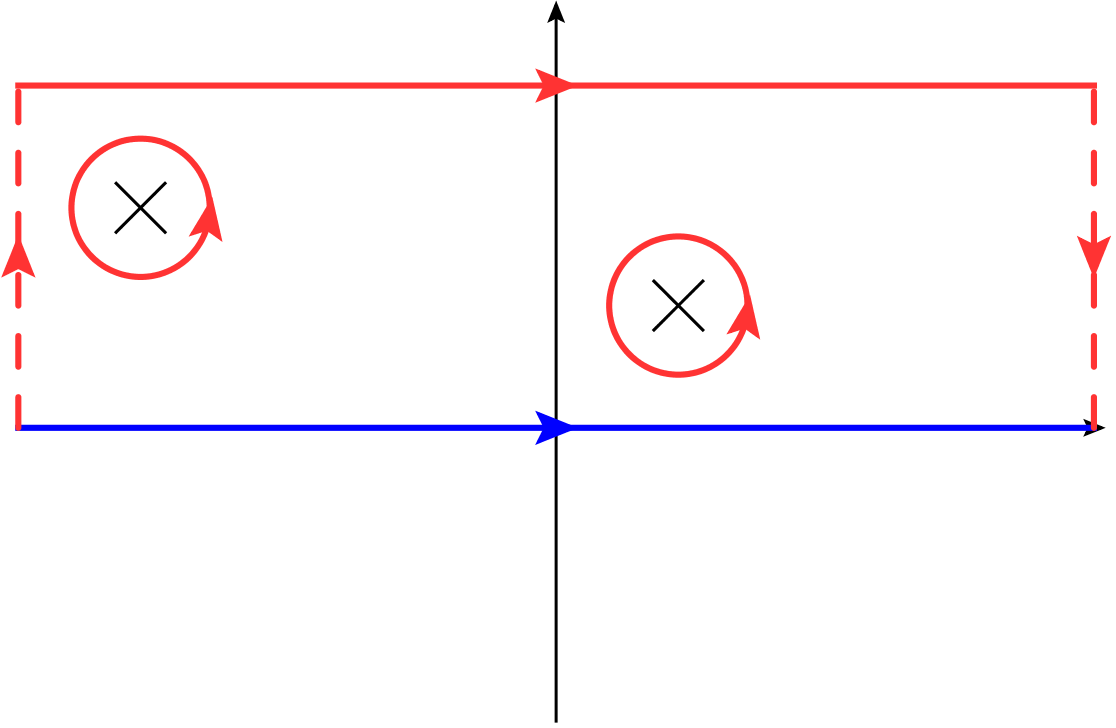}
\caption{Integration contour. The blue contour is the original one. The red contour is the result of the upwards shift. Possible poles of $f$ need to be accounted for and the two vertical segments need to be handled properly as well.}
\label{poisson}
\end{figure}

The Poisson summation formula on the real axis reads
\begin{equation}
\label{ps}
2\pi \sum_{k\in\mathbb{Z}}f(x+2\pi k) = \sum_{n\in\mathbb{Z}}e^{inx}\hat{f}(n).
\end{equation}
We are interested here in deriving the analogous formula for arbitrary \emph{complex} $x$. 
Let us evaluate
\begin{equation}
I = \sum_{n\in\mathbb{Z}}e^{i n z }\int_{\mathbb{R}}dx e^{-inx} f(x)
\end{equation}
where $z$ is an arbitrary complex number and $f$ is a complex function evaluated on the real axis.
Firstly, this equals
\begin{equation}
I = \sum_{n\in\mathbb{Z}}e^{i n z}\hat{f}(n),
\end{equation}
where $\hat{f}$ respresents the Fourier transform of $f$. Secondly, we shift the integration contour up to $+i\Im(z)$ as shown in figure \ref{poisson}.
Assuming no contributions from the short strips at infinity, then up to possible pole contributions, the resulting integral equals
\begin{equation}
I = \sum_{n\in\mathbb{Z}}e^{i n \Re(z) }\int_{\mathbb{R}}dx e^{-inx} f(x+i\Im(z)).
\end{equation}
Now, the sum over $n$ can be readily evaluated as
\begin{equation}
\sum_{n\in\mathbb{Z}}e^{i n \Re(z) }e^{-inx} = 2\pi \sum_{k\in\mathbb{Z}}\delta(-x+\Re(z) + 2\pi k).
\end{equation}
We obtain
\begin{equation}
I = 2\pi \sum_{k\in\mathbb{Z}}  f(\Re(z)+i\Im(z) + 2\pi k) = 2\pi \sum_{k\in\mathbb{Z}}  f(z + 2\pi k).
\end{equation}
Denote the collection of simple poles of $f$ whose imaginary part lies between 0 and $\Im(z)$ as $\mathcal{P}$, then the upwards contour shift also produces 
\begin{equation}
2\pi i \sum_{p_{i} \in \mathcal{P}}\text{Res}_{p_i}(f).
\end{equation}
Putting the pieces together, we finally obtain
\begin{equation}
\sum_{n\in\mathbb{Z}}e^{i n z}\hat{f}(n) = 2\pi \sum_{k\in\mathbb{Z}}\left[  f(z + 2\pi k) + i \sum_{p_{i} \in \mathcal{P}}\text{Res}_{p_i}(f)e^{ik(z-p_i)}\right].
\end{equation}
One sees that Poisson's summation formula still holds, up to the series of the second term.\\
In our case, it is precisely this series that we have neglected. 

\subsubsection*{Example}
Let us discuss a small example that will help us rewrite the above formula. We choose $f(z) = \delta(z)$. Obviously the two vertical segments do not contribute. The function also has no simple poles. Hence, the above reasoning yields
\begin{equation}
\sum_{n\in\mathbb{Z}}e^{inz} = 2\pi \sum_{k\in\mathbb{Z}}\delta(z+2\pi k),
\end{equation}
which is an analytic continuation of Dirac's comb function. We can utilize this formula to rewrite the general Poisson summation formula as
\begin{equation}
\label{ps2}
\boxed{
\sum_{n\in\mathbb{Z}}e^{i n z}\hat{f}(n) = 2\pi \sum_{k\in\mathbb{Z}}\left[  f(z + 2\pi k) + 2\pi i \sum_{p_{i} \in \mathcal{P}}\text{Res}_{p_i}(f)\delta(z-p_i + 2\pi k)\right]}.
\end{equation}
This formula is actually a quite nice example of the identity theorem from elementary complex analysis. Using this theorem, one immediately infers that Poisson's summation formula (\ref{ps}) should hold on the entire complex plane except on the poles and across branch cuts of the complex function $f$ (and their $2\pi k$ shifts). The above formula (\ref{ps2}) describes, in a distributional sense, what the correct formula is when incorporating the poles of the function $f$. Note that if branch points are present in $f$, the above formula does not hold, though it is clear what one should do to obtain the correct formula.

\subsection{Elaborate treatment}
\label{Ham3}
Equiped with this knowledge on the correct analytic continuation of Poisson's summation formula, let us now re-analyze the expression (\ref{expre}) in a more rigorous way. For clarity, let us lump together all $l$- and $b$-independent exponentials into a function $F$ and let us call $G=F\rho$ where $\rho$ is the density of states on $H_3^+$ as given by equation (\ref{dosh3}). For fixed $w$ and $p$, we are interested in:
\begin{align}
&\frac{1}{N}\sum_{l}\sum_{b}\sum_{q} \int_{\mathbb{R}^{+}}ds \int_{\mathbb{R}}dm e^{-4\pi\tau_2 s^2/(k-2)} e^{2\pi i \frac{b}{N}q}e^{-i\beta l m + lkw\beta}F(w,p,m,q)\rho(s,m,q) \\
\label{C77}
&= \sum_{l}\sum_{q\in N\mathbb{Z}} \int_{\mathbb{R}^{+}}ds \int_{\mathbb{R}}dm e^{-4\pi\tau_2 s^2/(k-2)} e^{-i\beta l m + lkw\beta}G(w,p,m,q,s) \\
&= \frac{1}{\beta}\sum_{l}\sum_{q\in N\mathbb{Z}} \int_{\mathbb{R}^{+}}ds e^{-4\pi\tau_2 s^2/(k-2)} e^{lkw\beta}\hat{G}(w,p,l,q,s),
\end{align}
where $\hat{G}$ is the Fourier transform of $G(w,p,\frac{m}{\beta},q,s)$. Using naive Poisson summation in $l$, we would get
\begin{align}
\frac{2\pi }{\beta}\sum_{n}\sum_{q\in N\mathbb{Z}} \int_{\mathbb{R}^{+}}ds e^{-4\pi\tau_2 s^2/(k-2)} F\left(w,p,-ikw+\frac{2\pi n }{\beta},q\right)\rho\left(s,-ikw+\frac{2\pi n }{\beta},q\right).
\end{align}
We see from this that we should substitute $m \to -ikw+\frac{2\pi n }{\beta}$ in both $F$ (representing the remaining $l$- and $b$-independent exponentials) and in $\rho$, given by expression (\ref{dosh3}). We hence obtain for the continuous states the following density of states
\begin{equation}
\label{doscone}
\mbox{{\small{$\displaystyle\rho(s,m,q) = 2\left[\frac{1}{2\pi}2\log(\epsilon) + \frac{1}{2\pi i}\frac{d}{2ds}\log\left(\frac{\Gamma(\frac{1}{2} - is - \frac{q}{2} - \frac{i\pi n}{\beta} -\frac{kw}{2})\Gamma(\frac{1}{2} - is -\frac{q}{2} + \frac{i\pi n}{\beta} +\frac{kw}{2})}{\Gamma(\frac{1}{2} + is -\frac{q}{2}- \frac{i\pi n}{\beta} -\frac{kw}{2})\Gamma(\frac{1}{2} + is -\frac{q}{2}+ \frac{i\pi n}{\beta} + \frac{kw}{2})}\right)\right]$}}}.
\end{equation}
This is again the result of our previous naive treatment in \ref{Ham2}. We now know that this is not entirely correct as possible poles might be present. It is known from earlier work on related models that discrete modes typically arise by crossing poles \cite{Maldacena:2000kv}\cite{Hanany:2002ev}\cite{Israel:2003ry}. In full generality, the computations that follow are quite tedious. We will hence first study the simple case where $\tau_1 = q = p = 0$ and $w>0$ to demonstrate the procedure and then slowly `turn up the heat' to work towards the general case.
\subsubsection*{Simplest case as a warm-up}
In this paragraph only we set $\tau_1 = q = p = 0$ and $w>0$. Right before the Poisson resummation, we have the expression (\ref{C77}):\footnote{An overall factor of $e^{\pi k w^2 \tau_2}$ was not written down here. We will reincorporate this factor in the end.}
\begin{align}
\label{C81}
\sum_l \int_{\mathbb{R}^{+}}ds \int_{\mathbb{R}}dm e^{-4\pi\tau_2 s^2/(k-2)} e^{- i \beta l  m + lkw\beta} \rho(s,m)e^{-2\pi i w m \tau_2}.
\end{align}
We now analyze this step by step. The integration over $m$ is on the real axis. Just like in the proof of the analytic continuation of Poisson's summation formula, we wish to shift the contour to imaginary value $-ikw$. The horizontal piece of the resulting contour and its analysis are what we have done above: they generate the continuous spectrum of states.\footnote{This is indeed simply obtained by substituting complex arguments in the real Poisson summation formula and this is what we did in the previous sections.} What we are interested in in this section, is the possibility of a pole in the complex $m$ plane. Can this occur? Obviously the exponentials in expression (\ref{C81}) have no poles. The density of states is given by
\begin{equation}
\label{dosresid}
\rho(s,m) = 2\frac{L}{2\pi}\left[\frac{1}{2\pi}2\log(\epsilon) + \frac{1}{2\pi i}\frac{d}{2ds}\log\left(\frac{\Gamma(\frac{1}{2} - is - im/2)\Gamma(\frac{1}{2} - is + im/2)}{\Gamma(\frac{1}{2} + is - im/2)\Gamma(\frac{1}{2} + is + im/2)}\right)\right].
\end{equation}
The first part is divergent, but has no poles as a function of $m$. This part hence entirely translates to the contour-shifted contribution. The second part however does allow poles. Firstly we split the logarithm in four parts, then we perform the derivative. The result is four terms of the form of a Digamma function, schematically:
\begin{equation}
\frac{\Gamma^{'}}{\Gamma} = \Psi.
\end{equation}
The Gamma function has no zeros. It has simple poles at all negative integers (including zero). The above combination hence has only simple poles when the imaginary part of $m$ equals $\pm 1$, $\pm 3$, $\pm 5$, $\hdots$. The poles then occur for
\begin{equation}
m = \pm 2s + (2n+1)i, \quad n \in \mathbb{Z}.
\end{equation}
The contour and some of the poles are illustrated in figure \ref{contour} below.\footnote{The two vertical contour segments are more problematic. If fact, the original complex function evaluated on the real axis which we started with, has an ill-defined limit for large real values of $m$. It is of the form $ \lim_{m\to\pm\infty}e^{imx}\ln(\left|Cm\right|)$ where the density of states has a logarithmic form for large values of $m$ (and we only care for the functional form of this equation) and $C$ is some constant. This same asymptotic form holds also on the two vertical contour segments. In a distributional sense, such limits are finite though and are equal to zero. In fact, it holds that $\lim_{n\to\infty} e^{inx}f(n) = 0$ as long as $f(n)$ is of order $\mathcal{O}(n^p)$ for large enough $n$ and for some $p\in\mathbb{R}$. 
For completeness, let us provide a small proof of the statement we need. To have $ \lim_{m\to\pm\infty}e^{imx}\ln(\left|Cm\right|) = 0$ for continuous $m$, it needs to hold for every subsequence so we focus on a subsequence $m_n$ for $n\in\mathbb{N}$ which satisfies $\text{lim}_{n\to\infty} m_n = \pm \infty$. For any testfunction $\varphi$ of compact support (say $L$), we then have
\begin{equation}
\left|\int dx e^{im_n x}\ln(\left|Cm_n\right|)\varphi(x)\right| = \left|\int dx e^{im_n x}\frac{\ln(\left|Cm_n\right|)}{m_n}\varphi'(x)\right| \leq \text{max}_{L}(\left|\varphi'\right|)L\left|\frac{\ln(\left|Cm_n\right|)}{m_n}\right|,
\end{equation}
which goes to zero as $n$ goes to infinity.}
\begin{figure}[h!!!!]
\centering
\includegraphics[width=8cm]{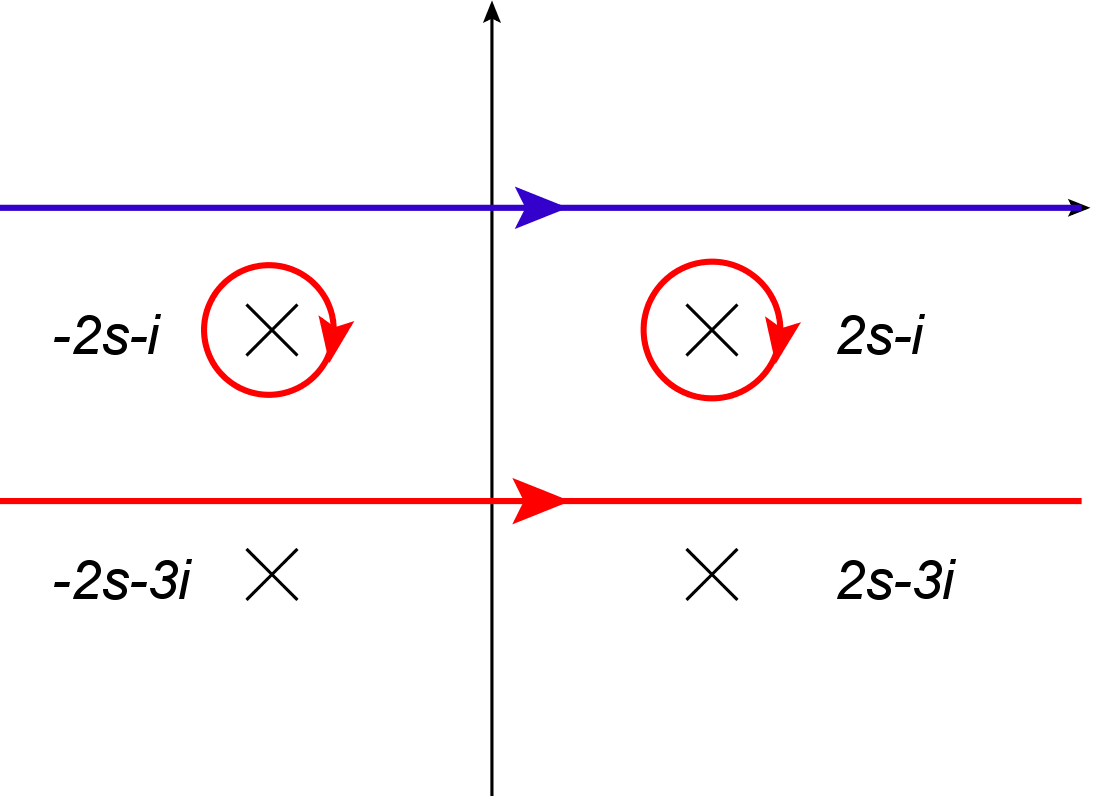}
\caption{The complex $m$ plane with the original integration contour in blue. The shifted contour is drawn in red. Two poles are crossed when $1< kw<3$ as is illustrated.}
\label{contour}
\end{figure}
We hence see that if $kw<1$, no poles are crossed and the analysis presented in \ref{Ham2} remains valid. However, when $kw>1$, at least two poles are present in the region and discrete states appear. Both poles have the same residue and one readily finds that the part of the residue coming from the density of states (\ref{dosresid}) equals\footnote{We again set $\frac{L}{P} = \beta$, with $P$ introduced in subsection \ref{Ham1}.} 
\begin{equation}
-2\pi i \text{Res}\rho(s,m) = -2\pi i 2\frac{\beta}{2\pi} \left(\frac{-2}{4\pi i}\right) = \frac{\beta}{\pi}.
\end{equation}
In the remainder of this paragraph, we focus on the case where exactly two poles are crossed. We generalize this in the next paragraphs. Next we sum the residues of both poles. These differ only in the sign preceding $s$, so schematically we can write
\begin{equation}
\int_{0}^{+\infty}ds e^{As}e^{-Bs^2} + \int_{0}^{+\infty}ds e^{-As}e^{-Bs^2} = \int_{-\infty}^{+\infty}ds e^{As}e^{-Bs^2},
\end{equation}
and both poles are taken care of simultaneously by simply integrating $s$ over the entire real axis. We obtain
\begin{equation}
\frac{\beta}{\pi}\sum_l \int_{\mathbb{R}}ds e^{-4\pi\tau_2 s^2/(k-2)} e^{- i \beta l  (2s-i) + lkw\beta} e^{-2\pi i w (2s-i) \tau_2}.
\end{equation}
The integral over $s$ is a simple Gaussian, yielding
\begin{equation}
\frac{\beta}{2\pi}\sqrt{\frac{k-2}{\tau_2}}\sum_l e^{\beta l(kw-1)} e^{-2\pi w \tau_2}e^{-\frac{(\beta l+2\pi w \tau_2)^2(k-2)}{4\pi\tau_2}}.
\end{equation}
A last Poisson resummation will yield the desired result. For the reader's comfort, we write down the Poisson resummation formula:
\begin{equation}
\label{PRformula}
\sum_{l\in\mathbb{Z}}\exp\left[-\pi a l^2 + 2\pi i b l\right] = a^{-1/2}\sum_{n\in\mathbb{Z}}\exp\left[-\pi\frac{(n-b)^2}{a}\right].
\end{equation}
One then finds
\begin{equation}
\sum_{n\in\mathbb{Z}} e^{-\frac{4\pi^3 n^2}{\beta^2(k-2)}\tau_2}e^{-\frac{4\pi^2 i (kw-1)n}{\beta (k-2)}\tau_2}e^{\frac{4\pi^2 i w n}{\beta}\tau_2}e^{\pi \frac{(kw-1)^2}{k-2}\tau_2}e^{-2\pi kw^2\tau_2}.
\end{equation}
From this, one needs to distill a factor $e^{\frac{\pi\tau_2}{k-2}}$ to serve as (part of) the central charge factor in the partition function. Extracting this and including the extra piece $e^{\pi k w^2 \tau_2}$ we inserted in the operator trace (\ref{expre}), one can see that the following conformal weights can be read off:
\begin{equation}
h = \bar{h} = -\frac{(kw-1)^2}{4(k-2)} + \frac{1}{4(k-2)} + \frac{\pi^2n^2}{\beta^2 (k-2)} + \frac{\pi i (kw-1) n }{\beta (k-2)} - \frac{\pi i w n }{\beta} + \frac{kw^2}{4}.
\end{equation}
We will rewrite this in a more clear way after we incorporate the other quantum numbers.

\subsubsection*{The general case for $w>0$}
\begin{figure}[h]
\includegraphics[width=15cm]{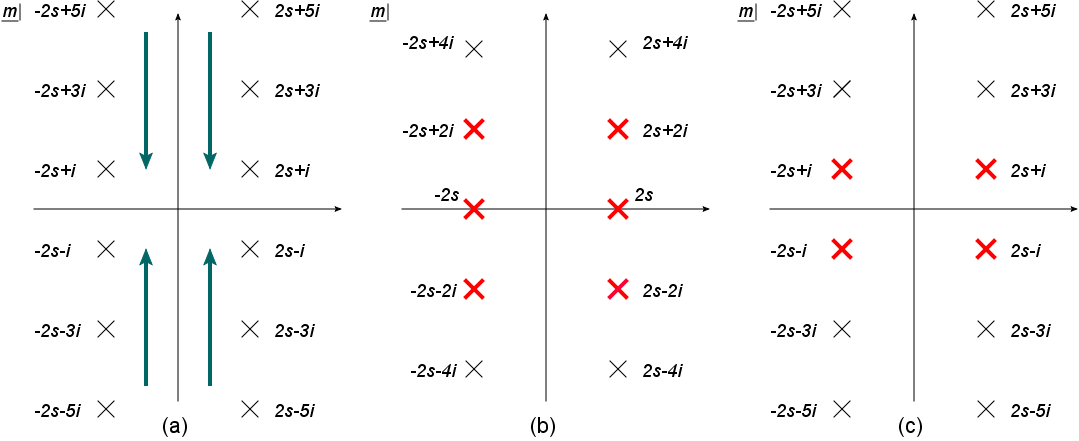}
\caption{(a) Poles in the complex $m$ plane for $q=0$. The big green arrows depict how the poles shift when $q>0$. (b) Poles when $q=3$. The bold red poles are doubly degenerate. The residues cancel out and the pole is effectively absent alltogether. General odd values of $q$ produce the same pattern of poles but with more (or less) poles doubly degenerate and hence absent. (c) Poles when $q=2$. All even values of $q$ produce the same pole pattern with again more poles doubly degenerate for increasing $q$.}
\label{polen}
\end{figure}

We now turn to the general case. We need to incorporate non-zero $\tau_1$, $p$ and $q$ quantum numbers, though we still focus on $w>0$. From now on, we also allow a general number of crossed poles. Let us first take a look at non-zero $q$ quantum numbers since these present the most elaborate modifications. The effect of $q$ is to shift the location of the poles of the Gamma function. The density of states is given by
\begin{equation}
\mbox{{\small{$\displaystyle\rho(s,m,q) = 2\frac{L}{2\pi}\left[\frac{1}{2\pi}2\log(\epsilon) + \frac{1}{2\pi i}\frac{d}{2ds}\log\left(\frac{\Gamma(\frac{1}{2} - is -q/2 - im/2)\Gamma(\frac{1}{2} - is -q/2+ im/2)}{\Gamma(\frac{1}{2} + is -q/2- im/2)\Gamma(\frac{1}{2} + is -q/2+ im/2)}\right)\right]$}}}.
\end{equation}
Poles can be found whenever 
\begin{align}
m &= \pm 2s + (-(2n+1)+q)i, \quad n \in \mathbb{N}, \\
m &= \pm 2s + ((2n+1)-q)i, \quad n \in \mathbb{N},
\end{align}
where $n$ equals $\left\{0, 1, 2 , \hdots\right\}$. The poles hence shift as shown in the figure \ref{polen}(a).\\
Poles originally in the upper half plane shift downwards (for positive $q$) and poles originally in the lower half plane shift upwards.\\
Since $q$ is an integer, one of two situations can occur. 
\begin{itemize}
\item{$q$ is odd. The poles go halfway in between where they are at $q=0$. However, depending on the value of $q$, several poles closest to the real axis become `degenerate'. Computing the residue at these double poles shows that they cancel. The situation is illustrated in figure \ref{polen}(b).}
\item{$q$ is even. The resulting set of poles is exactly equal to those with $q=0$. Likewise, multiply degenerate poles can occur and if they do, the residue becomes zero. The situation is illustrated in figure \ref{polen}(c).}
\end{itemize}
For $q$ negative, the poles in the upper and lower half plane move away from each other. The resulting picture of poles is actually the same as that for $-q$ (which is positive). Hence we can combine both case of $q$ and only consider negative $q$ and the resulting set of poles is then labeled by a positive (or zero) integer $l$, the quantum number associated to the $SL(2,\mathbb{R})$ discrete representation.
The net effect is thus to replace $+q/2 \to -\left|q\right|/2$. \\
The computation is not that hard, though it is important to keep track of all different contributions. Therefore let us first write down a list of every contribution we have. Firstly, we have explicit factors inserted in the operator trace (\ref{expre}). These are
\begin{align}
\label{remaining}
e^{\pi k w^2 \tau_2}e^{-p\tau_1 kw \beta}e^{-\frac{\beta^2kp^2}{4\pi}\tau_2}e^{lkw\beta}.
\end{align}
Secondly, the contributions from the $e^{2\pi i UJ}$ factors give
\begin{align}
\label{remaining2}
e^{2\pi i qw \tau_1}e^{-4\pi\tau_2\left(\frac{-ipq\beta}{4\pi}\right)}
\end{align}
and
\begin{align}
e^{-i\beta l m}e^{-2\pi i w m \tau_2}e^{p\beta m i \tau_1}.
\end{align}
The poles we cross are of the form $m = \pm2s - (odd)i - \left|q\right|i$ with $odd$ an odd integer $ = 1,3,5,\hdots$. We start with
\begin{align}
\sum_l \int_{\mathbb{R}}ds \int_{\mathbb{R}}dm e^{-4\pi\tau_2 s^2/(k-2)} e^{- i \beta l  m + lkw\beta} \rho(s,m)e^{-2\pi i w m \tau_2}e^{p\beta i m \tau_1}.
\end{align}
The first step is extracting the residues of this expression for the poles of the $m$-integral. After this, we perform the Gaussian integral. The result of this step is
\begin{align}
\frac{\beta}{2\pi}\sum_l e^{- \beta l  (odd + \left|q\right|) + lkw\beta} e^{-2\pi w (odd + \left|q\right|) \tau_2}e^{p\beta (odd + \left|q\right|) \tau_1}e^{-\frac{(k-2)}{4\pi\tau_2}(\beta l + 2\pi w \tau_2 - \beta p \tau_1)^2}.
\end{align}
Finally, we Poisson resum this expression. The parameters to be used in formula (\ref{PRformula}) are
\begin{align}
a &= \frac{\beta^2 (k-2)}{4\pi^2\tau_2}, \\
b &= \frac{\beta (kw- odd - \left|q\right|)}{2\pi i} - \frac{(k-2)\beta}{4\pi^2 i \tau_2}\left[2\pi w \tau_2 - \beta p \tau_1\right].
\end{align}
A careful, but straightforward analysis of the resulting factors in combination with the remaining prefactors written in (\ref{remaining}) and (\ref{remaining2}) shows that
\begin{align}
h &= -\frac{\tilde{j}(\tilde{j}-1)}{k-2} + \frac{qw}{2} - \frac{\pi i w n }{\beta} + \frac{kw^2}{4} - \frac{i \beta  pq}{4\pi} - \frac{pn}{2} + \frac{kp^2\beta^2}{4(2\pi)^2}, \\
\bar{h} &= -\frac{\tilde{j}(\tilde{j}-1)}{k-2} - \frac{qw}{2} - \frac{\pi i w n }{\beta} + \frac{kw^2}{4}  - \frac{i \beta  pq}{4\pi} + \frac{pn}{2} + \frac{kp^2\beta^2}{4(2\pi)^2},
\end{align}
where $\tilde{j} = \tilde{m} -l = \frac{kw}{2} - \frac{\left|q\right|}{2} - \frac{i\pi n }{\beta} - l$ where $l=0,1,2,\hdots$. The relation between the parameter $odd$, labeling the poles, and the $SL(2,\mathbb{R})$ parameter $l$ is $ l = \frac{odd -1}{2}$.

\subsubsection*{Negative $w$}
Now we briefly mention the differences for the case $w<0$. In this case, the contour needs to be shifted to the upper half plane. The pole also has its $q$-contribution reversed.\\
A first feature is the overall sign: the contour surrounding the poles is oriented in the opposite direction as before, but also the residue itself has the opposite sign. In all, no overall sign is present.\\
The changes with respect to the case $w>0$ are that for the pole contributions: $kw-odd \to kw+odd$ and $q\to -q$.\footnote{This swap of sign of $q$ is only present obviously for the factors originating from an $m$ quantum number, and not for the factors coming from $e^{2\pi i U J}$ contributions.}

\subsubsection*{Everything combined}
In all, the conformal weights in the most general case are given by
\begin{align}
h &= -\frac{\tilde{j}(\tilde{j}-1)}{k-2} + \frac{qw}{2} - \frac{\pi i w n }{\beta} + \frac{kw^2}{4} - \frac{i \beta  pq}{4\pi} - \frac{pn}{2} + \frac{kp^2\beta^2}{4(2\pi)^2}, \\
\bar{h} &= -\frac{\tilde{j}(\tilde{j}-1)}{k-2} - \frac{qw}{2} - \frac{\pi i w n }{\beta} + \frac{kw^2}{4}  - \frac{i \beta  pq}{4\pi} + \frac{pn}{2} + \frac{kp^2\beta^2}{4(2\pi)^2},
\end{align}
where $\tilde{j} = M - l= \frac{k\left|w\right|}{2} - \frac{\left|q\right|}{2} \pm \frac{i\pi n }{\beta} - l$ where $l=0,1,2,\hdots$ and the $\pm$ symbol equals $+$ if $w<0$ and $-$ if $w>0$. \\
A simple substitution $n \to -n$ allows us to compare these expressions to the continuous weights of equations (\ref{ads3spectrum3}) and (\ref{ads3spectrum4}): all terms are the same except the first one.\\
Since we have a set of discrete states, let us see whether they satisfy the `improved' unitarity constraints \cite{Maldacena:2000hw}: $ \frac{1}{2} < \tilde{j} < \frac{k-1}{2}$ with $\tilde{j}=\frac{k\left|w\right|}{2} - \frac{\left|q\right|}{2} \pm \frac{i\pi n }{\beta} - l$. Since $\tilde{j}$ is a complex quantity, we consider instead $\Re(\tilde{j})$ and it is this number that obeys the inequality in our case as we now show.\\
The first inequality $\frac{1}{2} < \Re(\tilde{j})$ corresponds precisely to the pole-crossing properties of the contour and is hence indeed satisfied in our case. \\
The second inequality $\Re(\tilde{j}) < \frac{k-1}{2}$ requires some input from the infinite product. We have previously argued that (due to brute force numerical computations) $\left|w\right| < 1/2$. It is clear that the worst case scenario for this inequality occurs when $\left|w\right| = 1/2$ and $q=l=0$. But then we have
\begin{equation}
\frac{k}{4} < \frac{k-1}{2} \quad \Leftrightarrow \quad k>2,
\end{equation}
which is obviously satisfied.\footnote{Note that the less strict upper bound $j < k/2$ which follows from the no-ghost theorem would yield an inequality in our case that is satisfied as long as $\left|w\right| < 1$.} Thus every discrete state we constructed as a pole that was crossed by the contour shift, satisfies indeed the unitarity constraints.\\
When taking a larger perspective on this derivation, we find it quite remarkable to find discrete representations, since our original starting point used only the complete set of continuous representations on $H_3^+$. Somehow, these wavefunctions `know' in advance what the discrete representations should look like.

\subsection{Chemical potential}
\label{chemapp}
For the sake of brevity, we will only discuss here the additional steps required compared to the derivations presented in the above analysis. \\
Firstly, we have
\begin{align}
U_{lp} &= -\frac{i\beta}{2\pi}(p\tau - l)(1+i\mu), \\
\Im(U_{lp}) &= - \frac{\beta}{2\pi}(p\tau_1-l) + \frac{\mu p \beta}{2\pi} \tau_2.
\end{align}
Two aspects should be taken care of to evaluate such partition functions. Firstly, one can relate the partition function to that of the conical spaces discussed in the previous subsections by taking $w \to \frac{\mu\beta}{2\pi}p$ in that analysis. The $q$ quantum number however runs over $\mathbb{Z}$ and not $N\mathbb{Z}$ in this case. Secondly, one should add an extra $l$-dependent exponential
\begin{equation}
\label{extra}
e^{-i\beta \mu ql},
\end{equation}
compared to the above analysis. The trace we should evaluate to agree with the path integral derivation is then
\begin{align}
\text{Tr}&\left[\exp\left(-(p\tau_1-l)k\frac{\mu p \beta}{2\pi}\beta+ \pi k\left(\frac{\mu p \beta}{2\pi}\right)^2\tau_2\right)\exp\left(-\tau_2\frac{\beta^2 k p^2}{4\pi}\right)\right. \nonumber \\
&\quad\quad\quad\quad \left.\times\exp\left(4\pi\tau_2\frac{\Delta+1/4}{k-2}\right)\exp(2\pi i (U_{lp} J^{3}_0 + \bar{U}_{lp} \overline{J}^{3}_0))\right] \\
 &=\frac{\beta\sqrt{k-2}}{8\pi\sqrt{\tau_2}}e^{- k\beta^2\left|l-p\tau\right|^2/4\pi\tau_2 + 2\pi \Im(U_{lp})^2/\tau_2 }\left|\sin(\pi U_{lp})\right|^{-2}.
\end{align}
Again we first analyze the continuous representations, obtained by the naive analytic continuation of Poisson's summation formula. The projection conditions now require
\begin{align}
&J_0^3 + \overline{J}_0^3 \in \mathbb{Z}, \\
&\frac{i\beta}{2\pi}(J_0^3 - \overline{J}_0^3) - \frac{\beta \mu}{2\pi}(J_0^3 + \overline{J}_0^3) + \frac{k\beta}{2\pi i}\frac{\mu p \beta}{2\pi} \in \mathbb{Z},
\end{align}
which is solved by
\begin{align}
m_{J} &= \frac{q}{2}(1-i\mu) + \frac{i\pi n }{\beta} + \frac{k\mu p \beta}{4\pi}, \\
\overline{m}_{J} &= \frac{q}{2}(1+i\mu) - \frac{i\pi n }{\beta} - \frac{k\mu p \beta}{4\pi},
\end{align}
for $q,n \in \mathbb{Z}$. Again these states are associated with the horizontal part of the shifted contour. The additional factor (\ref{extra}), although $l$-dependent, is of modulus one and hence it does not affect the location of the shifted contour. For the continuous states, this is relevant since the location of the shifted contour also dictates the substitution one needs to do in the density of states. Hence we see that here we can simply take the density of states (\ref{doscone}) with the replacement $w \to \frac{\mu\beta}{2\pi}p$.
The analysis of the resulting conformal weights is identical to the analysis for the angular orbifolds presented in \ref{Ham2}, except one extra term corresponding to the $\mp i \mu \frac{q}{2}$ in the above relations. This final term gives the corrections to the conformal weights
\begin{align}
h_{extra} &= -\frac{\beta \mu qp}{4\pi} - \frac{i\mu^2 qp\beta}{4\pi}, \\
\bar{h}_{extra} &= \frac{\beta \mu qp}{4\pi} - \frac{i\mu^2 qp\beta}{4\pi},
\end{align}
finally yielding
\begin{align}
h^{p}_{jqn} &= \frac{s^2 +1/4}{k-2} + i\frac{\mu n p}{2} - i\frac{qp\beta}{4\pi} + \frac{ p n}{2} + \frac{kp^2\beta^2}{4(2\pi)^2}(1+\mu^2) - i \frac{\mu^2\beta qp}{4\pi}+ h_{int}, \\
\bar{h}^{p}_{jqn} &= \frac{s^2 +1/4}{k-2} + i\frac{\mu n p}{2} - i\frac{qp\beta}{4\pi} - \frac{ p n}{2} + \frac{kp^2\beta^2}{4(2\pi)^2}(1+\mu^2) - i \frac{\mu^2\beta qp}{4\pi} + \bar{h}_{int}.
\end{align}
As a check, we see that $h - \bar{h} \in \mathbb{Z}$, a necessary condition for modular invariance. \\

\noindent Discrete states can be found by the same strategy as the one used before. Again, we only need to take a closer look at one contribution (\ref{extra}) while the remaining terms can be readily found from the conical spaces by taking $w \to \frac{\mu\beta}{2\pi}p$. 
One can then go through exactly the same computations as before to handle the discrete states. The extra exponential (\ref{extra}) simply goes along for the ride during the computations and only makes its appearance when the Poisson resummation needs to be applied. The new Poisson resummation parameters to be used in (\ref{PRformula}) are now
\begin{equation}
a = \frac{\beta^2(k-2)}{4\pi^2\tau_2}, \quad b = \frac{\beta(kw-odd - \left|q\right|- i\mu q))}{2\pi i } - \frac{(k-2)\beta}{4 \pi^2 i \tau_2}\left[2\pi w \tau_2 - \beta p \tau_1\right].
\end{equation}
The only difference is hence the replacement $\left|q\right| \to \left|q\right| + i\mu q$. Without going into details, we report the final result:
\begin{align}
h^{p}_{jqn} &= -\frac{\tilde{j}(\tilde{j}-1)}{k-2} - i\frac{\mu n p}{2} - i\frac{qp\beta}{4\pi} - \frac{ p n}{2} + \frac{kp^2\beta^2}{4(2\pi)^2}(1+\mu^2) - i \frac{\mu^2\beta qp}{4\pi}+ h_{int}, \\
\bar{h}^{p}_{jqn} &= -\frac{\tilde{j}(\tilde{j}-1)}{k-2} - i\frac{\mu n p}{2} - i\frac{qp\beta}{4\pi} + \frac{ p n}{2} + \frac{kp^2\beta^2}{4(2\pi)^2}(1+\mu^2) - i \frac{\mu^2\beta qp}{4\pi} + \bar{h}_{int},
\end{align}
where now $\tilde{j} = \frac{k\left|\mu p\right|\beta}{4\pi} - \frac{\left|q\right|}{2} - \frac{i\mu q}{2} \pm \frac{in\beta}{\pi} - l$. Again the imaginary part of $\tilde{j}$ is irrelevant for satisfying the unitarity constraints. We also note that the $+ i\mu q/2$ term does not contain an absolute value.

\end{document}